\documentclass[11pt]{article}
\usepackage[dvips]{graphicx}
\newcommand{\ket}[1]{\left | #1 \right \rangle}
\newcommand{\bra}[1]{\left \langle #1 \right |}

\def\openone{\leavevmode\hbox{\small1\kern-3.8pt\normalsize1}}
\def\RR{{\rm I\kern-.2emR}}

\newcommand{\beq}{\begin{equation}}
\newcommand{\eeq}{\end{equation}}
\newcommand{\beqa}{\begin{eqnarray}}
\newcommand{\eeqa}{\end{eqnarray}}

\def \bmatrix#1{ \left [ \matrix{#1} \right ] }

\begin{document}
\begin{center}
{\Large\bf Relative Entropy and Single Qubit
Holevo-Schumacher-Westmoreland Channel Capacity}\\
\bigskip
{\normalsize John Cortese}\\
\bigskip
{\small\it Institute for Quantum Information\\
Physics Department, 
California Institute of Technology 103-33,\\ 
Pasadena, CA 91125 U.S.A.}
\\[4mm]
\date{today}
\end{center}

\begin{center}
\today
\end{center}

\begin{abstract}
The relative entropy description of Holevo-Schumacher-Westmoreland
(HSW) classical channel capacities is applied
to single qubit quantum channels. A simple formula for the relative entropy 
of qubit density matrices in the Bloch sphere representation is derived.
The formula is combined with 
the King-Ruskai-Szarek-Werner qubit channel ellipsoid picture
to analyze several unital and non-unital qubit
channels in detail.
An alternate proof is presented 
that the optimal HSW signalling states for 
single qubit unital channels are those states
with minimal channel output 
entropy. The derivation is 
based on symmetries of the relative 
entropy formula, and the  
King-Ruskai-Szarek-Werner qubit channel ellipsoid picture. 
A proof is given that the average output density matrix of 
any set of optimal HSW signalling states for a 
( qubit or non-qubit ) quantum channel is unique.  

\end{abstract}


\tableofcontents

\pagebreak

\section{Introduction}

In 1999, Benjamin Schumacher and Michael Westmoreland published a paper entitled
\linebreak
{\em Optimal Signal Ensembles} \cite{Schumacher99}
that elegantly described the classical (product
state) channel capacity of quantum channels in terms of a function known as the
relative entropy. 
Building upon this view, we study single qubit channels, adding the
following two items to the Schumacher-Westmoreland analysis.

I) A detailed understanding of the convex hull shape of the set of 
quantum states
output by a channel. (The fact the set was convex has been known for
some time, but the detailed nature of the convex geometry was unknown
until recently.) 

II) A useful mathematical representation (formula) for the relative entropy
function, $\mathcal{D}(\, \rho \, \| \, \phi \, )$, when both $\rho$
and $\phi$ are single qubit density matrices. 

For single qubit channels, the work of King, Ruskai, Szarek, and Werner
has provided a concise description of the convex hull set\cite{Ruskai99a,rsw}.
In this paper, we
derive a useful formula for the relative entropy between qubit density matrices.
Combining this formula with the KRSW convexity information, we present 
from a relative entropy perspective 
several results, some previously known, and others new, related to 
the (product state) classical channel capacity of quantum channels. 
These include :

I) The average output density matrix for {\em any} optimal set of signalling
states that achieves the maximum classical channel capacity for a quantum channel 
is unique. For single qubit unital channels, Donalds equality leads to a symmetry which
tells us this average density matrix
must be $\frac{1}{2} \; \mathcal{I}$. 
This fact about the average density matrix 
allows us to conclude for unital qubit channels 
that the optimum signalling states are a subset of the states with minimum output
von Neumann entropy, as previously shown in \cite{Ruskai99a} .  
This symmetry also allows us to see why
only two orthogonal signalling states are needed to achieve the 
optimum classical channel capacity for single qubit unital channels,
and why the a priori probabilities for these two signalling 
states are $\frac{1}{2}$.  

II) The single qubit relative entropy formula allows us to understand 
geometrically why
the a priori probabilities for optimum signalling states for 
non-unital single qubit channels   
are not equal. 

III) Examples of channels which require non-orthogonal signalling states 
to achieve
optimal classical channel capacity are given. Such channels have been 
found before.
Here these channels are presented in a geometrical fashion based 
on the relative entropy formula
derived in Appendix A. 

\section{Background}
\subsection{Classical Communication over Classical and Quantum Channels}

\vspace{0.2in}

\setlength{\unitlength}{0.00033300in}%
\begingroup\makeatletter\ifx\SetFigFont\undefined
\def\x#1#2#3#4#5#6#7\relax{\def\x{#1#2#3#4#5#6}}%
\expandafter\x\fmtname xxxxxx\relax \def\y{splain}%
\ifx\x\y   
\gdef\SetFigFont#1#2#3{%
  \ifnum #1<17\tiny\else \ifnum #1<20\small\else
  \ifnum #1<24\normalsize\else \ifnum #1<29\large\else
  \ifnum #1<34\Large\else \ifnum #1<41\LARGE\else
     \huge\fi\fi\fi\fi\fi\fi
  \csname #3\endcsname}%
\else
\gdef\SetFigFont#1#2#3{\begingroup
  \count@#1\relax \ifnum 25<\count@\count@25\fi
  \def\x{\endgroup\@setsize\SetFigFont{#2pt}}%
  \expandafter\x
    \csname \romannumeral\the\count@ pt\expandafter\endcsname
    \csname @\romannumeral\the\count@ pt\endcsname
  \csname #3\endcsname}%
\fi
\fi\endgroup
\begin{picture}(15904,5856)(301,-7288)
\thicklines
\put(5551,-4486){\framebox(1725,975){}}
\put(8401,-4561){\framebox(2550,1050){}}
\put(7276,-3886){\vector( 1, 0){1050}}
\put(10951,-3961){\vector( 1, 0){675}}
\put(14401,-3961){\vector( 1, 0){1750}}
\put(4176,-3811){\vector( 1, 0){1375}}
\put(1801,-3811){\vector( 1, 0){2375}}
\put(4176,-6211){\framebox(10175,4725){}}
\put(11701,-4561){\framebox(1850,1050){}}
\put(13526,-3961){\vector( 1, 0){775}}
\put(15301,-2536){\makebox(0,0)[lb]{\smash{\SetFigFont{10}{12.0}{rm}Classical}}}
\put(15301,-3106){\makebox(0,0)[lb]{\smash{\SetFigFont{10}{12.0}{rm}Outputs }}}
\put(15526,-5026){\makebox(0,0)[lb]{\smash{\SetFigFont{10}{12.0}{rm}$Y_j$}}}
\put(751,-4951){\makebox(0,0)[lb]{\smash{\SetFigFont{10}{12.0}{rm}$X_i$}}}
\put(6826,-2461){\makebox(0,0)[lb]{\smash{\SetFigFont{10}{12.0}{rm}Quantum Channel Domain}}}
\put(12001,-4336){\makebox(0,0)[lb]{\smash{\SetFigFont{5}{6.0}{rm}( POVM ) }}}
\put(376,-3886){\makebox(0,0)[lb]{\smash{\SetFigFont{10}{12.0}{rm}Inputs}}}
\put(6826,-7261){\makebox(0,0)[lb]{\smash{\SetFigFont{14}{16.8}{rm}Classical Channel }}}
\put(301,-3361){\makebox(0,0)[lb]{\smash{\SetFigFont{10}{12.0}{rm}Classical }}}
\put(5701,-3961){\makebox(0,0)[lb]{\smash{\SetFigFont{10}{12.0}{rm}Encode }}}
\put(6076,-4336){\makebox(0,0)[lb]{\smash{\SetFigFont{10}{12.0}{rm}$\psi_i$}}}
\put(8701,-3886){\makebox(0,0)[lb]{\smash{\SetFigFont{10}{12.0}{rm}Quantum }}}
\put(8701,-4336){\makebox(0,0)[lb]{\smash{\SetFigFont{10}{12.0}{rm}Channel}}}
\put(12001,-3886){\makebox(0,0)[lb]{\smash{\SetFigFont{10}{12.0}{rm}Decode}}}
\end{picture}

\begin{center}
Figure 1: 
Transmission of classical information through a quantum channel. 
\end{center}

This paper discusses the transmission
of classical information over quantum channels
with no prior entanglement between the sender (Alice) and the
recipient (Bob). In such a scenario, classical information is encoded 
into a set of quantum states $\psi_i$. These states are transmitted
over a quantum channel. The perturbations encountered by the signals
while transiting the channel are described using the Kraus
representation formalism. A receiver at the channel output measures
the perturbed quantum states using a POVM set. The resulting classical
measurement outcomes represent the extraction of classical information
from the channel output quantum states.

There are two common criteria for measuring the quality 
of the transmission
of classical information over a channel, regardless of whether the
channel is classical or quantum. These criteria are 
the (Product State) Channel Capacity\cite{Holevo98a,Hausladen96a,Schumacher97c} 
and the Probability of Error (Pe).\cite{Fuchs96a} In this paper, we shall
focus on the first criterion, the Classical Information Capacity of a
Quantum Channel, $\mathcal{C}$. 

In determining the classical channel capacity,
we typically have an input signal constellation consisting of 
classical signals $x_i$\cite{Cover91a}. The classical channel capacity 
$\mathcal{C}$ is defined as \cite{Cover91a}:

$$
\mathcal{C} \;=\; 
\mbox{\Large Max}
_{\{all \;possible\; x_i\} } \quad  H(X) \;-\; H(\,X \, | \, Y
) 
$$

Here $H(X)$ is the Shannon entropy for the discrete random variable $X$. 
\linebreak 
$X \;\equiv \; \{ \; p_i \,=\, prob(x_i) \; \}\, , \, i \, = \, 1 \,,\, \cdots \,,\,  N$.  
The Shannon entropy $H(X)$ is defined as 
$H(X) \;=\; - \, \sum_{i=1}^N \; p_i \, \log ( \,p_i \, )$.  
For conditional random variables, we denote 
the probability of the random variable
$X$ given $Y$ as $p( X | Y )$. The corresponding
conditional Shannon entropy is defined as 
$H(X|Y) \;=\; - \, 
\sum_{i=1}^{N_X} \,  
\sum_{j=1}^{N_Y} \; p( x_i \, , \, y_j \, ) \;  \log [ \,p( \, x_i \,|\, y_j \, )\, ] $.  
Our entropy calculations shall be in bits, so $\log_2$ is used.
 
Suppose we have $|X|$ linearly independent and equiprobable input signals
$x_i$, and possible output signals $y_j$,  with $|Y| \geq |X|$. 
If there is no noise in the channel, 
then $\mathcal{C} \;=\; \log(|X|)$. Noise in the channel increases the
uncertainty in X given the channel output Y, and thus noise increases 
$H(\,X \, | \, Y)$, thereby decreasing $\mathcal{C}$ for fixed $ H(X)$.
Geometrically, the presence of random channel noise causes the channel
mapping $x_i \;\rightarrow\;y_j$ to change from a noiseless
one-to-one relationship, to a stochastic map. We say the possible
channel mappings of $x_i$ diffuse, occupying a region $\theta_i$
instead of a single unique state $y_j$.
As long as the regions $\theta_{i}$ have
disjoint support, the receiver can use Y to distinguish 
which X was sent. In this disjoint support
case, $ H(\,X \, | \, Y) \;\approx \; 0$
and $\mathcal{C} \;\approx\; H(X)$. This picture is frequently referred
to as sphere packing, since we view the diffused output signals as
roughly a sphere around the point in the output space where the
signals would
have been deposited had the channel introduced no perturbations. The
greater the channel noise, the greater the radius of the spheres. If
these spheres can be packed into a specified volume without
significant overlap, then the decoder can distinguish the input state
transmitted by determining which output sphere the decoded
signal falls into. 

For sending classical information over a quantum channel, we adhere to
the same picture. We seek to maximize $H(\, X\, )$ and 
minimize $H(\, X\,|\, Y\, )$, in
order to maximize the channel capacity $\mathcal{C}$. We encode each
classical input signal state $\{ \, x_i\, \} $ into a corresponding 
quantum state $\psi_i$. Sending $\psi_i$ through the channel,
the POVM decoder seeks to predict which $x_i$ was originally sent. 
Similar to the classical picture, the quantum channel will diffuse or
smear out the density matrix $\rho_i$ corresponding to the quantum
state $\psi_i$ as the quantum state passes through the channel.
The resulting channel output density matrix $\mathcal{E}(\rho_i)$ will have
support over a subspace $\phi_i$. As long as all the regions $\phi_i$ have
disjoint support,
the POVM based decoder will be able to distinguish which quantum state
$\rho_i$ entered the channel, and hence $H(X|Y) \; \approx \;  0$,
yielding $\mathcal{C} \;\approx\; H(X)$. 

For the classical capacity of quantum channels, we 
encode classical binary data into quantum states.
The product state classical capacity for a quantum channel 
maximizes channel
throughput by encoding a long block of m
classical bits $x_i$ into a long block consisting
of a tensor product of n single qubit
quantum states $\psi_j$ in an optimal manner which maximizes (product state) classical channel capacity. 

$$
\{ x_1, x_2, \; \cdots \; x_m \} \;\rightarrow \; 
\psi_1 \;\otimes\; \psi_2 \;\otimes\; \cdots \; \otimes \; \psi_n
$$

It has been widely conjectured, but not proven,
that the product state classical channel capacity of a quantum channel
is the classical capacity of a quantum channel. 

The Holevo-Schumacher-Westmoreland Theorem tells us 
that the classical product state channel capacity 
using the above encoding scheme is given by the Holevo 
quantity \mbox{\Large $\chi$} of the output signal ensemble, 
maximized over a single copy of all
possible input signal ensembles $\{ p_i\,,\, \rho_i\}$\cite{Nielsen00a}.

$$ \mathcal{C}_1 \;=\; \mbox{\Large Max}_{\{all \;possible\; p_i \;and \; \rho_i \}  } \quad
\mbox{\huge $\chi$}_{output} $$

$$ \qquad \;=\;
\mbox{\Large Max}_{\{all \;possible\; p_i \;and \; \rho_i \}  } \quad
\mathcal{S} \left ( \; \mathcal{E} \left (\sum_{i} \; p_i \, \rho_i \right )
\; \right ) \;-\; \sum_{i} \; 
p_i \, \mathcal{S}\left ( \; \mathcal{E}\left ( \rho_i \right ) \; \right ) $$

$\mathcal{S}(-)$ above is the von Neumann entropy.
The symbol $\mathcal{E}(\rho)$
represents the output density matrix obtained by presenting 
the density matrix $\rho$ at the channel input. Furthermore, 
the input signals $\rho_i$ can be
chosen to be pure states without affecting the 
maximization\cite{Nielsen00a}. 
Hereafter we shall call $\mathcal{C}_1$ defined above the
Holevo-Schumacher-Westmoreland (HSW) channel capacity.  

\vspace{0.2in}

\subsection{Relative Entropy and HSW Channel Capacity}
\label{schuwest}

An alternate, but equivalent, description of HSW channel capacity
can be made using relative entropy\cite{Schumacher99}.
The relative entropy $\mathcal{D}$ of two
density matrices, $\varrho$ and $\phi$, is defined 
as \cite{Schumacher99,Nielsen00a,Schumacher00a,Petz93} :

$$
\mathcal{D}( \, \varrho \,  \| \,  \phi \,  ) \;=\;
Tr \left [ \,  \varrho \, \log ( \, \varrho \, ) \;-\; \varrho \, \log( \,\phi \,) 
\right ]
$$

Here Tr[-] is the trace operator. 
Klein's inequality tells us that $\mathcal{D} \; \geq \; 0$,
with $\mathcal{D} \; \equiv \; 0$ iff $\varrho \;\equiv\;\phi$
\cite{Nielsen00a}. Note that we shall usually take our 
logarithms to be base 2.

To see how to represent $\chi$ in terms of $\mathcal{D}$,
consider the optimal signalling state ensemble 
$\{ \, p_k \,,\, \varrho_k \;=\; \mathcal{E}(\varphi_k)\; \}$.
Define $\varrho$ as $\sum_k \; p_k \, \varrho_k$.
Consider the following sum :

$$
\sum_k \; p_k \; \mathcal{D}( \, \varrho_k  \, \| \, \varrho \, ) \;=\; 
\sum_k \; \left \{  p_k \; Tr[ \,  \varrho_k \,  \log( \, \varrho_k \, ) \; ]  
\;- \; p_k \; Tr[ \; \varrho_k \; \log(\,\varrho \, ) \; ]  \right \}
$$

$$
=\quad \sum_k \; \left \{  p_k \; 
Tr \left [ \; \varrho_k \;  \log(\,\varrho_k \,) \; \right ]  \;\right \}
\;- \; 
Tr \left [ \; \sum_k \; \left \{ p_k \; \varrho_k \,  \log(\, \varrho\, ) \; 
\right \}  \right  ] 
$$

$$
\;=\; \sum_k \; \left \{  p_k \; Tr[ \; \varrho_k  \, \log( \, \varrho_k \, ) \; ]  \;\right \}
\;- \; Tr \left [ \; \varrho \, \log(\varrho) \;  \right ]   \quad=\quad 
\mathcal{S}(\, \varrho \, ) \;-\; \sum_k \; p_k \; 
\mathcal{S}( \, \varrho_k\, ) \;=\; \chi
$$

Thus, the HSW capacity $\mathcal{C}_1$ can be written as  

$$
\mathcal{C}_1 \;=\; 
\mbox{\Large Max}_{[all \;possible\; \{p_k\;,\; \varphi_k \} ] } \quad  
\sum_k \; p_k \; \mathcal{D} 
\left ( \, \mathcal{E}(\varphi_k) \,  || \, \mathcal{E}( \varphi )\, \right )
$$

\noindent
where the $\varphi_k$ are the quantum states input to the channel and
$\varphi \;=\; \sum_k \, p_k \, \varphi_k$.
We call an ensemble of channel output states 
$\{ \, p_k \,,\, \varrho_k \;=\; \mathcal{E}(\varphi_k)\; \}$ an optimal 
ensemble if this ensemble achieves $\mathcal{C}_1$.
Schumacher and Westmoreland proved the following five
properties related to optimal ensembles\cite{Schumacher99}.
\label{pgschuwest}

\vspace{0.2in}

I) $\mathcal{D}( \, \varrho_k \,  \| \,  \varrho \, ) \;=\; \mathcal{C}_1 \quad \forall \varrho_k$ in the optimal ensemble, and 
$\varrho \;=\; \sum p_k \, \varrho_k$. 

\vspace{0.2in}

II)
$\mathcal{D}( \, \xi \,  \| \,  \varrho \, ) \;\leq\; \mathcal{C}_1$
where 
$\{ \, p_k \,,\, \varrho_k \;=\; \mathcal{E}(\varphi_k)\; ,\;
\varrho \;=\; \sum p_k \, \varrho_k \; \}$ is an optimal ensemble, and
$\xi$ is {\em any} permissible channel output density matrix. 

\vspace{0.2in}

III) There exists at least one optimal ensemble 
$\{ \, p_k \,,\, \varrho_k \;=\; \mathcal{E}(\varphi_k)\; \}$ that
achieves $\mathcal{C}_1$. 

\vspace{0.2in}

IV) Let $\mathcal{A}$ be the set of possible channel output states
for a channel 
$\mathcal{E}$ corresponding to pure state inputs. Define 
$\mathcal{B}$ as the convex hull of the set of states 
$\mathcal{A}$. Then for $\varrho \;\in \mathcal{A}$ and  
$\xi \;\in $ $\mathcal{B}\;\equiv\; $ the convex hull of 
$\mathcal{A}$, we 
have \footnote{This result was originally derived in \cite{ohya}.} :

$$
\mathcal{C}_1 \;=\; 
\mbox{\Large Min}
_{\, \xi \, } \quad  \quad  
\mbox{\Large Max}
_{\,\varrho\,} \quad  \quad  
\mathcal{D} \left ( \; \varrho\;  \| \; \xi \;\right )
$$

\vspace{0.2in}

V) For every $\xi$ that satisfies the minimization in IV) above, there
exists an optimum signalling ensemble $\{ \; p_k \;,\; \rho_k \; \}$
such that $\xi \; \equiv \; \sum_k \; p_k \; \rho_k$.

\vspace{0.2in}

\subsection{The King - Ruskai - Szarek - Werner Qubit Channel Representation}

In this paper, we are primarily concerned with qubit channels, namely 
$\mathcal{E}(\varphi) \;=\; \varrho$, where $\varphi$ and $\varrho$ are qubit
density matrices. Several authors \cite{Ruskai99a,rsw} have developed a nice
picture of single qubit maps. Recall that single qubit density matrices can
be written in
the Bloch sphere representation. 
Let the density matrices $\varrho$ and $\varphi$ have 
the respective Bloch sphere representations :

$$
\varphi \; = \;  \frac{1}{2} \; ( \mathcal{I} \; + \; 
\vec{\mathcal{W}}_{\varphi}
\bullet \vec{\sigma} ) \; \quad \quad and \quad \quad
\; \varrho\; = \;  \frac{1}{2} \; 
( \mathcal{I} \; + \; \vec{\mathcal{W}_{\varrho}} \bullet \vec{\sigma} )  
$$

The symbol $ \vec{\sigma}$ means the vector of 2 x 2 Pauli matrices 

$$  \vec{\sigma} \; = \; 
\bmatrix{ \sigma_x \cr \sigma_y \cr \sigma_z }   
\;\;\;\;\; where 
\;\;\;\;\; \sigma_x \;=\; \bmatrix{ 0  & 1 \cr 1  & 0 },
\;\;\;\;\; \sigma_y \;=\; \bmatrix{ 0  & -i \cr i & 0 },
\;\;\;\;\; \sigma_z \;=\; \bmatrix{ 1  & 0 \cr 0 & -1 }
\;.$$

The Bloch vectors $\vec{\mathcal{W}}$ are real
three dimensional vectors
that have magnitude equal to one when representing a pure state density matrix,
and magnitude less than one for a mixed (non-pure) density matrix.

The King - Ruskai et al. qubit channel representation describes 
the channel as a mapping of input to output Bloch vectors.

$$
\bmatrix{ 1 \cr \widetilde{W_x}  \cr \widetilde{W_y}  \cr \widetilde{W_z} } 
\;=\;
\bmatrix{ 1 \quad  & 0 \quad & 0 \quad  & 0 \quad  \cr 
t_x & \lambda_x & 0 & 0 \cr
t_y & 0 & \lambda_y & 0 \cr
t_z & 0 & 0 & \lambda_z }
\quad 
\bmatrix{ 1 \cr W_x  \cr W_y  \cr W_z } 
$$

\label{channeldef}

All qubit channels have such a representation. 
The representation is unique up to a unitary rotation, and 
hence requires a choice of basis.  
The $t_k$ and $\lambda_k$ are real parameters which must satisfy certain
constraints in order to ensure the matrix above represents a completely positive
qubit map. ( Please see King - Ruskai for more details\cite{Ruskai99a}. )

From the King - Ruskai et al. qubit channel representation, we 
see that 
$\widetilde{\mathcal{W}_k} \;=\; t_k \;+\; \lambda_k \; \mathcal{W}_k$
or

$$
\mathcal{W}_k\;=\; 
\frac{  \; \widetilde{\mathcal{W}_k} \;-\; t_k \;}{\; \lambda_k \;}
$$

It has been shown that $\mathcal{C}_1$ can 
always be achieved using only pure input states\cite{Schumacher99}.
Therefore, all input signalling Bloch vectors obey 
$\; \| \, \vec{\mathcal{W}} \, \| \;=\; 1.$
Thus $\| \, \vec{\mathcal{W}} \, \|^2 \;=\; 1$, and 
$ \| \, \vec{\mathcal{W}} \, \|^2 \;=\; 1 \;=\; \mathcal{W}_x ^2 \;+\;  
\mathcal{W}_y ^2  \;+\;  \mathcal{W}_z ^2\; $ implies

$$
\left ( \; \frac{ \widetilde{\mathcal{W}_x} \; - \; t_x }{\lambda_x} \; \right ) ^2 \;+\;  
\left ( \; \frac{ \widetilde{\mathcal{W}_y} \; - \; t_y }{\lambda_y} \; \right ) ^2  \;+\;  
\left ( \; \frac{ \widetilde{\mathcal{W}_z} \; - \; t_z }{\lambda_z} \; \right ) ^2  \;=\; 1
$$  

The set of possible channel output states we shall be  
interested in is the set of channel outputs corresponding 
to pure state channel inputs. This set of states was
defined as $\mathcal{A}$ in section 2.2, and 
is the \emph{surface} of the ellipsoid shown above. The convex hull of 
the set of states $\mathcal{A}$ is the solid ellipsoid
defined as $\widetilde{\vec{\mathcal{W}}}$ such that

$$
\left ( \; \frac{ \widetilde{\mathcal{W}_x} \; - \; t_x }{\lambda_x} \; \right ) ^2 \;+\;  
\left ( \; \frac{ \widetilde{\mathcal{W}_y} \; - \; t_y }{\lambda_y} \; \right ) ^2  \;+\;  
\left ( \; \frac{ \widetilde{\mathcal{W}_z} \; - \; t_z }{\lambda_z} \; \right ) ^2  
\;\leq \; 1
$$  

\section{Relative Entropy In The Bloch Sphere Representation}

The key formula we shall use extensively is the relative entropy in the
Bloch sphere representation.
Here $\rho$ and $\phi$ have the respective Bloch sphere representations :

$$
\rho \; = \;  \frac{1}{2} \; ( \mathcal{I} \; + \; \vec{\mathcal{W}} \bullet \vec{\sigma} ) \; \; \; \; \; \; \phi \; = \;  \frac{1}{2} \; ( \mathcal{I} \; + \; \vec{\mathcal{V}} \bullet \vec{\sigma} )  
$$

We define $\cos(\theta)$ as :

$$
\cos(\theta) \;\;=\;\; 
\frac{\vec{\mathcal{W}}\bullet \vec{\mathcal{V}}}{ \; r \; q \; }
\;\;\;\;\; where \;\;\;\;\;
r \;=\; \sqrt { \vec{\mathcal{W}} \bullet \vec{\mathcal{W}}} 
\;\;\; and \;\;\; q \;=\; \sqrt { \vec{\mathcal{V}} \bullet \vec{\mathcal{V}} }
\;.
$$

In Appendix A, we prove the following formula for the relative entropy 
$\mathcal{D}(\, \varrho \, \| \,  \psi \, ) $ of two single qubit 
density matrices $\varrho$ and $\psi$ with Bloch sphere representations 
given above. 

$$
\mathcal{D}(\, \varrho \, \| \,  \psi \, ) \;=\; 
\frac{1}{2} \log_2 \left ( 1\;-\;r^2 \right ) \;+\; 
\frac{r}{2} \log_2 \left ( \frac{1\;+\;r}{1\;-\;r} \right )
\;-\; 
\frac{1}{2}
\;  \log_2 \left ( 1 \;-\; q^2\; \right ) 
\;-\; \frac{\; \vec{\mathcal{W}} \bullet \vec{\mathcal{V}} \; }
{\; 2 \; q\; } \; \log_2 \left ( \frac{  1 \;+\; q}{1 \;-\; q} \right)
$$

$$
\;=\; 
\frac{1}{2} \log_2 \left  ( 1\;-\;r^2 \right ) \;+\; 
\frac{r}{2} \log_2 \left ( \frac{1\;+\;r}{1\;-\;r} \right )
\;-\; 
\frac{1}{2}
\;  \log_2 \left ( 1 \;-\; q^2\; \right ) 
\;-\; \frac{r \; \cos ( \theta )  }
{\; 2 \; } \; \log_2 \left ( \frac{  1 \;+\; q}{1 \;-\; q} \right)
$$

where $\theta$ is the angle between $\vec{\mathcal{W}}$ 
and $\vec{\mathcal{V}}$, and $r$ and $q$ are as defined above. 

When $\phi$ in $\mathcal{D}( \, \rho \,  \| \,  \phi \, )$ is the maximally 
mixed state $\phi \;=\; \frac{1}{2} \, \mathcal{I}$, we have $q\, = \, 0$, and
$\mathcal{D}( \, \rho \,\| \,\phi \,)$ becomes the radially symmetric function 

$$
\mathcal{D}( \, \rho \,  \|  \, \phi  \, ) \;=\; 
\mathcal{D} \left ( \, \rho \,  \|  \, \frac{1}{2} \, \mathcal{I} \, \right ) \;=\; 
\frac{1}{2} \; \log_2 \left (  1 - r^2 \right ) \;+ \;  
\frac{r}{2} \; \log_2 \left (  \frac{1 + r}{1 - r } \right )  
\;=\; 1 \, - \, \mathcal{S}(\,\rho\,)\;.
$$

It is shown in Appendix A that
$\mathcal{D} \left ( \, \rho \, \| \,  \frac{1}{2} \; \mathcal{I} \, \right ) \;=\; 1 \;-\; \mathcal{S}(\rho)$,
where $\mathcal{S}(\rho)$ is the von Neumann entropy of $\rho$. 
In what follows, we shall often 
write $\mathcal{D}(\, \rho \, \| \, \phi \, )$ as 
$\mathcal{D}(\, \vec{\mathcal{W}} \, \| \, \vec{\mathcal{V}} \, )$, 
where $\vec{\mathcal{W}}$ and $\vec{\mathcal{V}}$ are 
the Bloch sphere vectors for $\rho$ and $\phi$ respectively.  

In what follows, we shall graphically determine the HSW channel
capacity from the intersection of contours of constant relative entropy
with the channel ellipsoid. To that end, and to help build intuition
regarding channel parameter tradeoffs, it is advantageous 
to obtain a rough idea of how the contours of constant
relative entropy $\mathcal{D}(\, \rho \, \| \, \phi \, )$ behave,
for fixed $\phi$, as $\rho$ is varied.  
Furthermore, it will turn out that due to symmetries in the relative 
entropy, we
frequently will only need to understand the relative entropy behavior
in a plane
of the Bloch sphere, which we choose to be the Bloch X-Y plane. 
In Figure 2, we plot a few contour lines for  
$\mathcal{D}( \, \rho  \,\| \, \phi \,=\, \frac{1}{2} \;
 \mathcal{I}  \, )$ in the X-Y Bloch sphere plane.  
In the figures that follow, we shall mark the location of 
$\phi$ with an asterisk.  
The contour values for $\mathcal{D}( \, \rho  \,\| \, \phi \,  )$ are 
shown in the plot title. The smallest value of 
$\mathcal{D}(\, \rho \, \| \, \phi \,)$ corresponds 
to the contour closest to the location of $\phi$.
The largest value of 
$\mathcal{D}(\, \rho \, \| \, \phi \, )$
corresponds to the outermost contour. 
For 
$\phi \;=\; \frac{1}{2} \, \mathcal{I}$, the location of 
$\phi$ is the Bloch sphere
origin. 

\begin{center}
\includegraphics*[angle=-90,scale=0.65]{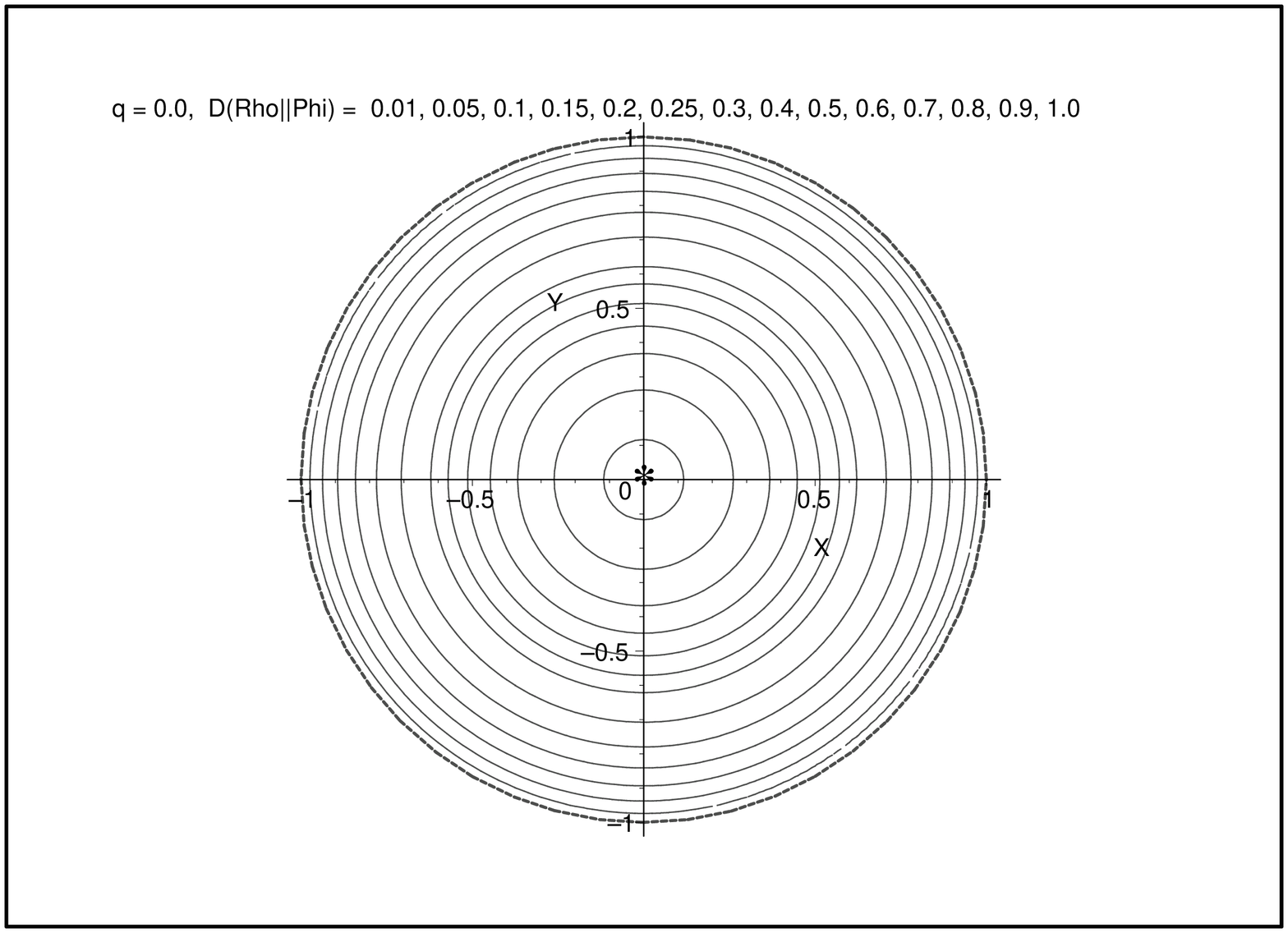}
\end{center}
\nopagebreak
\begin{center}
Figure 2: 
Contours of constant relative entropy $\mathcal{D}(\rho \| \phi )$
as a function of $\rho$ in the Bloch sphere X-Y plane  
for the fixed density matrix
$\phi \;=\; \frac{1}{2} \; \mathcal{I}$.
\end{center}

As an example of how these contour lines change as $\phi$ 
moves away from the maximally mixed state
$\phi \;=\; 	\frac{1}{2} \, \mathcal{I} $,
or equivalently as q becomes non-zero, we give
contour plots below for $q \; \ne \;0$. We let 
$\phi \;=\; \frac{1}{2} \; \{ \; \mathcal{I} \;+\; q \; \sigma_y \; \}$ with 
corresponding Bloch
vector $\vec{\mathcal{V}} \;=\; \bmatrix{ 0 \cr q \cr 0}$. 
The asterisk in these plots denotes the location of $\vec{\mathcal{V}}$. 
The dashed outer contour is a radius equal to one, indicating 
where the pure states lie.

\begin{center}
\includegraphics*[angle=-90,scale=0.54]{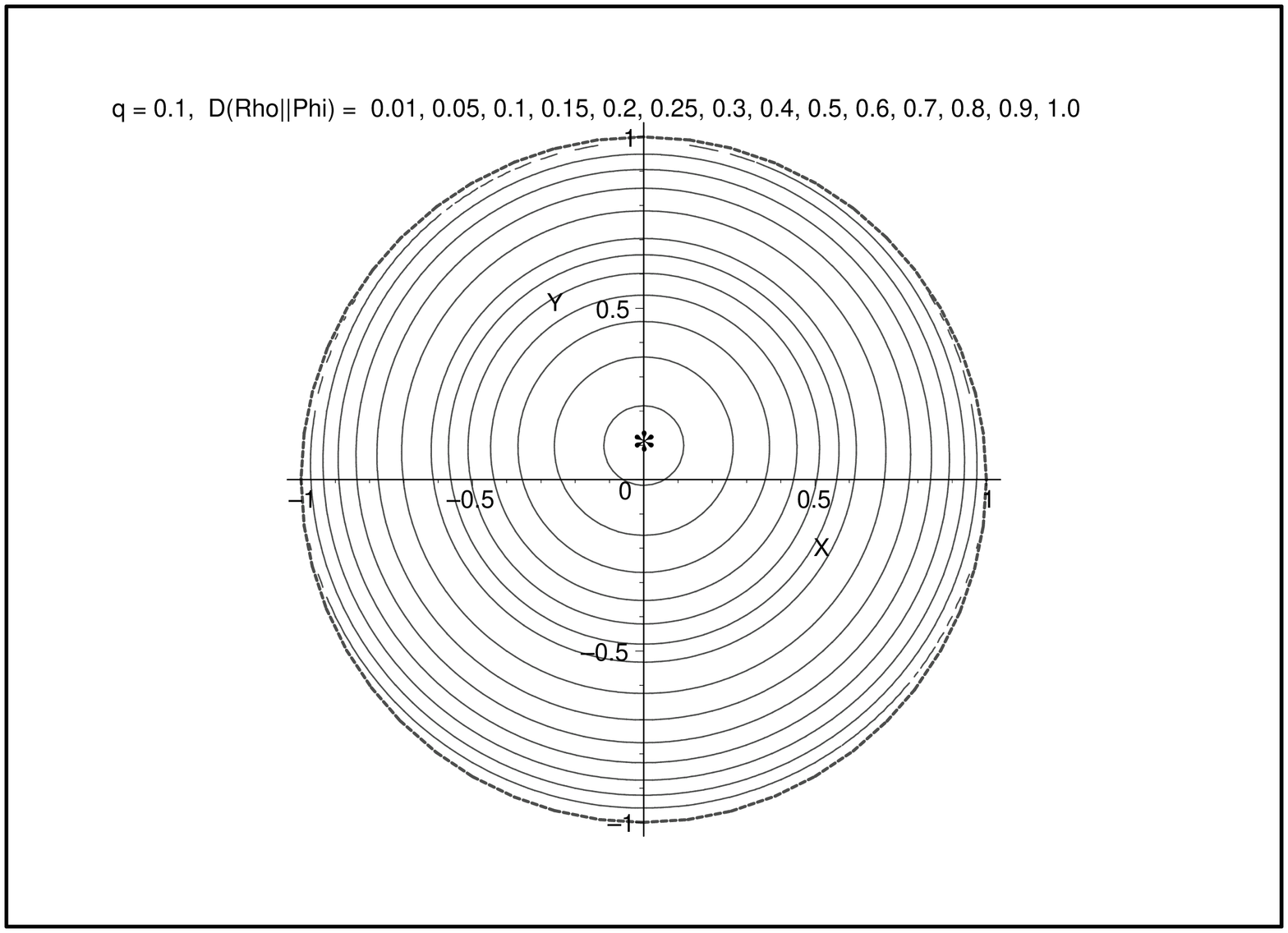}
\end{center}
\begin{center}
Figure 3: 
Contours of constant relative entropy $\mathcal{D}(\rho \| \phi )$
as a function of $\rho$ in the Bloch sphere X-Y plane  
for the fixed density matrix
$\phi \;=\; \frac{1}{2} \; \{ \; \mathcal{I}\; + \; 0.1 \, \sigma_y \; \}$.
\end{center}

\enlargethispage{0.2in}

\begin{center}
\includegraphics*[angle=-90,scale=0.54]{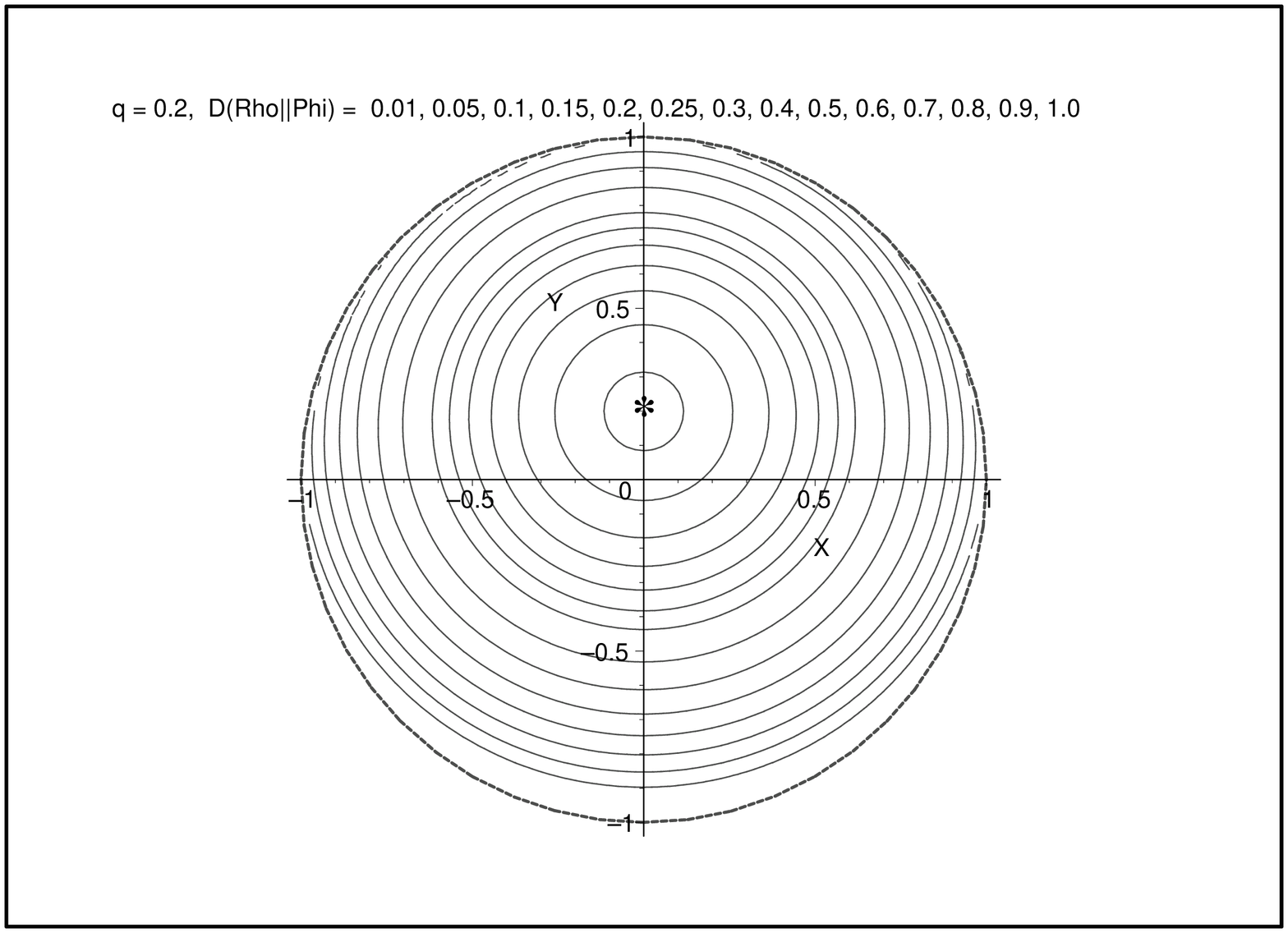}
\end{center}
\begin{center}
Figure 4: 
Contours of constant relative entropy $\mathcal{D}(\rho \| \phi )$
as a function of $\rho$ in the Bloch sphere X-Y plane  
for the fixed density matrix
$\phi \;=\; \frac{1}{2} \; \{ \; \mathcal{I}\; + \; 0.2 \, \sigma_y \; \}$.
\end{center}

\begin{center}
\includegraphics*[angle=-90,scale=0.54]{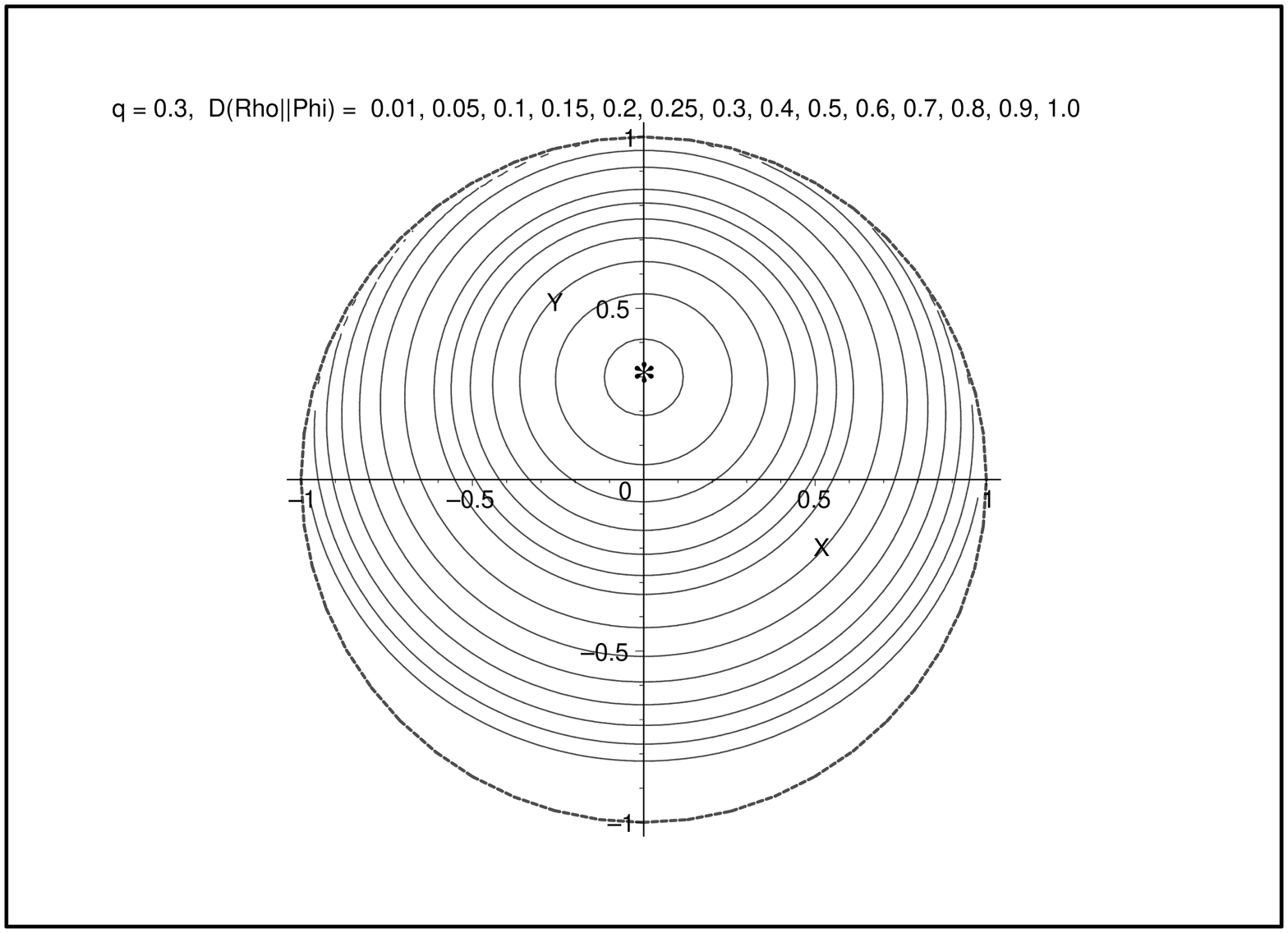}
\end{center}
\begin{center}
Figure 5:
Contours of constant relative entropy $\mathcal{D}(\rho \| \phi )$
as a function of $\rho$ in the Bloch sphere X-Y plane  
for the fixed density matrix
$\phi \;=\; \frac{1}{2} \; \{ \; \mathcal{I}\; + \; 0.3 \, \sigma_y \; \}$.
\end{center}

\enlargethispage{0.2in} 

\begin{center}
\includegraphics*[angle=-90,scale=0.54]{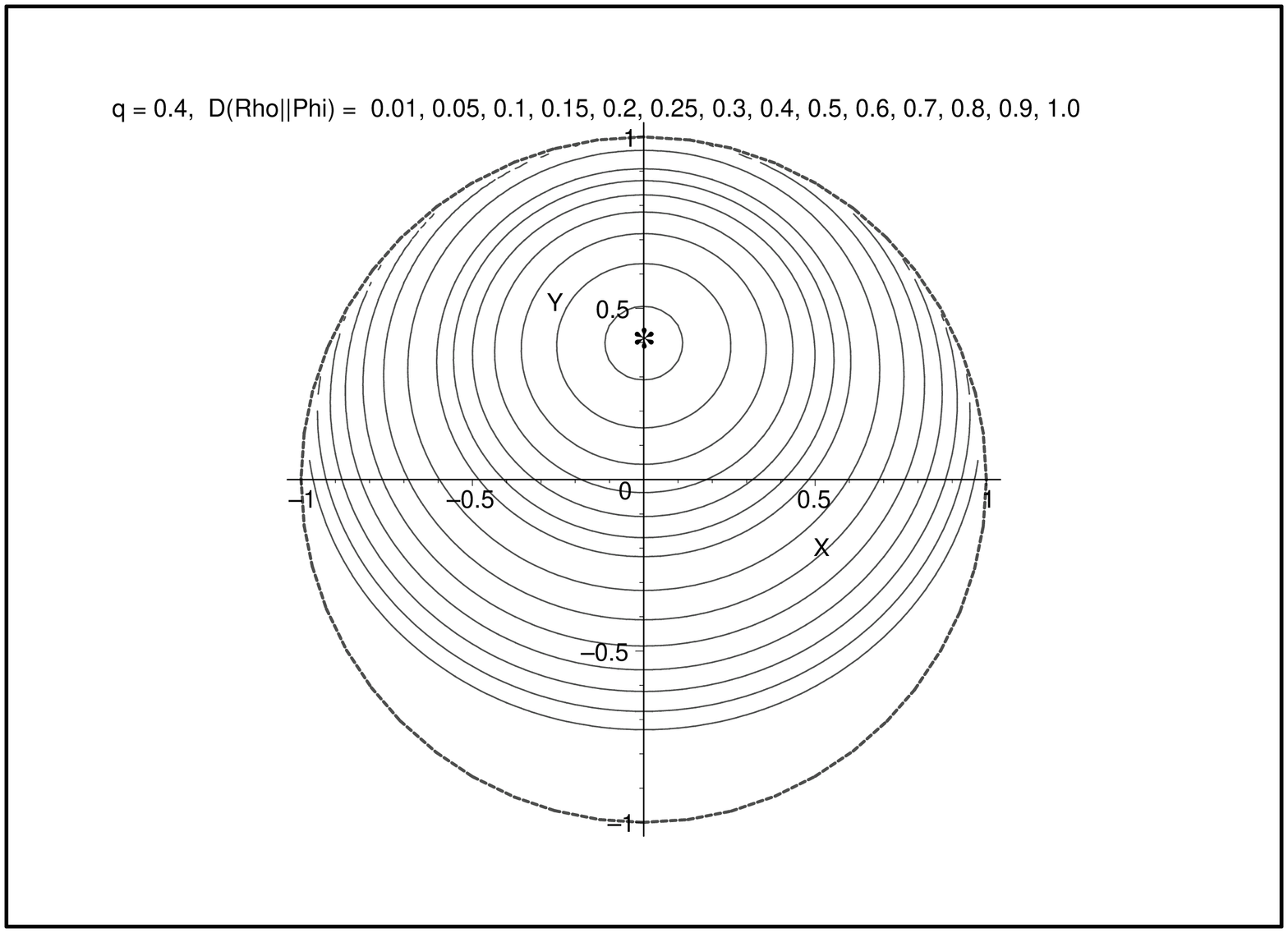}
\end{center}
\begin{center}
Figure 6:
Contours of constant relative entropy $\mathcal{D}(\rho \| \phi )$
as a function of $\rho$ in the Bloch sphere X-Y plane  
for the fixed density matrix
$\phi \;=\; \frac{1}{2} \; \{ \; \mathcal{I}\; + \; 0.4 \, \sigma_y \; \}$.
\end{center}

\begin{center}
\includegraphics*[angle=-90,scale=0.54]{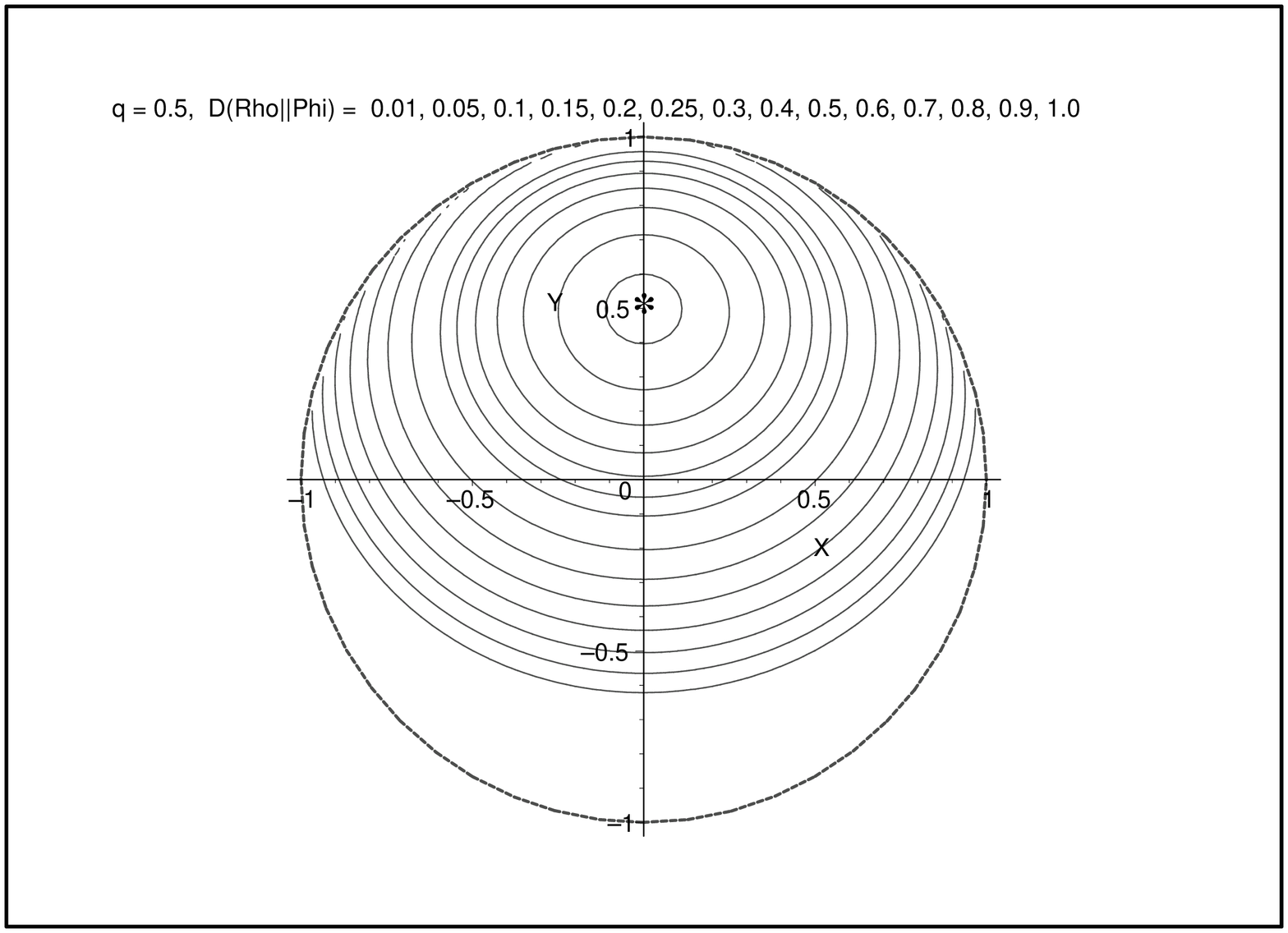}
\end{center}
\begin{center}
Figure 7:
Contours of constant relative entropy $\mathcal{D}(\rho \| \phi )$
as a function of $\rho$ in the Bloch sphere X-Y plane  
for the fixed density matrix
$\phi \;=\; \frac{1}{2} \; \{ \; \mathcal{I}\; + \; 0.5 \, \sigma_y \; \}$.
\end{center}

\enlargethispage{0.2in} 

\begin{center}
\includegraphics*[angle=-90,scale=0.54]{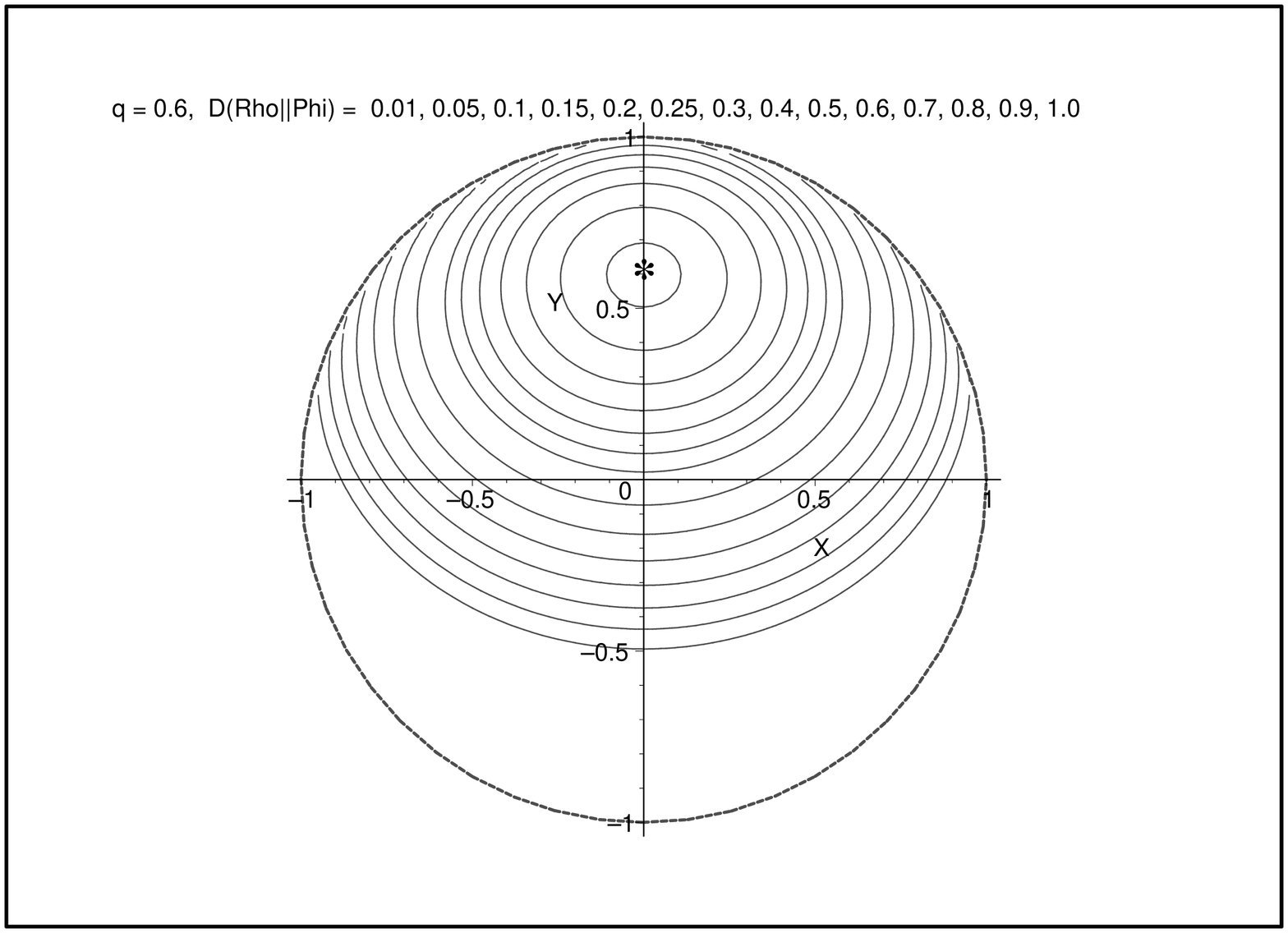}
\end{center}
\begin{center}
Figure 8:
Contours of constant relative entropy $\mathcal{D}(\rho \| \phi )$
as a function of $\rho$ in the Bloch sphere X-Y plane  
for the fixed density matrix
$\phi \;=\; \frac{1}{2} \; \{ \; \mathcal{I}\; + \; 0.6 \, \sigma_y \; \}$.
\end{center}

\begin{center}
\includegraphics*[angle=-90,scale=0.54]{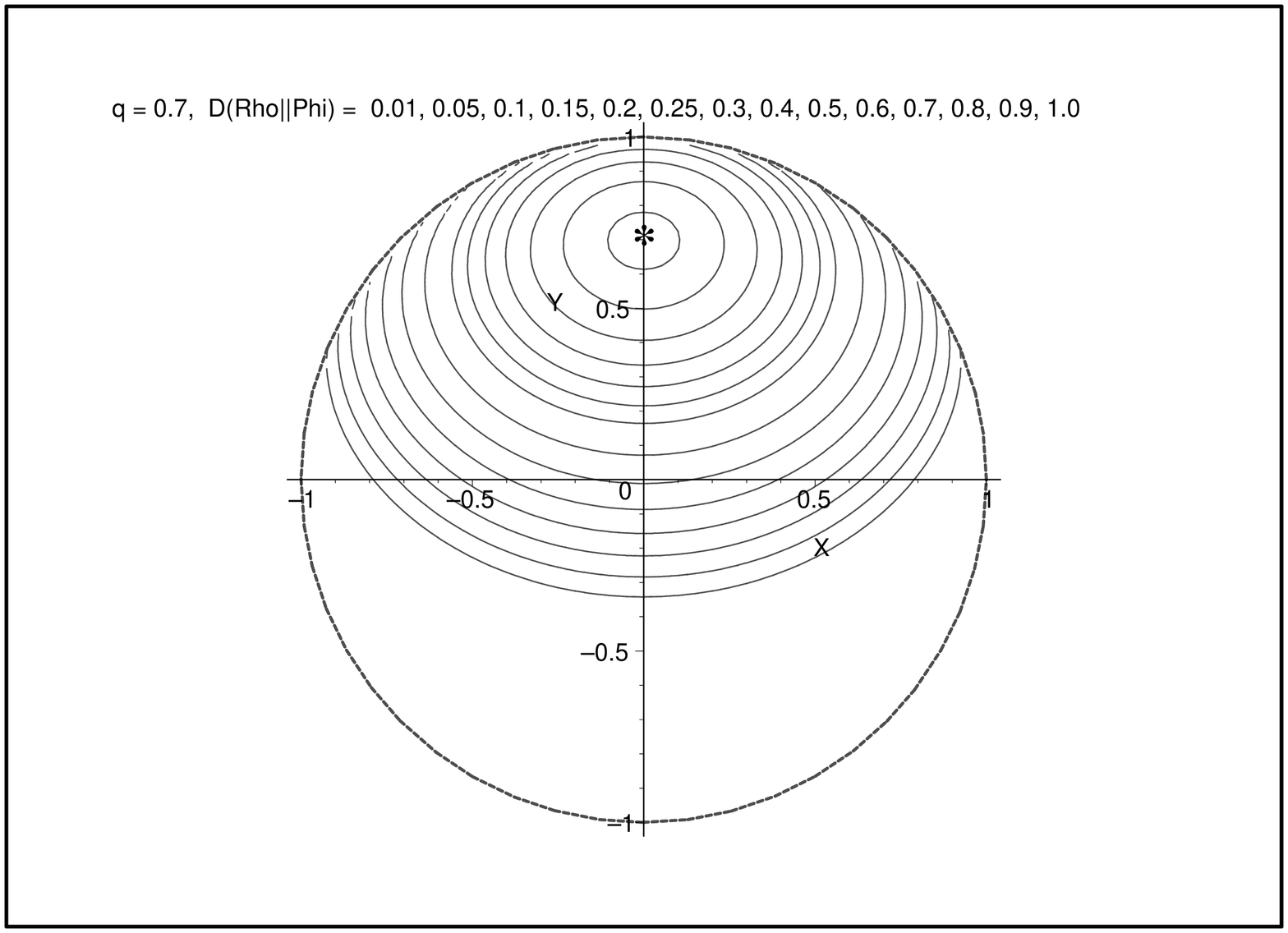}
\end{center}
\begin{center}
Figure 9:
Contours of constant relative entropy $\mathcal{D}(\rho \| \phi )$
as a function of $\rho$ in the Bloch sphere X-Y plane  
for the fixed density matrix
$\phi \;=\; \frac{1}{2} \; \{ \; \mathcal{I}\; + \; 0.7 \, \sigma_y \; \}$.
\end{center}

\enlargethispage{0.2in} 

\begin{center}
\includegraphics*[angle=-90,scale=0.54]{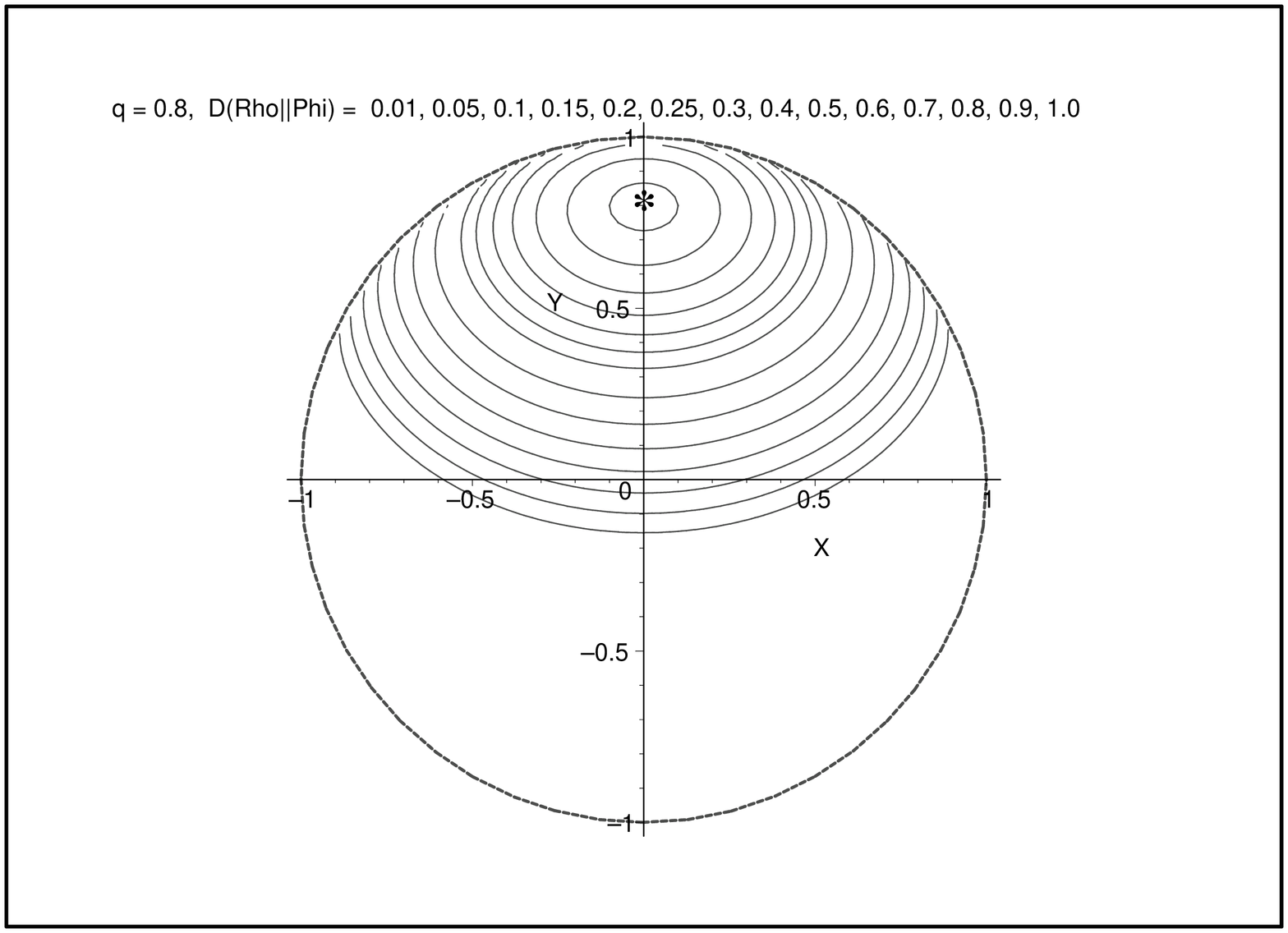}
\end{center}
\begin{center}
Figure 10:
Contours of constant relative entropy $\mathcal{D}(\rho \| \phi )$
as a function of $\rho$ in the Bloch sphere X-Y plane  
for the fixed density matrix
$\phi \;=\; \frac{1}{2} \; \{ \; \mathcal{I}\; + \; 0.8 \, \sigma_y \; \}$.
\end{center}

\begin{center}
\includegraphics*[angle=-90,scale=0.54]{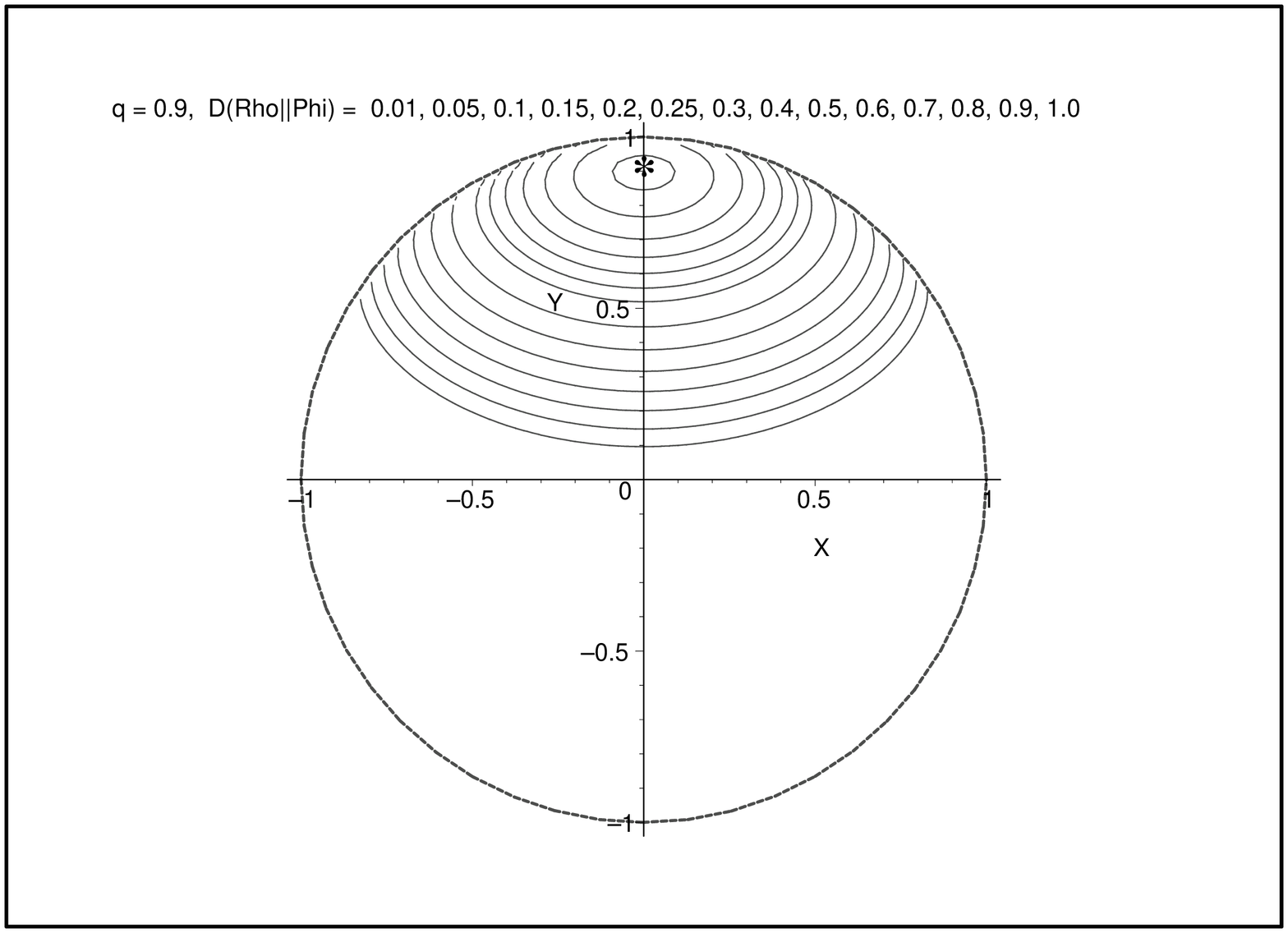}
\end{center}
\begin{center}
Figure 11:
Contours of constant relative entropy $\mathcal{D}(\rho \| \phi )$
as a function of $\rho$ in the Bloch sphere X-Y plane  
for the fixed density matrix
$\phi \;=\; \frac{1}{2} \; \{ \; \mathcal{I}\; + \; 0.9 \, \sigma_y \; \}$.
\end{center}

The two dimensional plots of $\mathcal{D}( \, \rho \, \| \, \phi \, )$ 
shown above tell us about the {\em three dimensional} nature of
$\mathcal{D}( \, \rho \, \| \, \phi \, )$. To see why, first note that we
can always rotate the Bloch sphere X-Y-Z axes to arrange for 
$\phi \; \equiv \; \vec{\mathcal{V}} \; \rightarrow \; \vec{q}$  to lie on the Y
axis, as the density matrices $\phi$ are shown in Figures 2 through 11 
above. 

Second, recall that our description of
$\mathcal{D}( \, \rho \, \| \, \phi \, )$ 
is a function of the three variables 
$\{ \, r \,,\, q\,,\, \theta \, \}$ only, which were defined above as the
length of the Bloch vectors corresponding to the density matrices 
$\rho$ and $\phi$ respectively, and the angle between these Bloch
vectors. 

$$
\mathcal{D}( \, \rho \, \| \, \phi \, ) \;\equiv\; f( \, r  
 \,,\, q\,,\, \theta \, )
$$
This means the two dimensional curves of constant 
$\mathcal{D}( \, \rho \, \| \, \phi \, )$
can be rotated about the Y axis as surfaces of
revolution, to yield three dimensional surfaces of constant 
$\mathcal{D}( \, \rho \, \| \, \phi \, )$. (In these two and 
three dimensional plots, the first argument of
$\mathcal{D}( \, \rho \, \| \, \phi \, )$, 
$\rho$, is being varied, while the second argument, $\phi$, is being held
fixed at a point on the Y axis.) 

	Our two dimensional plots above give us a good idea of the
three dimensional behavior of the surfaces of constant relative entropy
about the density matrix $\phi$ occupying the second slot in 
$\mathcal{D}( \, \cdots \, \| \, \cdots \, )$. 
A picture emerges of 
slightly warped ''eggshells'' nested like Russian dolls inside each other,
{\em roughly} centered on $\phi$.  
A mental picture of the behavior of 
$\mathcal{D}( \, \rho \, \| \, \phi \, )$
is useful because in what follows we shall 
superimpose the KRSW channel ellipsoid(s) onto Figures 2 through 11 above.
By moving $\vec{\mathcal{V}}$ (the asterisk) around  
in these pictures, we shall adjust the contours of constant 
$\mathcal{D}( \, \rho \, \| \, \phi \, )$, 
and thereby {\em graphically} determine the HSW channel capacity,
optimum (output) signalling states, and corresponding a priori
signalling probabilities. The resulting intuition we gain from these
pictures will help us understand channel parameter tradeoffs.

\section{Linear Channels}

Recall the KRSW specification of a qubit channel in terms of the six
real parameters 
$\{ \; t_x \,,\, t_y \,,\, t_z \,,\, \lambda_x \,,\, \lambda_y \,,\, \lambda_z \; \}$
as defined on page \pageref{channeldef} of this paper. 
A linear channel is one where 
$\lambda_x \;=\; \lambda_y \;=\; 0$, but
$\lambda_z \;\neq\; 0$. The shift quantities ${ \;t_k \; }$ can be any real
number, up to the limits imposed by the requirement that the map be
completely positive. For more details on the complete positivity
requirements of qubit maps, please see \cite{rsw}.  

A linear channel is a simple system that illustrates the basic ideas behind
our graphical approach to determining the HSW channel capacity $\mathcal{C}_1$.
Recall the relative entropy formulation for  $\mathcal{C}_1$.

$$
\mathcal{C}_1 \;=\; 
\mbox{\Large Max}_{[all \;possible\; \{p_k\;,\; \varphi_k \} ]} \quad  
\sum_k \; p_k \; \mathcal{D} \left ( \, \mathcal{E}(\, \varphi_k \,) \,  
|| \, \mathcal{E}( \, \varphi \, ) \,  \right )
$$

\noindent
where the $\varphi_k$ are the quantum states input to the channel and
$\varphi \;=\; \sum_k \, p_k \, \varphi_k$.
We call an ensemble of states 
$\{ \, p_k \,,\, \varrho_k \;=\; \mathcal{E}(\varphi_k)\; \}$ an optimal 
ensemble if this ensemble achieves $\mathcal{C}_1$.  

As discussed on page \pageref{pgschuwest}, 
Schumacher and Westmoreland showed the above maximization to determine
$\mathcal{C}_1$ is equivalent to the following min-max criterion :

$$
\mathcal{C}_1 \;=\; 
\mbox{\Large Min}
_{\, \phi\, } \quad  \quad  
\mbox{\Large Max}
_{\,\varrho_k\,} \quad  \quad  
\mathcal{D} \left ( \; \varrho_k \;  \| \; \phi \;\right )
$$

\noindent
where $\varrho_k$ is a density matrix on the surface of the channel
ellipsoid, and $\phi$ is a density matrix in the convex hull  
of the channel ellipsoid. For the linear channel,
the channel ellipsoid is a line
segment of length $2 \; \lambda_z$ centered on
$\{\; t_x \,,\, t_y \,,\,  t_z \; \}$. 
Thus, both $\varrho_k$ and $\phi$ must lie somewhere along this line segment. 
Furthermore, Schumacher and Westmoreland tell us that
$\phi$ must be expressible as a convex combination of the $\varrho_k$ which
satisfy the above min-max\cite{Schumacher99}.

To graphically implement the min-max criterion, we overlay the channel
ellipsoid on the contour plots of relative entropy previously found. 
We wish to determine the location of the optimum $\phi$ and the
optimum relative entropy contour that achieves the min-max. 
The generic overlap scenarios are shown below, labeled Cases 1 - 5. 
From our plots of relative entropy, we know that contours of 
relative entropy are {\em roughly} circular about $\phi$. We denote
the location of $\phi$ below by an asterisk ( {\textbf *} ).  
The permissible $\varrho_k$ are those density matrices at the intersection 
of the relative entropy contour and the channel ellipsoid, here a line
segment.  

	Let us examine the five cases shown below, seeking the optimum 
$\phi$ and the relative entropy contour corresponding to $\mathcal{C}_1$
(the circles below), by eliminating those cases 
which do not make sense in light of the minimization-maximization above. 

\begin{center}

\setlength{\unitlength}{0.00033300in}%
\begingroup\makeatletter\ifx\SetFigFont\undefined
\def\x#1#2#3#4#5#6#7\relax{\def\x{#1#2#3#4#5#6}}%
\expandafter\x\fmtname xxxxxx\relax \def\y{splain}%
\ifx\x\y   
\gdef\SetFigFont#1#2#3{%
  \ifnum #1<17\tiny\else \ifnum #1<20\small\else
  \ifnum #1<24\normalsize\else \ifnum #1<29\large\else
  \ifnum #1<34\Large\else \ifnum #1<41\LARGE\else
     \huge\fi\fi\fi\fi\fi\fi
  \csname #3\endcsname}%
\else
\gdef\SetFigFont#1#2#3{\begingroup
  \count@#1\relax \ifnum 25<\count@\count@25\fi
  \def\x{\endgroup\@setsize\SetFigFont{#2pt}}%
  \expandafter\x
    \csname \romannumeral\the\count@ pt\expandafter\endcsname
    \csname @\romannumeral\the\count@ pt\endcsname
  \csname #3\endcsname}%
\fi
\fi\endgroup
\begin{picture}(12691,3339)(368,-5863)
\thicklines
\put(1126,-3961){\circle{1500}}
\put(7126,-3961){\circle{1500}}
\put(9901,-3886){\circle{1500}}
\put(12301,-3961){\circle{1500}}
\put(4351,-3961){\circle{2348}}
\put(1501,-2986){\line( 0,-1){1875}}
\put(9901,-2761){\line( 0,-1){2400}}
\put(12301,-3211){\line( 0,-1){1500}}
\put(4351,-3511){\line( 0,-1){900}}
\put(7201,-2536){\line( 0,-1){1875}}
\put(976,-5761){\makebox(0,0)[b]{\smash{\SetFigFont{14}{16.8}{rm}Case 1 }}}
\put(3901,-5836){\makebox(0,0)[b]{\smash{\SetFigFont{14}{16.8}{rm}Case 2}}}
\put(7201,-5761){\makebox(0,0)[b]{\smash{\SetFigFont{14}{16.8}{rm}Case 3}}}
\put(9901,-5761){\makebox(0,0)[b]{\smash{\SetFigFont{14}{16.8}{rm}Case 4}}}
\put(12301,-5761){\makebox(0,0)[b]{\smash{\SetFigFont{14}{16.8}{rm}Case 5}}}
\put(1126,-3961){\makebox(0,0)[b]{\smash{\SetFigFont{14}{16.8}{rm}*}}}
\put(7201,-3961){\makebox(0,0)[b]{\smash{\SetFigFont{14}{16.8}{rm}*}}}
\put(9901,-4036){\makebox(0,0)[b]{\smash{\SetFigFont{14}{16.8}{rm}*}}}
\put(12301,-4036){\makebox(0,0)[b]{\smash{\SetFigFont{14}{16.8}{rm}*}}}
\put(4351,-4036){\makebox(0,0)[b]{\smash{\SetFigFont{14}{16.8}{rm}*}}}
\end{picture}

\end{center}

\begin{center}
Figure 12: 
Scenarios for the intersection of the optimum
relative \linebreak entropy contour with a linear channel ellipsoid. 
\end{center}

Case 1 is not an acceptable
configuration because $\phi$ does not lie inside the channel
ellipsoid, meaning for the linear channel, $\phi$ does not lie
on the line segment. Case 2 is not acceptable because there are no 
permissible $\varrho_k$, since the relative entropy contour does not intersect
the channel ellipsoid line segment anywhere. Case 3 is not acceptable because  
Schumacher and Westmoreland tell us that $\phi$ must be expressible
as a convex combination of the $\varrho_k$ density matrices which satisfy the
above min-max requirement. There is only one permissible $\varrho_k$ density
matrix in Case 3, and since, as seen in the diagram for Case 3,
$\phi\;\neq\; \varrho_1$, we do not have an
acceptable configuration. Case 4 at first appears acceptable. However, 
here we do not achieve the maximization in the min-max relation,
since we can do better by using a relative entropy contour with a larger
radii. Case 5 is the ideal situation. The relative entropy contour
intersects both of the line segment endpoints. Taking a larger radius relative
entropy contour does not give us permissible $\varrho_k$, since we would
obtain Case 2 with a larger radii. For Case 5, if we moved $\phi$ as we
increased the relative entropy contour, we would
obtain Case 3, again an unacceptable configuration. In Case 5, using 
the two $\varrho_k$ that lie at the intersection
of the relative entropy contour and 
the channel ellipsoid line segment, we can form a convex combination of
these $\varrho_k$ that equals $\phi$. Case 5 is the best we can do, meaning
Case 5 yields the largest radius relative entropy contour which satisfies
the Schumacher-Westmoreland requirements. The
value of this largest radii relative entropy contour is the HSW channel
capacity we seek, $\mathcal{C}_1$. 
  
We now restate Case 5 in Bloch vector notation.
We shall associate 
the Bloch vector $\vec{\mathcal{V}}$ with $\phi$, and
the Bloch vectors $\vec{\mathcal{W}}_k$ with the $\varrho_k$ density
matrices.
For the linear channel, from our analysis above which resulted in
Case 5, we know that $\vec{\mathcal{V}}$ must lie on the line segment 
between the two endpoint vectors 
$\vec{\mathcal{W}}_{+}$ and 
$\vec{\mathcal{W}}_{-}$. 
(Note that from here on, 
we shall drop the tilde $\widetilde{\;}$ 
we were previously using to denote
channel output Bloch vectors, as almost all the
Bloch vectors we shall talk about below are channel output Bloch
vectors. The few instances when this is not the case shall be obvious.) 

For a general linear channel, the KRSW ellipsoid channel parameters satisfy 

\begin{center}
$\{ \; t_x \,\neq\, 0
\;,\; t_y \,\neq\, 0\;,\; 
t_z \,\neq\, 0\;,\;
\lambda_x \, = \, 0\;,\; 
\lambda_y \, = \, 0\;,\; 
\lambda_z \,\neq\, 0\; \}$.
\end{center}

Thus, we can explicitly determine the Bloch vectors 
$\vec{\mathcal{W}}_{+}$ and
$\vec{\mathcal{W}}_{-}$, which we write below.

$$
\rho_{+} \;\rightarrow \;
 \vec{\mathcal{W}}_{+} \;=\; 
\bmatrix{ t_x \cr t_y \cr t_z \; + \; \lambda_z  }
,\;\;\; and \;\;\; 
\rho_{-} \;\rightarrow \;
\vec{\mathcal{W}}_{-} \;=\; 
\bmatrix{ t_x \cr t_y \cr t_z \; - \; \lambda_z  }
$$

Note that the $t_k$ and $\lambda_z$ are real numbers,
and any of them may be negative.

The Bloch vector $\vec{\mathcal{V}}$ however requires more work. 
We parameterize the Bloch sphere vector 
$\vec{\mathcal{V}}$
corresponding to $\phi$ by the real number  
$\alpha$, specifying a position for
$\vec{\mathcal{V}}$ along the line segment between
$\vec{\mathcal{W}}_{+}$  and $\vec{\mathcal{W}}_{-}$.

$$
\phi \; \rightarrow \; \vec{\mathcal{V}}  \;=\; 
\bmatrix{ t_x \cr t_y \cr t_z \; + \; \alpha \, \lambda_z  }
$$

Here $\alpha \; \in \; [-1,1]$.  
Now recall that the Schumacher-Westmoreland
maximal distance property 
( see property $\#$ I 
in Section \ref{schuwest} )
tells us that $D(\rho_{+} || \phi) \;=\; D(\rho_{-} || \phi)$.
To find $\vec{\mathcal{V}}$, 
we shall apply the formula we have derived 
for relative entropy in the Bloch
representation to $D(\rho_{+} || \phi) \;=\; D(\rho_{-} || \phi)$,
and solve for $\alpha$. The details are in Appendix B. 

\subsection{A Simple Linear Channel Example}

To illustrate the ideas presented above,
we take as a simple example the linear channel with
channel parameters :
$\{ \; 
t_x\;=\;0\;,\;  
t_y\;=\;0 \;,\;  
t_z\;=\;0.2\;,\;  
\lambda_x\;=\;0 \;,\;  
\lambda_y\;=\;0 \;,\;  
\lambda_z\;=\;0.4\; \}$. 
Because the channel is linear with
$t_x\;=\;t_y\;=\;0$, we shall be able to easily
solve for $\vec{\mathcal{V}}$ and $\mathcal{C}_1$.

We define the real numbers $r_{+}$ and $r_{-}$ as the Euclidean distance
in the Bloch sphere from the Bloch sphere origin to the
Bloch vectors $\vec{\mathcal{W}}_{+}$  and $\vec{\mathcal{W}}_{-}$.
That is, $r_{+}$ and $r_{-}$ are the magnitudes of the Bloch vectors 
$\vec{\mathcal{W}}_{+}$  and $\vec{\mathcal{W}}_{-}$ defined above.
For the channel parameter numbers given, we find 
$r_{+} \;=\; \|\; 0.4 \;+\; 0.2\;\|\;=\; 0.6$ 
and $r_{-} \;=\; \| \; 0.2 \;-\; 0.4\;\|\;=\; 0.2$.  
We similarly define $q$ to be the magnitude of the Bloch
vector $\vec{\mathcal{V}}$.

To find $\vec{\mathcal{V}}$, 
we shall apply the formula we have derived 
for relative entropy in the Bloch
representation to 
$D(\rho_{+} || \phi) \;=\; D(\rho_{-} || \phi)$, or in Bloch sphere
notation,
$D(\vec{\mathcal{W}}_{+} \, || \, \vec{\mathcal{V}}\,) \;=\; 
D(\vec{\mathcal{W}}_{-} \, || \, \vec{\mathcal{V}}\,)$.
The formula for relative entropy derived in Appendix A is :

$$
\mathcal{D}(\, \varrho_k \, \| \,  \phi \, ) \;=\; 
\frac{1}{2} \, \log_2 \left ( 1\;-\;r_k^2 \right ) \;+\; 
\frac{r_k}{2} \, \log_2 \left ( \frac{1\;+\;r_k}{1\;-\;r_k} \right )
\;-\; 
\frac{1}{2}
\;  \log_2 \left ( 1 \;-\; q^2\; \right ) 
\;-\; \frac{\; \vec{\mathcal{W}}_k \, \bullet \,  \vec{\mathcal{V}} \; }
{\; 2 \; q\; } \; \log_2 \left ( \frac{  1 \;+\; q}{1 \;-\; q} \right)
$$

$$
\;=\; 
\frac{1}{2} \, \log_2 \left  ( 1\;-\;r_k^2 \right ) \;+\; 
\frac{r_k}{2} \, \log_2 \left ( \frac{1\;+\;r_k}{1\;-\;r_k} \right )
\;-\; 
\frac{1}{2}
\;  \log_2 \left ( 1 \;-\; q^2\; \right ) 
\;-\; \frac{r_k \; \cos ( \theta_k )  }
{\; 2 \; } \; \log_2 \left ( \frac{  1 \;+\; q}{1 \;-\; q} \right)
$$

where $\theta_k$ is the angle between $\vec{\mathcal{W}}_k$ 
and $\vec{\mathcal{V}}$. 
Intuitively, one notes that the nearly circular
relative entropy contours about $\phi \;\equiv\; \vec{\mathcal{V}}$
tells us that  $\vec{\mathcal{V}} \; \approx \; 
\frac{\; \vec{\mathcal{W}}_{+} \;+\; \vec{\mathcal{W}}_{-} \; }{2}$.
Given the channel parameter numbers, this fact about 
$\vec{\mathcal{V}}$,
together with the linear nature of the channel ellipsoid,
tell us that $\theta_{+} \;=\; 0$ and 
$\theta_{-} \;=\; \pi$, so that 
$\cos( \, \theta_{+}\, ) \;=\; 1$ and 
$\cos( \, \theta_{-}\, ) \;=\; -1$.
Using this information about the 
$\theta_k$, and the identity 

$$
\tanh^{(-1)} [ \; x \; ] \;=\;  
\frac{1}{2} \, \log \left ( \frac{1\;+\;x}{1\;-\;x} \right )
$$

the relative entropy equality relation
between the two endpoints of the linear channel can be solved for q.

$$
q_{optimum} \;=\; \tanh \left [ \; 
\frac{ \frac{1}{2} \; \ln \left [ \frac{1 - r_{+}^2}{1 - r_{-}^2} \right ] \;+\;
r_{+} \; \tanh^{(-1)} [ r_{+} ] \;- \; r_{-} \; \tanh^{(-1)} [ r_{-} ] }{r_{+} \;+\;r_{-} } \; \right ] \;=\; 0.2125. 
$$

Thus, 

$$
\vec{\mathcal{W}}_{+} \;=\; \bmatrix{ 0 \cr 0 \cr 0.6 }
,\;\;\;\vec{\mathcal{W}}_{-} \;=\; \bmatrix{ 0 \cr 0 \cr -0.2 }, \;\;\; and\;\;\;
\vec{\mathcal{V}} \;=\; \bmatrix{ 0 \cr 0 \cr 0.2125 }
$$

The corresponding density matrices are :

$$
\rho_{+}  \;=\; \frac{1}{2} \; (\; \mathcal{I} \;+\; \vec{\mathcal{W}_{+}}
\bullet \vec{\sigma} \; ) ,
\;\;\;\rho_{-}  \;=\; \frac{1}{2} \; (\; \mathcal{I} \;+\; \vec{\mathcal{W}_{-}}
\bullet \vec{\sigma} \; ) ,
\;\;\;\phi \;=\; \frac{1}{2} \; (\; \mathcal{I} \;+\; \vec{\mathcal{V}}
\bullet \vec{\sigma} \; ) 
$$

This yields 
$\mathcal{D}(\, \rho_{+} \, \| \, \phi \, ) \;=\; 
\mathcal{D}( \, \rho_{-} \, \| \,  \phi \,  ) \;=\; 0.1246$.
Thus, the HSW channel capacity $\mathcal{C}_1$ is 0.1246.
The location of the two density matrices
$\rho_{+}$ and $\rho_{-}$ are shown in Figure 13
below as {\textbf O}. 

Furthermore, the Schumacher-Westmoreland analysis tells us that the two states
$\rho_{+}$ and $\rho_{-}$  must average to $\phi$, in the sense that if $p_{+}$
and $p_{-}$ are the a priori probabilities of the two output signal states, then
$p_{+} \; \rho_{+} \;+\; p_{-} \; \rho_{-} \;=\; \phi$. In our Bloch sphere
notation, this relationship becomes 
$p_{+} \; \vec{\mathcal{W}}_{+} \;+\; p_{-} \; 
\vec{\mathcal{W}}_{-} \;=\;  \vec{\mathcal{V}}$. The asterisk 
({\textbf * }) in Figure 13
below shows the position of $ \vec{\mathcal{V}}$.  

Another relation relating the a priori probabilities
$p_{+}$ and $p_{-}$ is $p_{+} \;+\; p_{-} \;=\;1$.  
Using these two equations, we can solve for 
the a priori probabilities
$p_{+}$ and $p_{-}$. For our example, 

$$
p_{+} \vec{\mathcal{W}}_{+} \;+\;  p_{-} \vec{\mathcal{W}}_{-}
\;=\; p_{+} \bmatrix{ 0 \cr 0 \cr  0.6 } \;+\;  
p_{-} \bmatrix{ 0 \cr 0 \cr -0.2 }
\;=\; \vec{\mathcal{V}} \;=\; \bmatrix{ 0 \cr 0 \cr 0.2125 }
$$

Solving for $p_{+}$ and $p_{-}$ yields
$p_{+} \;=\; 0.5156$ and $p_{-} \;=\; 0.4844$.

Note that here we have found the optimum {\em output} signal states
$\rho_{+}$ and $\rho_{-}$. From these one can find the optimum 
{\em input} signal states by finding the states $\varphi_{+}$ and 
$\varphi_{-}$ which map to the respective optimum output states
$\rho_{+}$ and $\rho_{-}$. In our example above, these are 
$\varphi_{+} \; \rightarrow \; \vec{\mathcal{W}}_{+}^{Input} \;=\; \bmatrix{ 0 \cr 0 \cr 1 }$ and 
$\varphi_{-} \; \rightarrow \; \vec{\mathcal{W}}_{-}^{Input} \;=\; 
\bmatrix{ 0 \cr 0 \cr -1 }$.

\begin{center}
\includegraphics*[angle=-90,scale=0.6]{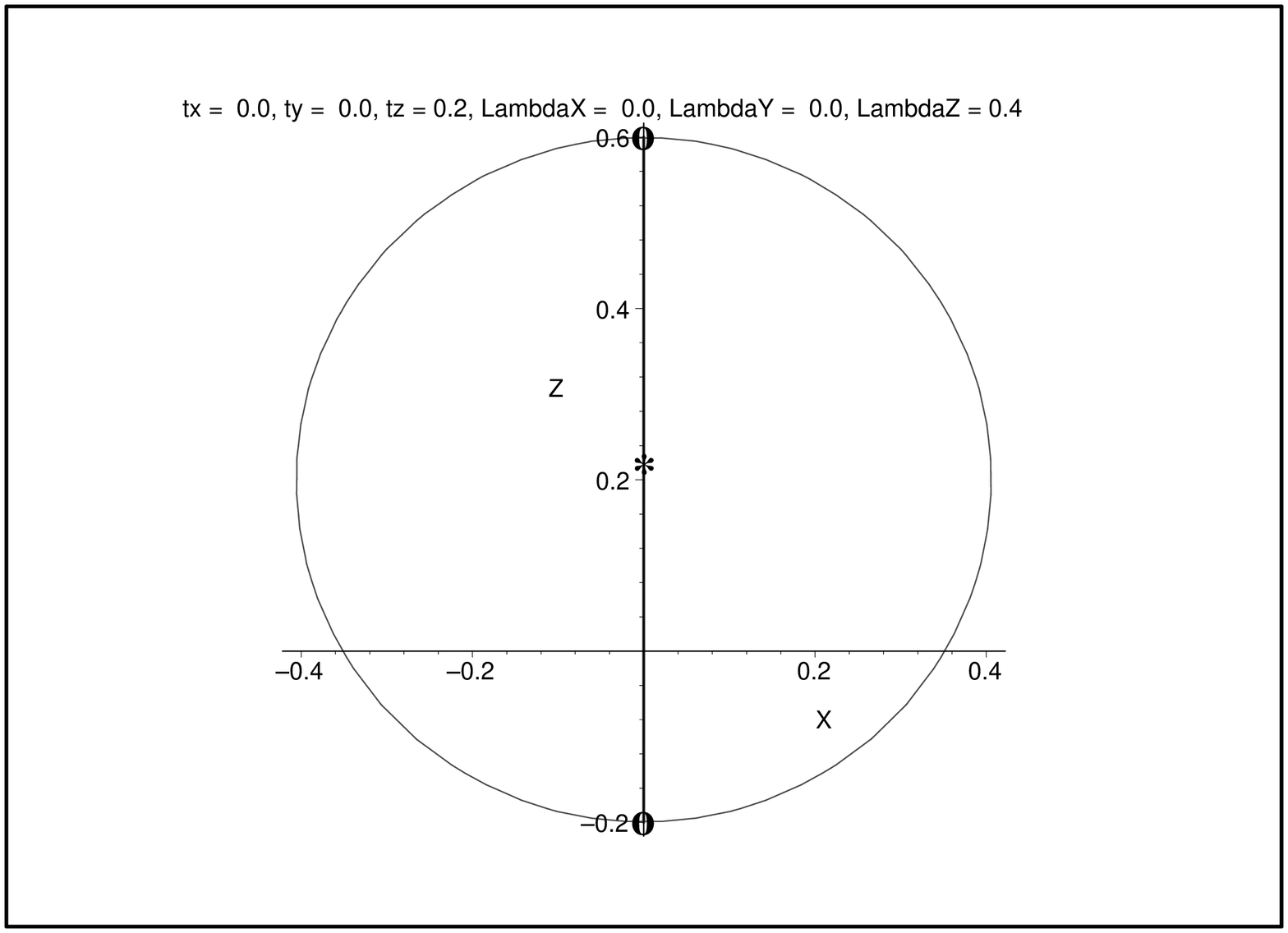}
\end{center}

\begin{center}
Figure 13: 
The intersection in the Bloch sphere X-Z plane of
a linear channel ellipsoid and the optimum relative
entropy contour. The optimum output signal
states are shown as {\textbf O}. 
\end{center}

For the general linear channel, where any or all of the $t_k$
can be non-zero, we can reduce the capacity calculation to
the solution of a single, one dimensional transcendental equation. 
( Please see Appendix B for the full derivation. )

Define 

$$
r_{+}^2 \; =\; t_x^2 \;+\; t_y^2 \;+\; ( \; t_z \;+\; \lambda_z \; )^2 
$$ 

$$
q^2 \; =\; t_x^2 \;+\; t_y^2 \;+\; ( \; t_z \;+\; \beta \, \lambda_z \; )^2 
$$ 

$$
r_{-}^2 \; =\; t_x^2 \;+\; t_y^2 \;+\; ( \; t_z \;-\; \lambda_z \; )^2 
$$ 

The two quantities $r_{+}$ and $r_{-}$ are the Euclidean 
distances from the Bloch sphere origin to the signalling 
states $\rho_{+} \;\equiv \; \vec{\mathcal{W}}_{+}$ and 
$\rho_{-} \;\equiv \; \vec{\mathcal{W}}_{-}$ respectively. 
The quantity $q$ is the Euclidean distance from
the Bloch sphere origin to the density matrix 
$\phi \;\equiv \; \vec{\mathcal{V}}$.
We define the three Bloch vectors 
$\vec{r}_{+}$,  $\vec{q}$ and $\vec{r}_{-}$ in Figure 14 below, and
refer to their respective magnitudes as 
$r_{+}$,  $q$, and $r_{-}$.

\begin{center}

\setlength{\unitlength}{0.00033300in}%
\begingroup\makeatletter\ifx\SetFigFont\undefined
\def\x#1#2#3#4#5#6#7\relax{\def\x{#1#2#3#4#5#6}}%
\expandafter\x\fmtname xxxxxx\relax \def\y{splain}%
\ifx\x\y   
\gdef\SetFigFont#1#2#3{%
  \ifnum #1<17\tiny\else \ifnum #1<20\small\else
  \ifnum #1<24\normalsize\else \ifnum #1<29\large\else
  \ifnum #1<34\Large\else \ifnum #1<41\LARGE\else
     \huge\fi\fi\fi\fi\fi\fi
  \csname #3\endcsname}%
\else
\gdef\SetFigFont#1#2#3{\begingroup
  \count@#1\relax \ifnum 25<\count@\count@25\fi
  \def\x{\endgroup\@setsize\SetFigFont{#2pt}}%
  \expandafter\x
    \csname \romannumeral\the\count@ pt\expandafter\endcsname
    \csname @\romannumeral\the\count@ pt\endcsname
  \csname #3\endcsname}%
\fi
\fi\endgroup
\begin{picture}(9000,6738)(1201,-7015)
\thicklines
\put(1801,-3961){\line( 2, 1){7780}}
\put(1801,-3961){\line( 5,-2){7758.621}}
\put(1801,-3961){\line( 6, 1){7783.784}}
\put(9571,-101){\line( 0,-1){7000}}
\put(1201,-4561){\makebox(0,0)[lb]{\smash{\SetFigFont{12}{14.0}{rm}Bloch}}}
\put(1201,-4956){\makebox(0,0)[lb]{\smash{\SetFigFont{12}{14.0}{rm}Sphere}}}
\put(1201,-5351){\makebox(0,0)[lb]{\smash{\SetFigFont{12}{14.0}{rm}Origin}}}
\put(10201,-2761){\makebox(0,0)[lb]{\smash{\SetFigFont{12}{14.0}{rm}q}}}
\put(10201,-361){\makebox(0,0)[lb]{\smash{\SetFigFont{12}{14.0}{rm}$r_{+}$}}}
\put(10201,-6961){\makebox(0,0)[lb]{\smash{\SetFigFont{12}{14.0}{rm}$r_{-}$}}}
\end{picture}

\end{center}

\begin{center}
Figure 14: 
Definition of the Bloch vectors 
$\vec{r}_{+}$, 
$\vec{q}$, and  
$\vec{r}_{-}$ used in the derivation below. 
\end{center}

We solve the transcendental equation below for $\beta$.

$$
\frac{ 4 \; \lambda_z \; ( t_z \;+\; \beta \lambda_z \;) \; 
\tanh^{(-1)} ( \, q \, ) \; }{q} \;=\; 
2 \; r_{+} \; \tanh^{(-1)} ( r_{+} ) \;  - \; 2 \,
r_{-} \; \tanh^{(-1)} ( r_{-} ) \;+\; \ln ( \, 1\, - \, r_{+}^2 \,  )  \; 
-  \; \ln(\, 1 \, - \, r_{-}^2 \, )
$$

Note that $q$ is a function
of $\beta$, while $r_{+}$ and $r_{-}$ are not. Thus, the
right hand side remains constant while $\beta$ is varied. The smooth
nature of the functions of $\beta$ on the left hand side allow a solution
for $\beta$ to be found fairly easily. 

As in our simpler linear channel example above, we have  

$$
\vec{\mathcal{W}_{+}} \;=\; \bmatrix{ t_x \cr t_y \cr t_z \;+\; \lambda_z }
,\;\;\;
\vec{\mathcal{W}_{-}} \;=\; \bmatrix{ t_x \cr t_y  \cr t_z \;-\; \lambda_z },
\;\;\; and\;\;\;
\vec{\mathcal{V}} \;=\; \bmatrix{ t_x \cr t_y \cr t_z + \beta \, \lambda_z }
\;.
$$

where $\beta \; \in ( \,-1,1\,)$. The corresponding density 
matrices are : 

$$
\rho_{+}  \;=\; \frac{1}{2} \; (\; \mathcal{I} \;+\; \vec{\mathcal{W}_{+}}
\bullet \vec{\sigma} \; ) ,
\;\;\;\rho_{-}  \;=\; \frac{1}{2} \; (\; \mathcal{I} \;+\; \vec{\mathcal{W}_{-}}
\bullet \vec{\sigma} \; ) ,
\;\;\;\phi \;=\; \frac{1}{2} \; (\; \mathcal{I} \;+\; \vec{\mathcal{V}}
\bullet \vec{\sigma} \; ) 
$$

The channel capacity $\mathcal{C}_{1}$ is found from the relations

$$
D(\rho_{+} || \phi ) \;=\; D(\rho_{-} || \phi ) \;=\; \chi_{optimum} 
\;=\; \mathcal{C}_{1}
\;.
$$

The a priori signaling probabilities are found by solving the 
simultaneous probability equations $p_{+} \;+\; p_{-} \;=\;1$, and 

$$
p_{+} \vec{\mathcal{W}_{+}} \;+\;  p_{-} \vec{\mathcal{W}_{-}}
\;=\; p_{+} \bmatrix{ t_x \cr t_y \cr  t_z \;+\; \lambda_z } \;+\;  
p_{-} \bmatrix{ t_x \cr t_y \cr t_z \;-\; \lambda_z }
\;=\; \vec{\mathcal{V}} \;=\; \bmatrix{ t_x \cr t_y \cr t_z \;+\; \beta
\, \lambda_z }
$$

This leads to a second probability equation of
$p_{+} \;-\; p_{-} \;=\;\beta$, yielding:

$$
p_{+} \;=\; \frac{ 1 \;+\; \beta }{2} 
\;\;\;\;\;\;\;
\;\;\;\;\;\;\;
and 
\;\;\;\;\;\;\;
\;\;\;\;\;\;\;
p_{-} \;=\; \frac{ 1 \;-\; \beta }{2} 
$$

\subsection{A More General Linear Channel Example}

In the simple linear channel example above, we used 
$\{ \; t_x \,=\, t_y \,=\, 0\; , \; \lambda_x \, = \, \lambda_y \, =\; 0\;\}$.  
This choice yielded a 
rotational symmetry about the Z - axis which assured us the 
location of the optimum average output density matrix 
$\rho \;=\; p_{+} \, \rho_{+} \;+\; p_{-} \, \rho_{-}$
was on the Z - axis. We used this fact to advantage in predicting the
angles $\theta_{\{+\,,\,-\}}$, 
where $\theta_{\{+\,,\,-\}}$ was the angle between 
$\vec{\mathcal{W}}_{\{+\,,\,-\}}$ 
and $\vec{\mathcal{V}}$. 
Since we knew $\vec{\mathcal{W}}_{\{+\,,\,-\}}$ 
lay on the Z - axis, we found
$\theta_{+} \;=\; 0$ and 
$\theta_{-} \;=\; \pi$, simplifying the 
$\cos\left (\,\theta_{\{+\,,\,-\}}\, \right )$ terms in the relative entropy 
expressions for 
$D(\rho_{+} || \phi)$ and $D(\rho_{-} || \phi)$.
In general, we do not have values of $\pm \, 1$ for 
$\cos\left (\,\theta_{\{+\,,\,-\}} \, \right )$, and this complicates finding 
a solution for the linear channel relation 
$D(\rho_{+} || \phi) \;=\; D(\rho_{-} || \phi)$.

A more general linear channel example is one where the 
parameters $\{ \; t_x \;,\; \; t_y \;,\; \; t_z \; \}$
are all non-zero.
Consider the parameter set  
$\{ \; 
t_x \, = \, 0.1, \; 
t_y \, = \, 0.2, \; 
t_z \, =  \,0.3, \; 
\lambda_x \, = \, 0,\; 
\lambda_y \, = \, 0,\; 
\lambda_z \, = \, 0.4\; \}$.  
Solving the transcendental equation derived in Appendix B yields 
$\beta \;=\; 0.0534$ and
$\vec{\mathcal{V}} \;=\; \bmatrix{ 0.1 \cr 0.2 \cr 0.3214 }$.
Using the density matrix $\phi$ calculated from the Bloch vector
$\vec{\mathcal{V}}$ gives us a HSW channel capacity 
$\mathcal{C}_1$ of
$D(\rho_{+} || \phi ) \;=\; D(\rho_{-} || \phi ) \;=\; 0.1365$.

As discussed above, $p_{+} \;+\; p_{-} \;=\;1$, 
and $p_{+} \;-\; p_{-} \;=\;\beta$.
Solving for $p_{+}$ and $p_{-}$ yields
$p_{+} \;=\; 0.5267$ and $p_{-} \;=\; 0.4733$.

The optimum {\em input} Bloch vectors are :

$$
\varphi_{+} \; \rightarrow \; \vec{\mathcal{W}}_{+}^{Input} \;=\; \bmatrix{ 0 \cr 0 \cr 1 } \qquad  and  \qquad 
\varphi_{-} \; \rightarrow \; \vec{\mathcal{W}}_{-}^{Input} \;=\; 
\bmatrix{ 0 \cr 0 \cr -1 }\; .
$$

The optimum {\em output } Bloch vectors are :

$$
\rho_+ \;=\; \mathcal{E}(\, \varphi_+\, )
\; \rightarrow \; \vec{\mathcal{W}}_{+}^{Output} \;=\; \bmatrix{ 0.1 \cr 0.2 \cr 0.7 } \qquad  and  \qquad 
\rho_- \;=\; \mathcal{E}(\, \varphi_- \, )
\; \rightarrow \; \vec{\mathcal{W}}_{-}^{Output} \;=\; 
\bmatrix{ 0.1 \cr 0.2 \cr -0.1 }\; .
$$ 

Below we show in Figure 15 and Figure 16 the
$\{x,z\}$ and $\{y,z\}$ slices of the {\em linear}
channel ellipsoid. One
can see that the relative entropy curve 
$D( \, \rho \, \| \, \phi \, ) \;=\; \mathcal{C}_1 \;=\; 0.1365$ touches
the ellipsoid at two locations in both cross sections.  (The $\{x,y\}$
cross section is trivial.)

\begin{center}
\includegraphics*[angle=-90,scale=0.53]{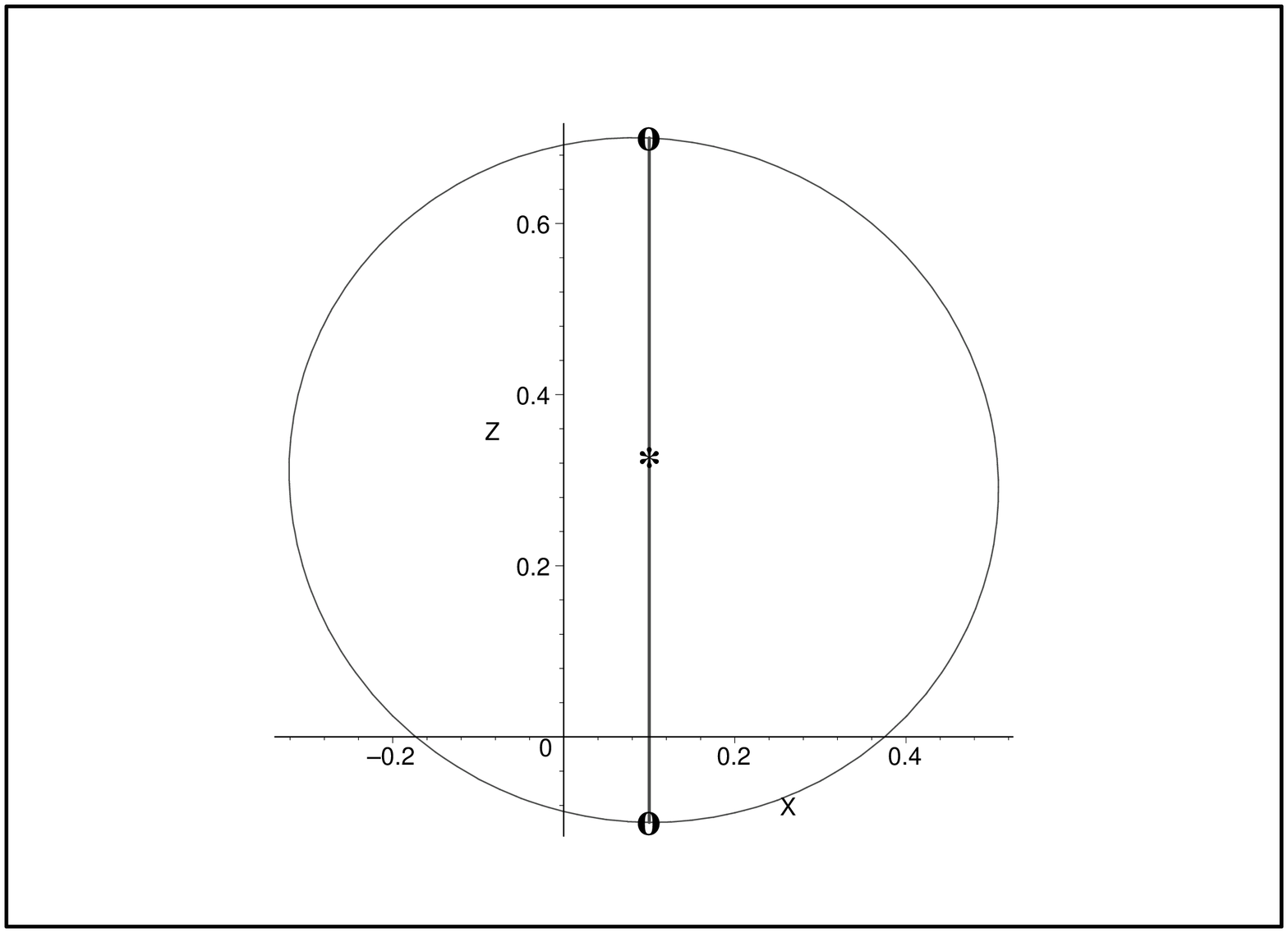}
\end{center}

\begin{center}
Figure 15: 
The intersection in the Bloch sphere X-Z plane of
a linear channel ellipsoid and the optimum relative
entropy contour. The optimum output signal
states are shown as {\textbf O}. 
\end{center}

\begin{center}
\includegraphics*[angle=-90,scale=0.53]{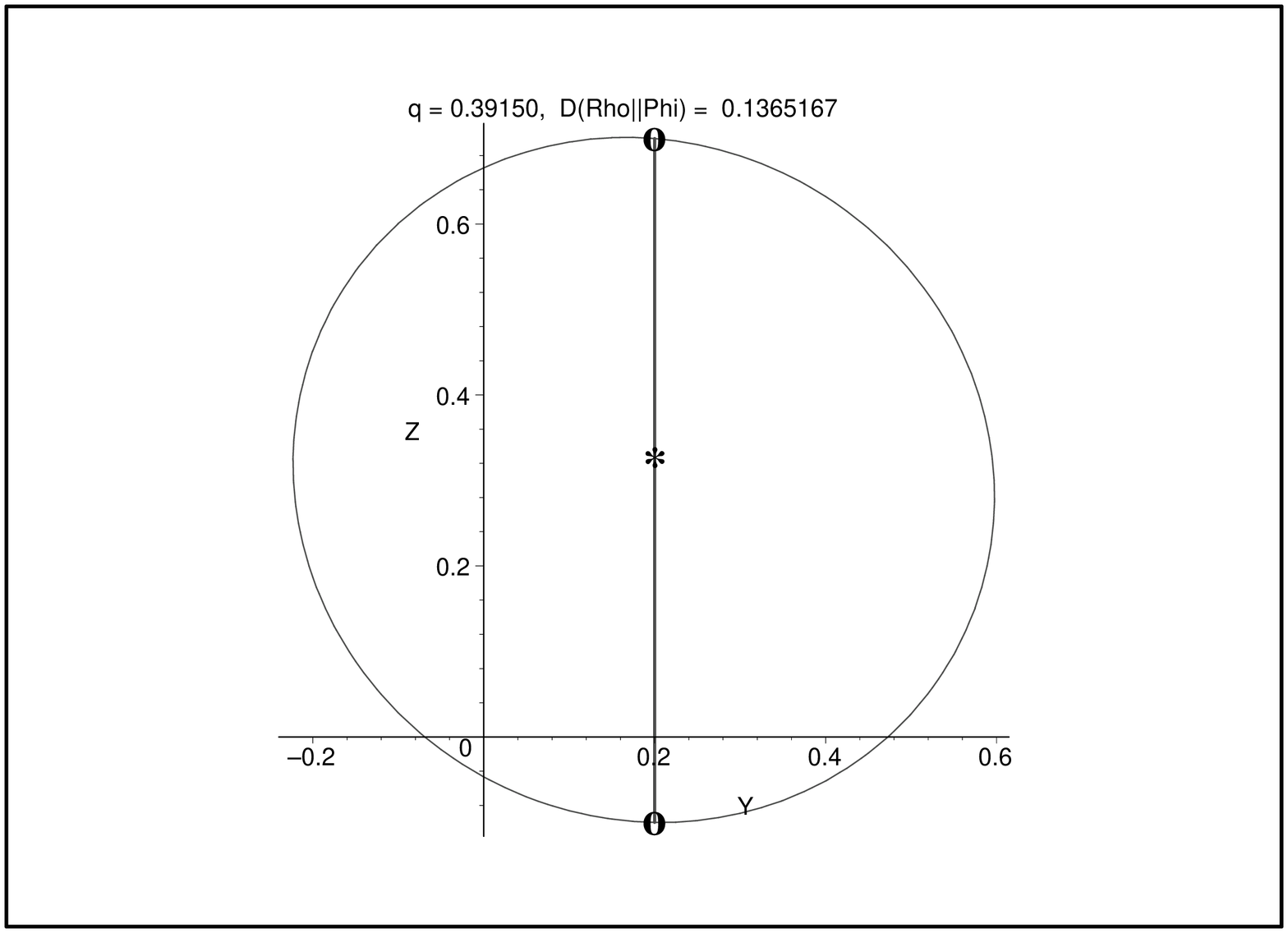}
\end{center}

\begin{center}
Figure 16: 
The intersection in the Bloch sphere Y-Z plane of
a linear channel ellipsoid and the optimum relative
entropy contour. The optimum output signal
states are shown as {\textbf O}. 
\end{center}

\section{Planar Channels}

A planar channel is a quantum channel where 
two $\lambda_k$ are non-zero, and one $\lambda_k$ is zero.
For a planar channel, the $\{ \; t_k \; \}$ can have any values 
allowed by complete
positivity. A planar channel restricts the possible output 
density matrices to lie in the plane in the Bloch sphere
which is specified by the non-zero $\lambda_k$.
In comparison to the linear channels discussed above, the 
planar channels additional output degree of freedom  (planar has 
two non-zero $\lambda_k$ versus a single linear non-zero $\lambda_k$) 
means a slightly different approach to determining $\mathcal{C}_1$ than
that discussed for linear channels must be developed. As for linear channels, 
we seek
to find the optimum density matrix $\phi \;\equiv \; \vec{\mathcal{V}}$ 
interior to the ellipsoid
which minimizes the distance to the
most "distant", in a relative entropy sense, point(s) on the ellipsoid
surface.  
We shall find the optimum $\vec{\mathcal{V}}$ in two ways : graphically and
iteratively.
Both approaches utilize the following theorem from Schumacher and
Westmoreland\cite{Schumacher99}.

{\em Theorem : }

$$
\mathcal{C}_1 \quad = \quad Min_{\phi}
\quad Max_{\rho} \quad D(\, \rho \, \| \, \phi \, ) 
$$

The maximum is taken over the {\em surface} of the ellipsoid, and the
minimum is taken over the {\em interior} of the ellipsoid.
In order to apply the min max formula above for $\mathcal{C}_1$ for 
planar channels, we need a result about the uniqueness of 
the average output ensemble density matrix 
$\rho \;=\; \sum_k \; p_k \, \rho_k$ for different optimal 
ensembles $\{ \; p_k \;,\; \rho_k \; \}$. 

\subsection{Uniqueness Of The Average Output Ensemble Density Matrix}

The question we address is if there exists two optimum
signalling ensembles, $\{ \; p_k \;,\; \rho_k \; \}$ and
$\{ \; p'_k \;,\; \rho'_k \; \}$ of channel output states,
whether the two resulting average density matrices,
$\rho \;=\; \sum_k \; p_k \, \rho_k$ and 
$\rho' \;=\; \sum_k \; p'_k \, \rho'_k$ are equal.

{\em Theorem : } The density matrix $\phi$ which achieves
the minimum in the min-max formula above for $\mathcal{C}_1$ is unique.

{\em Proof :}

From property V in Section 2.2, we know the 
optimum $\phi$ which attains the minimum above
must correspond to the average of a set of signal states of an optimum
signalling ensemble. We shall prove the uniqueness of $\phi$ by 
postulating there are two optimum signal ensembles, with possibly
different average density matrices, $\sigma$ and $\xi$. We will then prove
that $\sigma$ must equal $\xi$, thereby implying $\phi$ is unique. 

Let $\{ \, \alpha_i \,,\, \rho_i \, \}$ be an optimum signal ensemble,
with probabilities $\alpha_i$ and density matrices $\rho_i$, where
$\alpha_i \;\geq \; 0$ and $\sum_i \; \alpha_i \;=\; 1$. 
Define $\sigma \;=\; \sum_i \; \alpha_i \; \rho_i$. 
By property I in Section 2.2, we know that 
$ \mathcal{D}(\, \rho_i \, \| \, \sigma \, ) \;=\; \chi_{optimum} \;=\;
\mathcal{C}_1 \quad \forall \; i$.

Now consider a second, optimum signal ensemble 
$\{ \, \beta_j \,,\, \phi_j \, \}$
differing in at least one 
density matrix $\rho_i$ and/or one probability $\alpha_i$ from the
optimum ensemble  $\{ \, \alpha_i \,,\, \rho_i \, \}$.
Define $\xi \;=\; \sum_j \; \beta_j \; \phi_j$. 
Consider the quantity 
$\sum_i \; \alpha_i \; \mathcal{D}( \, \rho_i \, \| \, \xi \, )$.
Let us apply Donald's equality, which is discussed in Appendix C.

$$
\sum_i \; \alpha_i \; \mathcal{D}( \, \rho_i \, \| \, \xi \, ) \;=\;
\mathcal{D}( \, \sigma \, \| \, \xi \, ) \;+\;
\sum_i \; \alpha_i \; \mathcal{D}( \, \rho_i \, \| \, \sigma \, )
$$

Since 
$\mathcal{D}( \, \rho_i \, \| \, \sigma \, ) \;=\; \chi_{optimum} \quad \forall \; i$,  and $\sum_i \; \alpha_i \;=\; 1$,
we obtain :

$$
\sum_i \; \alpha_i \; \mathcal{D}( \, \rho_i \, \| \, \xi \, ) \;=\;
\mathcal{D}( \, \sigma \, \| \, \xi \, ) \;+\; \chi_{optimum}
$$

From property II in Section 2.2, 
since $\xi$ is the average of a set of optimal signal
states $\{ \, \beta_j \,,\, \phi_j \, \}$, we know that
$\mathcal{D}( \, \rho_i \, \| \, \xi \, ) \;\leq\; \chi_{optimum} \; \forall \, i$.
Thus
$\sum_i \; \alpha_i \; \mathcal{D}( \, \rho_i \, \| \, \xi \, )
\;\leq\;\chi_{optimum}$. 
Combining this inequality constraint on 
$\sum_i \; \alpha_i \; \mathcal{D}( \, \rho_i \, \| \, \xi \, )$
with what we know about 
$\sum_i \; \alpha_i \; \mathcal{D}( \, \rho_i \, \| \, \xi \, )$ 
from Donald's equality, we obtain the two relations :

$$
\sum_i \; \alpha_i \; \mathcal{D}( \, \rho_i \, \| \, \xi \, ) \;=\; 
\mathcal{D}( \, \sigma \, \| \, \xi \, ) \;+\; \chi_{optimum}
\quad \quad and \quad \quad 
\sum_i \; \alpha_i \; \mathcal{D}( \, \rho_i \, \| \, \xi \, ) \;\leq\; 
\chi_{optimum}
$$

From Klein's inequality, 
we know that 
$\mathcal{D}( \, \sigma \, \| \, \xi \, )\; \geq \; 0$, with equality
iff  $\sigma \, \equiv \, \xi$.
Thus, the only way the equation

$$
\sum_i \; \alpha_i \; \mathcal{D}( \, \rho_i \, \| \, \xi \, ) \;=\; 
\mathcal{D}( \, \sigma \, \| \, \xi \, ) \;+\; \chi_{optimum}
$$

can be satisfied is if 
we have $\sigma \, \equiv \, \xi$, for then 
$\mathcal{D}( \, \sigma \, \| \, \xi \, )\; = \; 0$
and we have  

$$
\sum_i \; \alpha_i \; \mathcal{D}( \, \rho_i \, \| \, \xi \, ) \;=\; 
\mathcal{D}( \, \sigma \, \| \, \xi \, ) \;+\; \chi_{optimum}
\;=\; \chi_{optimum}
$$

and

$$
\sum_i \; \alpha_i \; \mathcal{D}( \, \rho_i \, \| \, \xi \, ) \;=\; 
\sum_i \; \alpha_i \; D( \, \rho_i \, \| \, \sigma  \, ) \;=\; 
\sum_i \; \alpha_i \; \chi_{optimum} \;=\; \chi_{optimum}
$$

Therefore, only in the case where  $\sigma \, \equiv \, \xi$ is 
Donald's equality satisfied.
Since $\sigma$ and $\xi$ were the average output density 
matrices for two different, but 
arbitrary optimum signalling ensembles, we conclude 
the average density matrices of all optimum signalling 
ensembles must be equal, thereby implying $\phi$ is unique. 

$\bigtriangleup$ - {\em End of Proof}. 

Note that although we are primarily concerned with qubit channels in this
paper, only generic properties of the relative entropy were used in the
above proof of uniqueness, and therefore the result holds for {\em all} 
channels.

\subsection{Graphical Channel Optimization Procedure}

We shall now describe a graphical technique for finding 
$\phi_{optimum} \;\equiv \; \vec{\mathcal{V}}_{optimum}$.
Recall the contour surfaces of constant relative entropy 
for various values of $\vec{\mathcal{V}}$ shown previously. 
We seek to adjust the location of $\vec{\mathcal{V}}$ inside the
channel ellipsoid such that the largest possible contour value
$\mathcal{D}_{max} \;=\; \mathcal{D}(\, \vec{\mathcal{W}} \,\|\, \vec{\mathcal{V}} \, )$
touches the ellipsoid surface, and the
remainder of the $\mathcal{D}_{max}$ contour surface lies entirely outside the
channel ellipsoid. Our linear channel example illustrated this idea.
In that example, the $\mathcal{D}_{max}$ contour
intersects the ''ellipsoid'' at $r_{+}$ and
$r_{-}$, and otherwise lies outside the line segment 
between $r_{+}$ and $r_{-}$ representing the
convex hull of $\mathcal{A}$. 
(Recall from the discussion of the Schumacher and
Westmoreland paper in Section 2.2 that the points on the
ellipsoid surface were defined as the set
$\mathcal{A}$, and the interior of the 
ellipsoid, where $\vec{\mathcal{V}}$ 
lives, is the convex hull of $\mathcal{A}$.)

A good place to start is with  
$\vec{\mathcal{V}}_{initial} \;=\; \bmatrix{ t_x \cr t_y \cr t_z }$.
We then "tweak" $\vec{\mathcal{V}}$ 
as described above to find $\vec{\mathcal{V}}_{optimum}$.
Note that $\vec{\mathcal{V}}_{optimum}$ 
should be near $\vec{\mathcal{V}}_{initial}$ because of 
the {\em almost} radial symmetry of $\mathcal{D}$ about
$\vec{\mathcal{V}}$ as seen in Figures 2 through 11. 

This technique is graphically implementing 
property IV
in Section 2.2. In Bloch sphere notation, we have :

$$
\mathcal{C}_1 \;=\; 
\mbox{\Large Min}
_{\vec{\mathcal{V}}} \quad  \quad  
\mbox{\Large Max}
_{\vec{\mathcal{W}}} \quad  \quad  
\mathcal{D} \left ( \; \vec{\mathcal{W}} \;  \| \; \vec{\mathcal{V}}\;\right )
$$

\noindent
where $\vec{\mathcal{W}}$ is on the channel ellipsoid surface 
and $\vec{\mathcal{V}}$ is in the interior of the ellipsoid. 
Moving $\vec{\mathcal{V}}$ from the optimum position described above
will increase 
$\;Max _{\vec{\mathcal{W}}} \quad  
\mathcal{D} \left ( \; \vec{\mathcal{W}} \;  \| \; \vec{\mathcal{V}}\;\right )$,
since a larger contour value of $\mathcal{D}$ would then intersect the
channel ellipsoid surface, thereby {\em increasing} 
$\; Max _{\vec{\mathcal{W}}} \quad  
\mathcal{D} \left ( \; \vec{\mathcal{W}} \;  \| \; \vec{\mathcal{V}}\;\right )$.
Yet $\vec{\mathcal{V}}$ should be adjusted to {\em minimize} 
$\; Max _{\vec{\mathcal{W}}} \quad  
\mathcal{D} \left ( \; \vec{\mathcal{W}} \;  \| \; \vec{\mathcal{V}}\;\right )$.

\subsection{Iterative Channel Optimization Procedure}

For the iterative treatment, we outline an
algorithm which converges to $\vec{\mathcal{V}}_{optimum}$.
First, we need a lemma. 

{\em Lemma :} Let $\vec{\mathcal{V}}$ and $\vec{\mathcal{W}}$ 
be any two Bloch sphere vectors. Define a third Bloch sphere vector
$\vec{\mathcal{U}}$ as :

$$
\vec{\mathcal{U}} \;=\; ( \; 1 \; - \alpha \; ) \; \vec{\mathcal{W}} \;+\; 
\alpha \; \vec{\mathcal{V}}
$$ 

where $\alpha \;\in \; (0,1)$. 
Then 

$$
\mathcal{D}(\, \vec{\mathcal{W}} \,  \|  \, \vec{\mathcal{U}} \, ) \; < \; 
\mathcal{D}( \, \vec{\mathcal{W}} \, \| \,  \vec{\mathcal{V}} \, )
$$ 

{\em Proof :} By the joint convexity property of the relative entropy 
\cite{Nielsen00a} :
$$
\mathcal{D}( \, \{\, \alpha \,\rho_1 +\, ( \, 1 \,-\, \alpha \,) \, \rho_2  \, 
\}\, \| \,  \{\,
\alpha \,\phi_1 +\, ( \, 1 \,-\, \alpha \,) \, \phi_2 \,  \} \,) \;
\leq \; 
\alpha \; \mathcal{D}( \,\rho_1 \,  \| \, \phi_1 \, ) \; + \;
( \, 1 \, - \, \alpha \, ) \; \mathcal{D}( \,\rho_2 \,  \| \, \phi_2 \, )
$$
where $\alpha \;\in \; (0,1)$. 
Let $\rho_1 \,=\, \rho_2 \,\equiv \, \vec{\mathcal{W}}$,   
$\phi_1 \,\equiv \, \vec{\mathcal{V}}$ and  
$\phi_2 \,\equiv \, \vec{\mathcal{W}}\, $ with
$\, \vec{\mathcal{U}} \,=\, ( \; 1 \; - \alpha \; ) \; \vec{\mathcal{W}} \;+\; 
\alpha \; \vec{\mathcal{V}}$. 
We obtain :
$$
\mathcal{D}(\, \vec{\mathcal{W}} \,  \|  \, \vec{\mathcal{U}} \, ) \; = \; 
\mathcal{D}( \, \vec{\mathcal{W}} \, \| \, 
\alpha \,\vec{\mathcal{V}} +\, ( \, 1 \,-\, \alpha\,) \, 
\vec{\mathcal{W}} \, ) \;
\leq \; 
\alpha \; \mathcal{D}( \,\vec{\mathcal{W}} \,  \| \, \vec{\mathcal{V}} \, ) \;
\; + \;( \, 1 \, - \, \alpha \, ) 
\; \mathcal{D}( \,\vec{\mathcal{W}} \,  \| \, \vec{\mathcal{W}} \, ) \;
$$

But $\mathcal{D}( \,\vec{\mathcal{W}} \,  \| \, \vec{\mathcal{W}} \, ) \; = \; 0$,
by Klein's inequality\cite{Nielsen00a}. Thus,
 
$$
\mathcal{D}(\, \vec{\mathcal{W}} \,  \|  \, \vec{\mathcal{U}} \, ) \; \leq \; 
\alpha \; \mathcal{D}( \,\vec{\mathcal{W}} \,  \| \, \vec{\mathcal{V}} \, ) 
\; < \; 
\mathcal{D}( \,\vec{\mathcal{W}} \,  \| \, \vec{\mathcal{V}} \, )
$$

since $\alpha \;\in \; (0,1)$. 

$\bigtriangleup$ - {\em End of Proof}.  

We use the lemma above to guide us in iteratively adjusting $\vec{\mathcal{V}}$ 
to converge towards $\vec{\mathcal{V}}_{optimal}$.
Consider  
$\mathcal{D}( \,\vec{\mathcal{W}} \,  \| \, \vec{\mathcal{V}} \, )$, where
$\vec{\mathcal{W}} \,\in\, \mathcal{A}$ and 
$\vec{\mathcal{V}} \;\in \; \mathcal{B} \;\equiv $
the convex hull of $\mathcal{A}$.
We seek to find $\mathcal{C}_1 $ in an iterative fashion. 
We do this by holding $\vec{\mathcal{V}}$ fixed, and  
finding one of the 
$\vec{\mathcal{W}'} \,\in\, \mathcal{A}$ 
which maximizes 
$\mathcal{D}( \,\vec{\mathcal{W}} \,  \| \, \vec{\mathcal{V}} \, )$. 
From our lemma above, if we now move $\vec{\mathcal{V}}$ towards 
$\vec{\mathcal{W}'}$, we shall cause 
$\mathcal{D}_{max}(\, \vec{\mathcal{V}}\, ) \,=\, Max_{\vec{\mathcal{W}}} \; 
\mathcal{D}( \,\vec{\mathcal{W}} \,  \| \, \vec{\mathcal{V}} \, )$ to
decrease. We steadily decrease 
$\mathcal{D}_{max}( \,\vec{\mathcal{V}} \, )$
in this manner until we reach a point 
where any movement of 
$\vec{\mathcal{V}}$ will increase 
$\mathcal{D}_{max}( \,\vec{\mathcal{V}} \, )$. 
Our uniqueness theorem above tells us there is only one
$\vec{\mathcal{V}}_{optimal}$. Our lemma above tells us we cannot 
become stuck in a local minima in moving towards 
$\vec{\mathcal{V}}_{optimal}$.  Thus, when we reach the point where 
any movement of $\vec{\mathcal{V}}$ will increase 
$\mathcal{D}_{max}( \,\vec{\mathcal{V}} \, )$,
we are done and have found 
$\vec{\mathcal{V}}_{final } \;=\; \vec{\mathcal{V}}_{optimum}$.

To summarize, we find the optimum $\vec{\mathcal{V}}$ using the following algorithm.

1) Generate a random starting point 
$\vec{\mathcal{V}}_{initial}$
in the interior of the ellipsoid ( $\in \;  \mathcal{B}$ ).
( In actuality, since the contour surfaces of constant relative entropy
are {\em roughly} spherical about $\vec{\mathcal{V}}$, a good place to 
start is
$\vec{\mathcal{V}}_{initial} \;=\; \bmatrix{ t_x \cr t_y \cr t_z }$ .)

2) Determine the set of points $\{ \; \vec{\mathcal{W}'}\; \}$
on the ellipsoid surface most distant, in a relative
entropy sense, from our 
$\vec{\mathcal{V}}$.
This maximal distance is  
$\mathcal{D}_{max}( \, \vec{\mathcal{V}} \, )$ defined above as
$\mathcal{D}_{max}( \, \vec{\mathcal{V}} \, ) 
\;=\; Max_{ \vec{\mathcal{W}'}} \; 
\mathcal{D}( \, \vec{\mathcal{W}'} \, \| \, 
\vec{\mathcal{V}}\, )$.

3) Choose at random one Bloch sphere vector from our 
maximal set of points $\{ \; \vec{\mathcal{W}'}\; \}$. 
Call this selected point $\widehat{\vec{\mathcal{W}'}}$.  
In the 3 real dimensional Bloch sphere space, 
make a small step from $\vec{\mathcal{V}}$ towards the surface point
vector, $\widehat{\vec{\mathcal{W}'}}$. 
That is, update $\vec{\mathcal{V}}$ as follows :

$$
\vec{\mathcal{V}}_{new} \;=\; 
( \,1\;-\; \epsilon\, ) \, \vec{\mathcal{V}}_{old} \;+\; 
\epsilon\, \, \widehat{\vec{\mathcal{W}'}}
$$

4) Loop by going back to step 2) above, using our new, updated
$\vec{\mathcal{V}}_{new}$, and continue to loop until  
$\mathcal{D}_{max}$ is no longer changing.

This algorithm converges 
to $\phi_{optimum} \;\equiv\; \vec{\mathcal{V}}_{optimum}$, because we
steadily proceed downhill minimizing 
$Max _{\vec{\mathcal{W}}} \quad  
\mathcal{D} \left ( \; \vec{\mathcal{W}} \;  \| \; \vec{\mathcal{V}}\;\right )$,
and our lemma above tells us we can never get stuck in a local minima.

\subsection{Planar Channel Example}

We demonstrate the iterative algorithm above with a planar channel example.  
Let 
\linebreak
$\{ \; 
t_x \;= \; 0.3\; ,\;
t_y \;= \; 0.1\; ,\;   
t_z \;= \; 0\; , \,\;  
\lambda_x \;= \; 0.4\; ,\;   
\lambda_y \;= \; 0.5\; , \; 
\lambda_z \;= \; 0\;\}$.
The iterative algorithm outlined above yields 
$\vec{\mathcal{V}}\;=\; \bmatrix{ 0.3209 \cr 0.1112 \cr 0 }$
and a HSW channel capacity 
$\mathcal{C}_1 \;=\; \mathcal{D}_{optimum} \;=\; 0.1994$. 
Shown below in Figure 17 
is a plot of the planar channel ellipsoid (the inner
curve), and the curve of constant relative entropy
$\mathcal{D}( \, \rho \, \| \, \phi \, )   \;  = \; \mathcal{D}_{optimum}$ centered 
at $\vec{\mathcal{V}}$, which is marked with an asterisk {\textbf *}.
One can see that the $\mathcal{D}_{max}$ curve  
intersects the ellipsoid curve at two points, marked with {\textbf O},
and these two points are the optimum channel output signals $\rho_i$.


The optimum input and output signalling states for this channel
were determined as described in Appendix E and are :

$$
P_1 \;=\; 0.4869 , \quad \quad
\vec{W}_1^{Input} \;=\; \bmatrix{ -0.0207 \cr -0.9998 \cr 0 }, \quad \quad 
\vec{W}_1^{Output} \;=\; \bmatrix{ 0.2917 \cr -0.3999 \cr 0 } 
\;.
$$

$$
P_2 \;=\; 0.5131,  \quad \quad
\vec{W}_2^{Input} \;=\; \bmatrix{ 0.1215 \cr 0.9926 \cr 0 } , \quad \quad 
\vec{W}_2^{Output} \;=\; \bmatrix{ 0.3486 \cr 0.5963 \cr 0 } 
\;.
$$

These signal states yield an average channel output Bloch vector $\vec{\mathcal{V}}$ of 

$$
\vec{\mathcal{V}} \;=\; 
P_1 \, \cdot \, \vec{W}_1^{Output} \;+\; P_2 \, \cdot \, \vec{W}_2^{Output}
\;=\; \bmatrix{ 0.3209 \cr 0.1113 \cr 0 }\; 
.
$$

Figure 17 below shows the location of the channel ellipsoid ( the inner
dashed curve ), the contour of constant relative entropy
( the solid curve ) for $\mathcal{D} \, =\, 0.1994$,
the location of the two optimum input pure states $\rho_i^{Input}$, 
(the two {\textbf O} states on the circle of radius one), and the two 
optimum output signal states $\rho_i^{Output}$,
also denoted by {\textbf O}, on the channel
ellipsoid curve. Note that the optimum input signalling 
states are non-orthogonal. 

\begin{center}
\includegraphics*[angle=-90,scale=0.6]{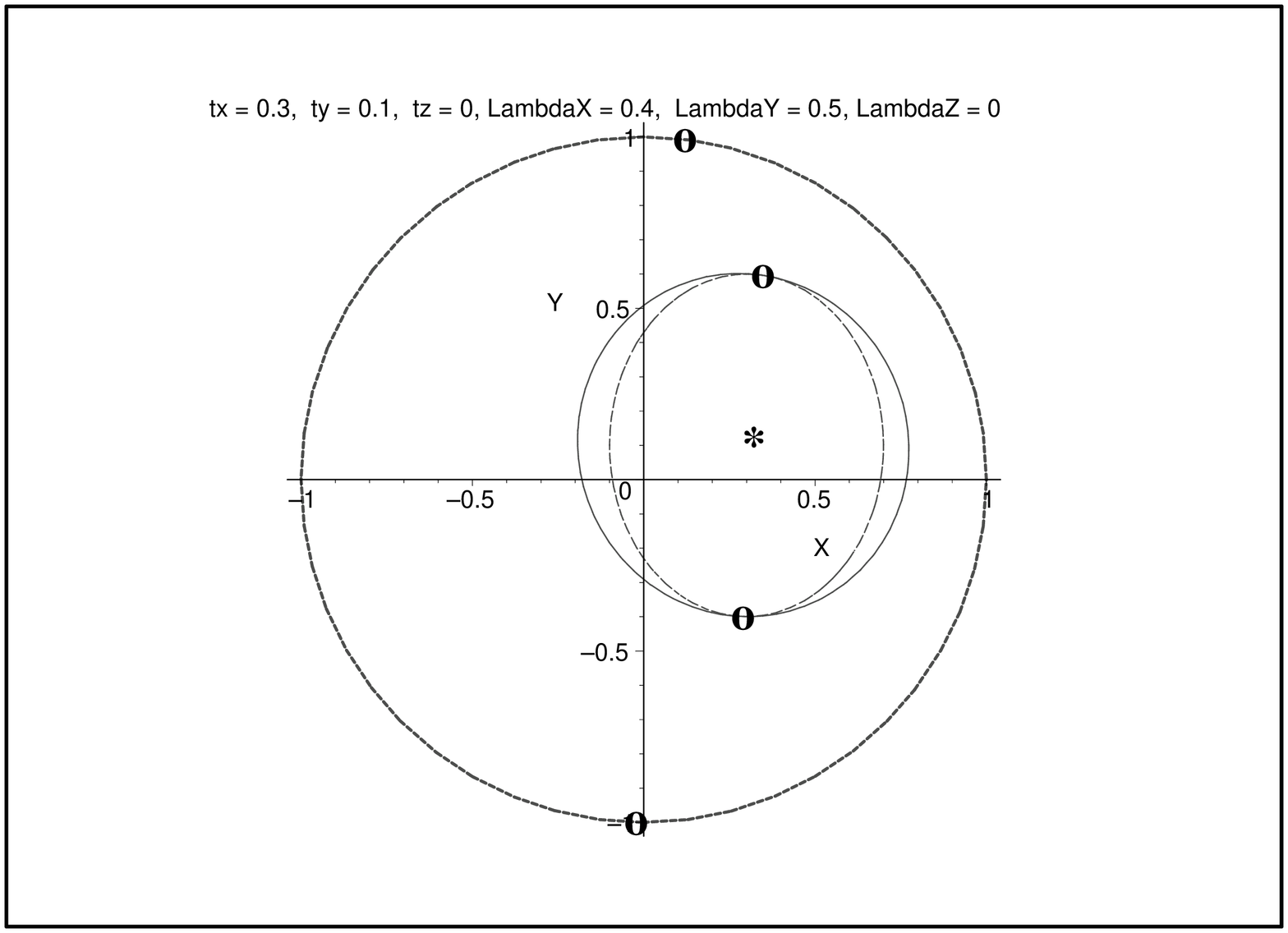}
\end{center}

\begin{center}
Figure 17: 
The intersection in the Bloch sphere X-Y plane of
a planar channel ellipsoid (the inner dashed curve)
and the optimum relative
entropy contour (the solid curve). The two
optimum input signal states (on the outer bold dashed
Bloch sphere boundary curve) and the two optimum output signal
states (on the channel ellipsoid  {\em and} the optimum relative entropy 
contour curve) are shown as {\textbf O}. 
\end{center}

Another useful picture is how the relative entropy changes as we make
our way around the channel ellipsoid. We consider the Bloch X-Y plane
in polar coordinates $\{ \, r \, , \, \theta \, \}$, where
we measure the angle $\theta$ with respect to the
origin of the Bloch X-Y plane axes.
( Note that $\theta$ only fully
ranges over $[0 , 2 \pi ]$ when the origin of
the Bloch sphere lies inside the channel ellipsoid. )
The horizontal line at the
top of the plot is the channel capacity $\mathcal{C}_1 \;=\; 0.1994$. 
Note that the two relative entropy peaks 
correspond to the locations of the two output optimum signalling states. 

\begin{center}
\includegraphics*[angle=-90,scale=0.6]{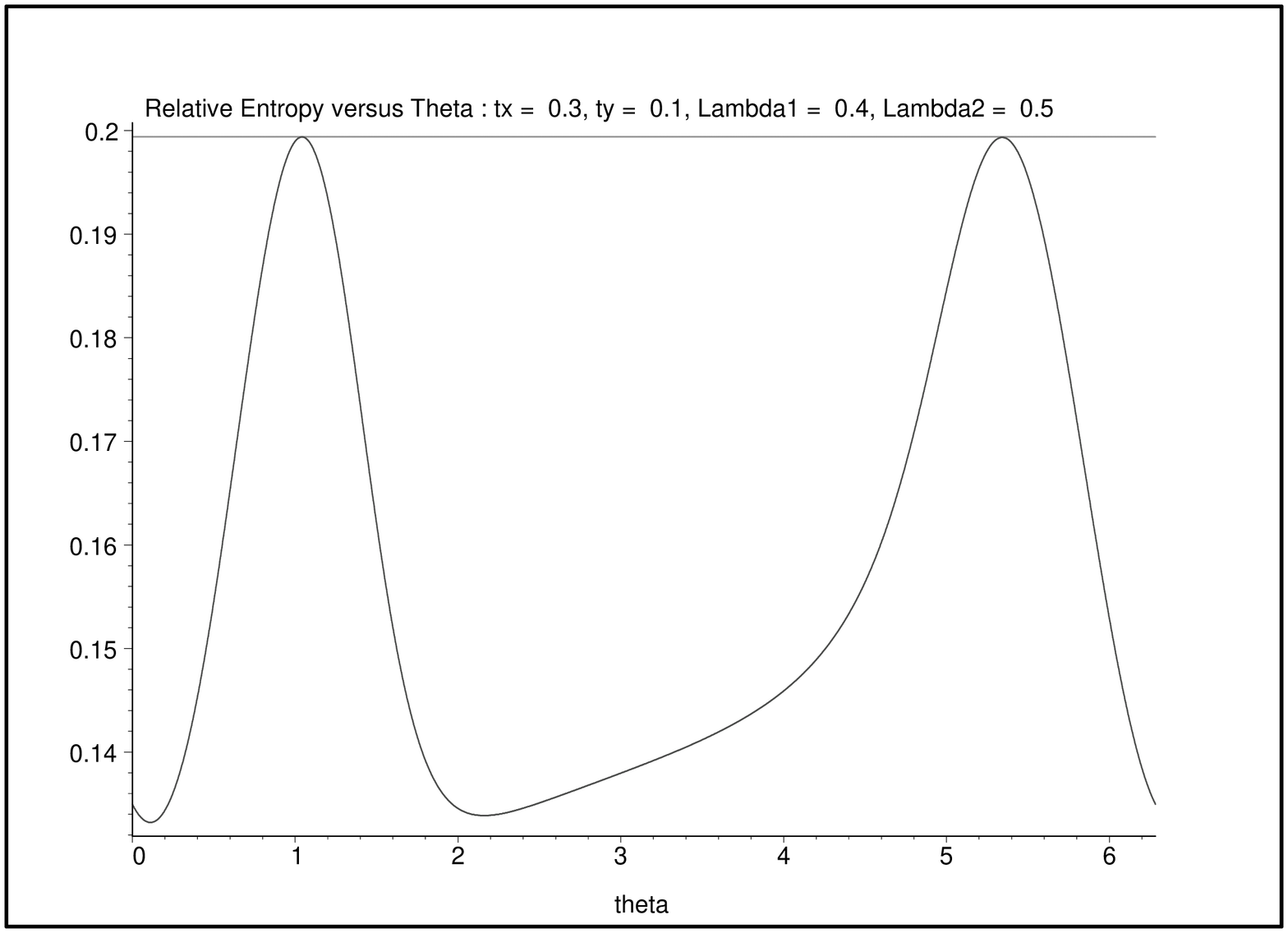}
\end{center}

\begin{center}
Figure 18: 
The change in $\mathcal{D}( \; \rho \; \| \; \phi \,\equiv \, $ 
{\textbf *} ) as we move $\rho$ around the channel ellipsoid.
The angle theta is with respect to the Bloch sphere origin. 
\end{center}

For this channel, the optimum channel capacity is achieved using an
ensemble consisting of only
two signalling states. Davies theorem tells us that for
single qubit channels, an optimum ensemble need contain at most four
signalling states. 
Using the notation of \cite{King01},
we call $C_2$ the optimum output $\mathcal{C}_1$ HSW channel
capacity attainable 
using only two input signalling states, 
$C_3$ is the optimum output $\mathcal{C}_1$ HSW channel capacity 
attainable using only
three input signalling states, and
$C_4$ is the optimum output $\mathcal{C}_1$ HSW channel capacity 
attainable using only
four input signalling states. Thus, for this channel, we see that
$C_2 \;=\; C_3 \;=\; C_4$. That is, for this channel,
allowing more than two signalling
states in your optimal ensemble does not yield additional channel 
capacity over an optimal ensemble with just two signalling states. 

\vspace{0.2in}

\section{Unital Channels}

Unital channels are quantum channels that map the identity to 
the identity : $\mathcal{E}(\mathcal{I}) \;=\; \mathcal{I}$. Due to this
behavior, unital channels possess certain symmetries. In the ellipsoid 
picture, King and Ruskai \cite{Ruskai99a} have shown that for unital
channels, the $\{\,t_k\,\}$ are zero. This yields an ellipsoid centered 
at the origin of the Bloch sphere. The resulting symmetry of such an
ellipsoid will allow us to draw powerful conclusions. 

First, recall that we know there
exists at least one optimal signal ensemble, $\{\, p_i \,,\,\rho_i\, \}$,
which attains the HSW channel capacity $\mathcal{C}_1$. 
( See property III 
in Section \ref{schuwest}. )
Now consider the
symmetry evident in the formula we have derived for the relative entropy
for two single qubit density operators. We have :

$$
\mathcal{D}(\, \rho \, \| \, \phi \, ) \;=\; 
\mathcal{D} ( \, \vec{\mathcal{W}} \, \| \, \vec{\mathcal{V}} \, ) \; 
= \; f( \, r \, , \, q \, , \, \theta \, ) 
$$

where $r \;=\; \| \, \vec{\mathcal{W}} \, \|$, 
$q \;=\; \| \, \vec{\mathcal{V}} \, \|$, and
$\theta$ is the angle between 
$\vec{\mathcal{W}}$ and $\vec{\mathcal{V}}$. Thus, 
if $\rho_i \, \in \, \mathcal{A}$ and $\phi\,\in\,\mathcal{B}$,
with
$\mathcal{D}(\, \rho_i \, \| \, \sigma \,)\;=\; 
\mathcal{D}(\, \vec{\mathcal{W}}_i \, \| \, \vec{\mathcal{V}}\,)\;=\; 
\chi_{optimum} \;=\; \mathcal{C}_1
$,
then acting in $\mathcal{R}^3$, reflecting 
$\rho_i \, \equiv \, \vec{\mathcal{W}}_i$ and
$\sigma \, \equiv \, \vec{\mathcal{V}}$ through the Bloch sphere origin to
obtain
$\rho_i' \, \equiv \, \vec{\mathcal{W}}'_i$ and
$\sigma' \, \equiv \, \vec{\mathcal{V}}'$, yields elements 
of $\mathcal{A}$ and $\mathcal{B}$ respectively. 
Furthermore, these transformed density matrices will also satisfy 
$
\mathcal{D}(\, \rho_i' \, \| \, \sigma' \,)\;=\; 
\mathcal{D}(\, \vec{\mathcal{W}}'_i \, \| \, \vec{\mathcal{V}}'\,)\;=\; 
\chi_{optimum} \;=\; \mathcal{C}_1
$,
because $r$, $q$, and $\theta$ remain the same when we reflect through the
Bloch sphere origin. That is, 
the symmetry of the 
unital channel ellipsoid about the Bloch sphere origin, corresponding
to the density matrix $\frac{1}{2} \, \mathcal{I}$, together with the
symmetry present in the qubit relative entropy formula yields a symmetry
for the optimal signal ensemble $\{\, p_i \,,\,\rho_i\, \}$, where 
$\sigma \;=\; \sum_i \, p_i \, \rho_i $, or equivalently
$\vec{\mathcal{V}} \;=\; \sum_i \, p_i \, \vec{\mathcal{W}}_i$.
This symmetry indicates that for every optimal signal ensemble 
$\{\, p_i \,,\,\rho_i\, \}$, there exists another ensemble, 
$\{\, p'_i \,,\,\rho'_i\, \}$, obtained by reflection through the Bloch
sphere origin. Since we know there exists at least one optimal 
signal ensemble, we must conclude that if
$\sigma \;=\; \sum_i \, p_i \, \rho_i \;\neq \; \frac{1}{2} \,
\mathcal{I}$, then two optimal ensembles exist with 
$\sigma \;\neq \; \sigma'$. However, by our uniqueness proof above, we are
assured that 
$\sigma \;=\; \sum_i \, p_i \, \rho_i$ is a unique density matrix,
regardless of the states $\{ \, p_i \, , \, \rho_i \, \}$ used, as long
as the 
states $\{ \, p_i \, , \, \rho_i \, \}$ are an optimal ensemble. 
Thus we must conclude that 
$\sigma \;=\; \sum_i \, p_i \, \rho_i \;\equiv \; \frac{1}{2} \, \mathcal{I}$,
since only the density matrix $\frac{1}{2} \, \mathcal{I}$ maps into itself upon
reflection through the Bloch sphere origin. 
Summarizing these observations, we can state the following.

{\em Theorem :}

For all unital qubit channels, and all optimal signal ensembles 
$\{\, p_i \,,\,\rho_i\, \}$, the average density matrix 
$\sigma \;=\; \sum_i \, p_i \, \rho_i \;\equiv \; \frac{1}{2}
\,\mathcal{I}$.

\vspace{0.2in}

In Appendix A, it is shown that

$$
\mathcal{D} \left (\, \rho \, \| \, \frac{1}{2} \,\mathcal{I}\, \right )\;=\; 
1 \;-\; \mathcal{S}(\,\rho\, )
$$

where $\mathcal{S}(\,\rho\, )$ is the von Neumann entropy of the 
density matrix $\rho$. Thus, our relation for the HSW channel capacity 
$\mathcal{C}_1$ becomes :

$$
\mathcal{C}_1 \;=\; 
\sum_I \; p_i \, \mathcal{D} \left ( \, \rho_i \, \| \, \frac{1}{2} \,\mathcal{I}
\,\right) \;=\;  1 \;-\; \sum_i \;p_i \, \mathcal{S}(\,\rho_i\, )
$$

To maximize $\mathcal{C}_1$, we seek to minimize the 
$\sum_i \; \mathcal{S}(\,\rho_i\, )$, subject to the constraint that the 
$\rho_i$ satisfy $\sum_i \; p_i \, \rho_i\;=\; \frac{1}{2} \,
\mathcal{I}$, for some set of a priori probabilities $\{\, p_i\, \}$. 
Recall that $\mathcal{S}(\,\rho\, ) \;\equiv\; \mathcal{S}(\,r\, )$
is a strictly decreasing function of $r$, where 
$r$ is the magnitude of the Bloch vector
corresponding to $\rho$. 
(Please see the plot below of 
$\mathcal{S}(\,\rho\, ) \;\equiv\; \mathcal{S}(\,r\, )$.)

\begin{center}
\includegraphics*[angle=-90,scale=0.6]
{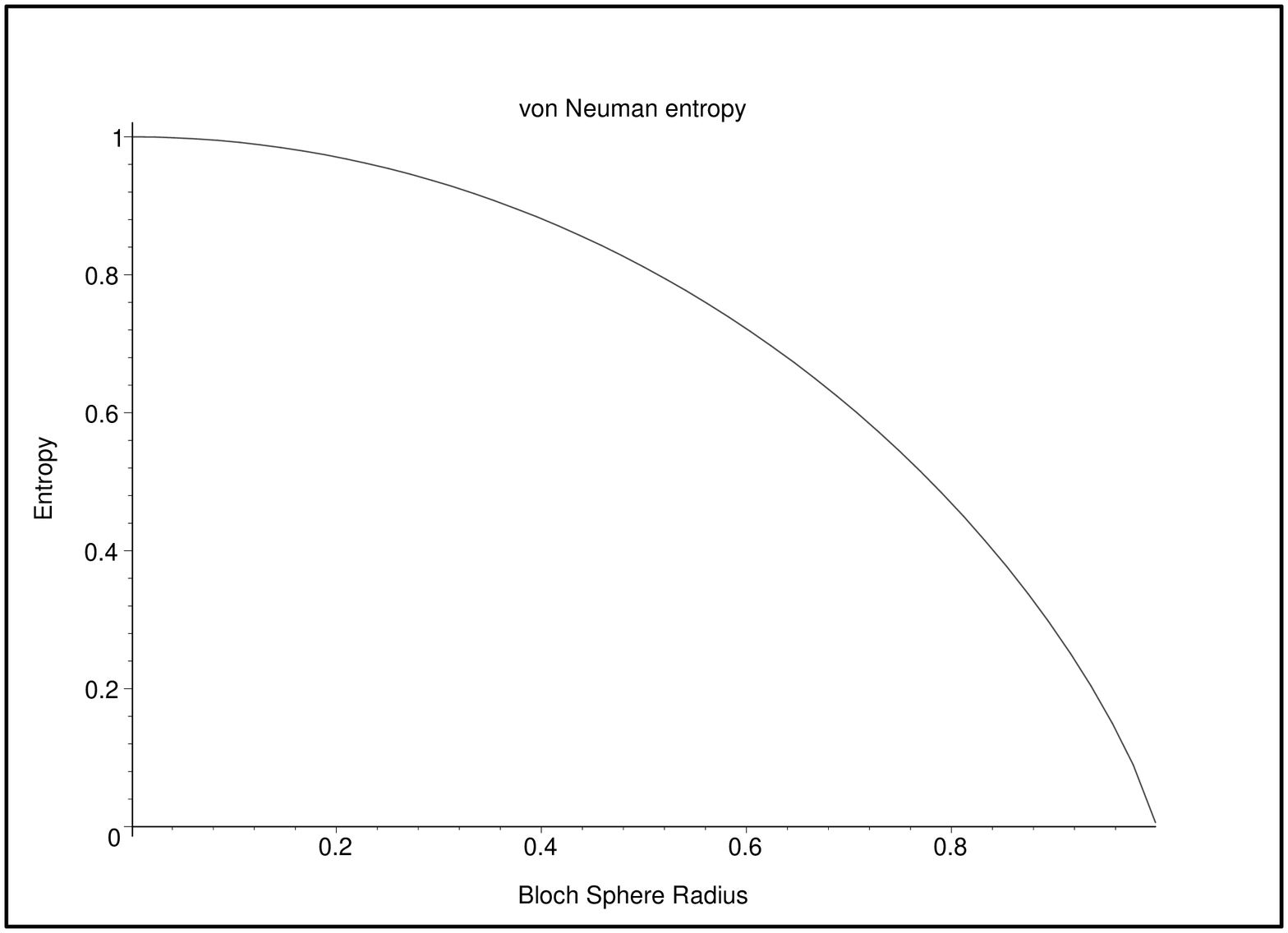}
\end{center}

\begin{center}
Figure 19: 
The Von Neumann entropy $\mathcal{S}(\rho)$ for a single qubit $\rho$ 
\linebreak 
as a function of the Bloch sphere radius $ r \;\in [0,1]$.  
\end{center}

Thus we seek to find a set of $\rho_i$ which lie
most distant, in terms of {\em Euclidean} distance in $\mathcal{R}^3$,
from the ellipsoid origin, and for which a convex combination
of these states equals the Bloch sphere origin. 

Let us examine a few special cases. For the unital channel
ellipsoid, consider the case where the major axis is unique in 
length, and has total length $2 \, \lambda^{major \, axis}$. Let
$\rho_{+}$ and $\rho_{-}$ be the states lying at the end of the major
axis. By the symmetry of the ellipsoid, we have 

$$
\frac{1}{2} \, \rho_{+} \;+\; 
\frac{1}{2} \, \rho_{-} \;=\; \
\frac{1}{2} \; \mathcal{I}
$$

Furthermore, the magnitude of the corresponding Bloch sphere vectors 
$r_{+} \;=\; \| \, \vec{\mathcal{W}}_{+}\,\|$ and 
$r_{-} \;=\; \| \, \vec{\mathcal{W}}_{-}\,\|$ are equal,
$r_{+} \;=\; r_{-} \;=\; 1 \;-\; 
\left | \, \lambda^{major \, axis} \, \right |$. 

Above, we use $|\cdots|$ around $\lambda^{major \, axis}$
because 
$\lambda^{major \, axis}$ can be a negative quantity in the King - Ruskai
et al. formalism. 
Using this value of $r\,=\,r_{+}\,=\,r_{-}$ yields for $\mathcal{C}_1$ :

$$
\mathcal{C}_1 \;=\; 1 \;-\;   
2 \;  \left ( \; \frac{1}{2} \; \mathcal{S}(r) \; \right ) \;=\;
1 \;-\; 
\mathcal{S}\left ( 
\; \left | \, \lambda^{major \, axis} \, \right | \; \right  ) 
\; .
$$

If the major axis is not the unique axis of maximal length, then any
set of convex probabilities and states $\{\,p_i\,,\, \rho_i\,\}$ such that 
the states lie on the major {\em surface} and 
$\; \sum_i \, p_i \, \rho_i \;\equiv \; \frac{1}{2} \, \mathcal{I}\; $
will suffice. 

Thus we reach the same conclusion obtained by King and Ruskai in an
earlier paper\cite{Ruskai99a}. 
Summarizing, we can state the following.

{\em Theorem :}

The optimum output signalling states for unital qubit channels correspond
to the minimum output von Neumann entropy states.

\vspace{0.2in}

Furthermore, we can also conclude :

{\em Theorem :}

For unital qubit channels, the channel capacities consisting of 
signal state ensembles with two, three and four signalling states are 
equal. 
Furthermore, the optimum HSW channel capacity can be attained with a,
possibly non-unique, pair
of equiprobable ($p_1 \,=\, p_2 \,=\, \frac{1}{2}$)
signalling states arranged opposite one another with
respect to the Bloch sphere origin. 

{\em Proof :}

Using the notation above, $C_2 \;=\; C_3 \;=\; C_4$. From the
geometry of the centered channel ellipsoid, we can 
always use just two signalling states with the minimum output entropy
to convexly reach $\frac{1}{2} \; \mathcal{I}$. Thus, utilizing 
more than two signaling
states will not yield any channel capacity improvement beyond using two
signalling states. The equiprobable nature of the two signalling states
derives from the symmetry of the signalling states on the channel
ellipsoid, in that one signalling state being the reflection of the
other signalling state through the Bloch sphere origin means the states may be
symmetrically added to yield an average state corresponding to the Bloch
sphere origin. It is this reflection symmetry which makes the two
signalling states equiprobable.

$\bigtriangleup$ - {\em End of Proof}. 

The last three theorems were previously proven by King and Ruskai in
section 2.3 of \cite{Ruskai99a}. Here we have merely shown their results
in the relative entropy picture.

\subsection{The Depolarizing Channel}

The depolarizing channel is a unital channel with $\{\, t_k\,=\, 0\,\}$ and
$\{\, \lambda_k\,=\, \frac{4\,x\,-\,1}{3}\,\}$, as discussed in more detail
in Appendix D. The parameter $x \; \in \; [\,0\,,\,1\,]$. 
Using the analysis above, we can conclude that :

$$
\mathcal{C}_1 \;=\; 1 \;-\;   
2 \;  \left ( \; \frac{1}{2} \; \mathcal{S}(r) \; \right ) \;=\;
1 \;-\; 
\mathcal{S}\left ( 
\; \left | \, \lambda^{major \, axis} \, \right | \; \right  ) 
\;=\; 1 \;-\; 
\mathcal{S}\left ( 
\; \left | \, \frac{ 4 \, x \;-\; 1}{3}\, \right | \; \right  ) 
$$

$$
\;=\; 
\frac{\;1 \;+\; \left | \, \frac{ 4 \, x \;-\; 1}{3}\, \right | \; }{2}
\; \log_{2} \left (\;1 \;+\; \left | \, \frac{ 4 \, x \;-\; 1}{3}\, 
\right | \; \right)
\;+\; 
\frac{\;1 \;-\; \left | \, \frac{ 4 \, x \;-\; 1}{3}\, \right | \; }{2}
\; \log_{2} \left (\;1 \;-\; \left | \, \frac{ 4 \, x \;-\; 1}{3}\, 
\right | \; \right)
$$

We plot $\mathcal{C}_1$ below.

\begin{center}
\includegraphics*[angle=-90,scale=0.6]
{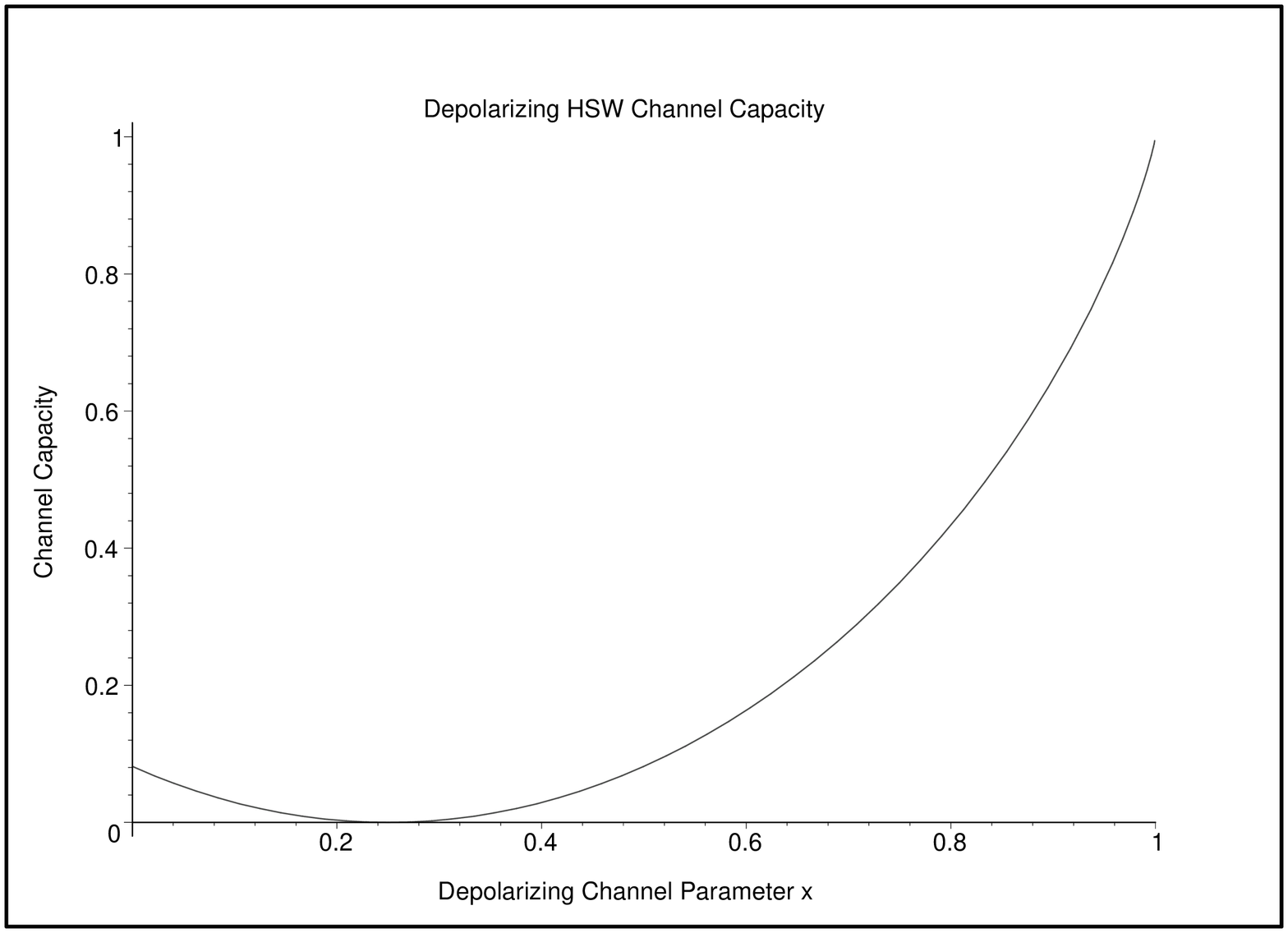}
\end{center}

\begin{center}
Figure 20: 
The Holevo-Schumacher-Westmoreland classical channel 
capacity for the depolarizing channel as a function of 
the depolarizing channel parameter {\em x}. 
\end{center}

\subsection{The Two Pauli Channel}

The Two Pauli channel is a unital channel with $\{\, t_k\,=\, 0\, \}$ and
$\{\lambda_x\,=\, \lambda_y\,=\, x\,\}$, and
$\{\lambda_z\,=\, \, 2\,x\,-\,1\,\}$, 
as discussed in more detail
in Appendix D. The parameter $x \; \in \; [\,0\,,\,1\,]$. 
The determination of the major axis/surface is tricky due
to the need to take into account the {\em absolute value}
of the $\lambda_k$. We plot
below the absolute value of the $\lambda_k$. The dotted curve below
corresponds to the absolute value of $\lambda_x$ and $\lambda_y$.
The V-shaped solid curve corresponds to the absolute value of 
$\lambda_z$.

\begin{center}
\includegraphics*[angle=-90,scale=0.6]
{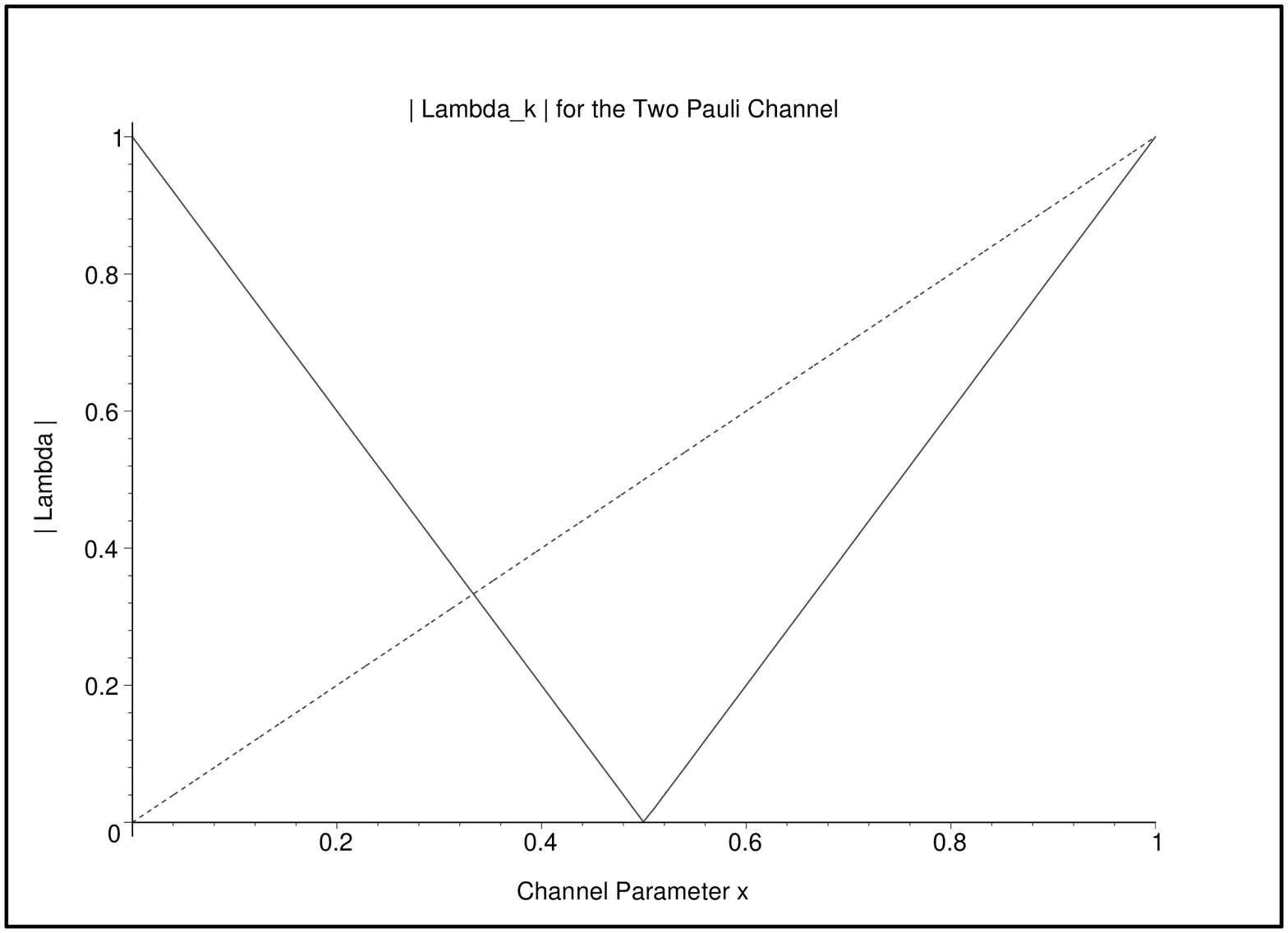}
\end{center}

\begin{center}
Figure 21: 
Calculating the length of the major axis of 
the channel ellipsoid for the two pauli channel as a function of the two 
pauli channel parameter {\em x}.
\end{center}

The intersection point occurs at $x  \,=\, \frac{1}{3}$. Thus
$\lambda_z$ is the major axis for $x \,\leq\,\frac{1}{3}$ and the
$\{\, \lambda_x \,, \,\lambda_y\,\}$ surface is the major axis surface
for $x \,\geq\,\frac{1}{3}$. The Bloch sphere radius
corresponding to the minimum entropy states is $1 \, - \, 2 \,x$ for 
$x \,\leq\,\frac{1}{3}$ and $x$ for $x \,\geq\,\frac{1}{3}$.

Using our analysis above, 
we can conclude that 
for $x \,  \leq \, \frac{1}{3}$, we have :

$$
\mathcal{C}_1 \;=\; 1 \;-\;   
2 \;  \left ( \; \frac{1}{2} \; \mathcal{S}(r) \; \right ) \;=\;
1 \;-\; 
\mathcal{S}\left ( 
\; \left | \, \lambda_z \, \right | \; \right  ) 
\;=\; 1 \;-\; 
\mathcal{S}\left ( \; 1 \, - \, 2  \, x \, \; \right  ) 
$$

$$
\;=\;  1 
\;+\;  
x \; \log_{2} \left (\; x \; \right)
\;+\; 
( \;1 \;-\; x \; ) \; 
\; \log_{2} \left (\;1 \;-\; x \; \right)
$$

while for $x \,  \geq \, \frac{1}{3}$, we have :

$$
\mathcal{C}_1 \;=\; 1 \;-\;   
2 \;  \left ( \; \frac{1}{2} \; \mathcal{S}(r) \; \right ) \;=\;
1 \;-\; 
\mathcal{S}\left ( 
\; \left | \, \lambda_x \, \right | \; \right  ) 
\;=\; 1 \;-\; 
\mathcal{S} ( \; x \; ) 
$$

$$
\;=\; 
\frac{\;1 \;+\; x \; }{2}
\; \log_{2} (\; 1 \;+\; x \; )
\;+\; 
\frac{\;1 \;-\; x \; }{2}\;
\; \log_{2} (\;1 \;-\;x \; )
$$

We plot 
$\mathcal{C}_1$ below, using 
the appropriate function in their allowed ranges of $x$.

\begin{center}
\includegraphics*[angle=-90,scale=0.6]
{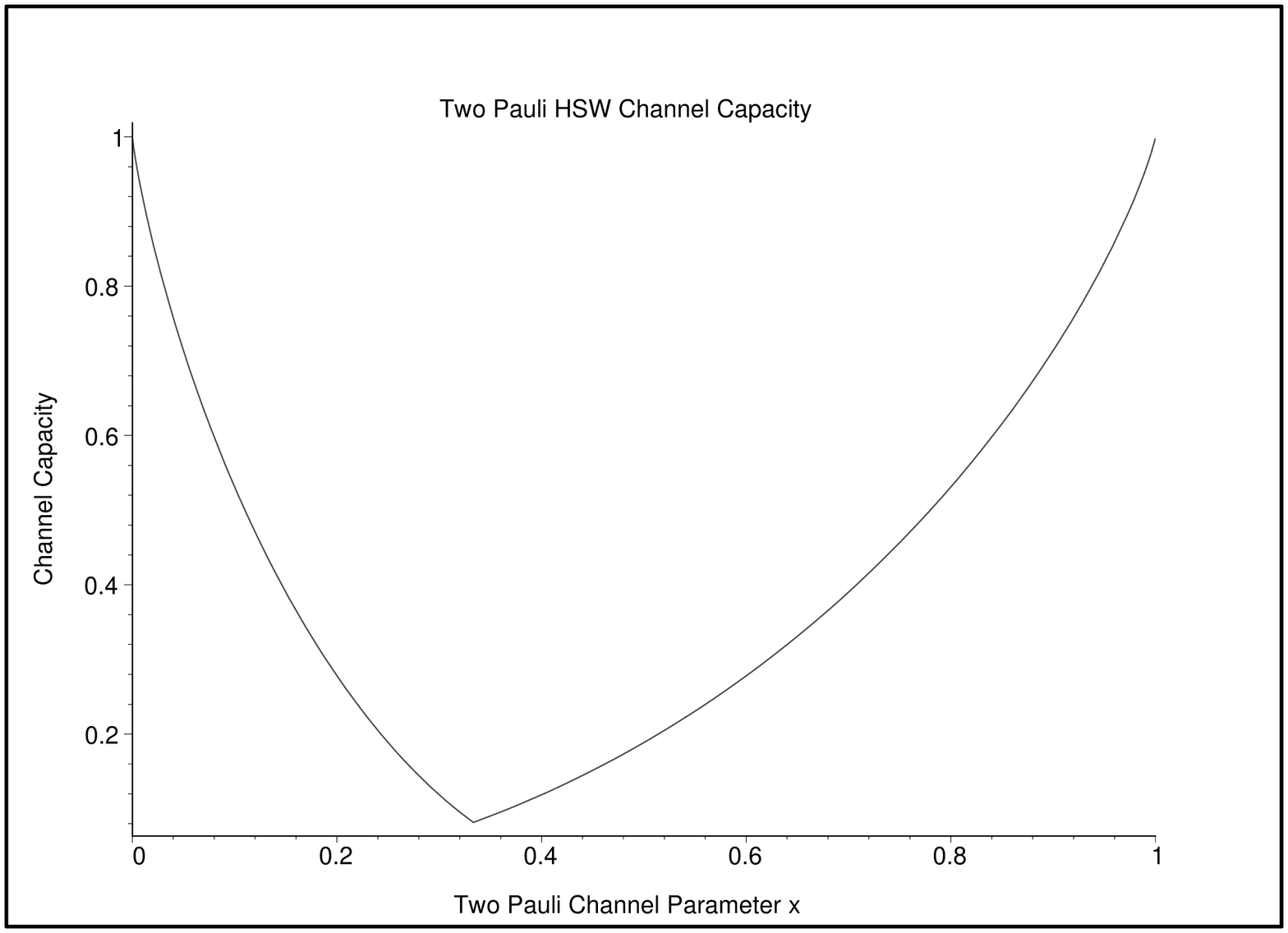}
\end{center}

\begin{center}
Figure 22: 
The Holevo-Schumacher-Westmoreland classical channel 
capacity for the two pauli channel as a function of 
the two pauli channel parameter {\em x}. 
\end{center}

Note the symmetry evident in the plots. Examining our graph above
for $\mathcal{C}_1$, one sees that 
for $ 0 \, \leq \,  \alpha \,\leq \, \frac{1}{3}$,
we have $\mathcal{C}_1( \, \frac{1}{3} \,-\, \alpha \,) \;\equiv \; 
\mathcal{C}_1( \, \frac{1}{3} \,+\, 2 \, \alpha \,)$. This symmetry is
also readily seen from the relations for $\mathcal{C}_1$ in the two
allowed ranges of $x$ (less than and greater than $\frac{1}{3}$). 

For $ x \, \le \, \frac{1}{3}$, setting $ x \;=\; \frac{1}{3} \,-\, \alpha$,

$$
\mathcal{C}_1^{-} (\alpha) \;=\; 1 \, + \, 
\frac{ 1 - 3 \alpha }{3} \; \log_2 \left ( \; \frac{ 1-3\alpha}{3} \right )
\;+\;
\frac{2+3\alpha}{3} \log_2 \left ( \frac{ 2+3\alpha}{3} \right ) 
$$

For $ x \, \ge \, \frac{1}{3}$, setting $ x \;=\; \frac{1}{3} \,+\,2 \alpha$,

$$
\mathcal{C}_1^{+} (\alpha) \;=\; 
\frac{ 4 + 6 \alpha }{6} \; \log_2 \left ( \; \frac{ 4+6\alpha}{3} \right )
\;+\; 
\frac{2-6\alpha}{6} \log_2 \left ( \frac{ 2-6\alpha}{3} \right ) 
$$

$$
\;=\; 
\left ( \; \frac{ 4 + 6 \alpha }{6} \; 
\;+\; 
\frac{2-6\alpha}{6} 
\right ) \; +\; 
\frac{ 4 + 6 \alpha }{6} \; \log_2 \left ( \; \frac{ 2+3\alpha}{3} \right )
\;+\; 
\frac{2-6\alpha}{6} \log_2 \left ( \frac{ 1-3\alpha}{3} \right ) 
$$

$$
\;=\; 
1 \; +\; 
\frac{ 2 + 3 \alpha }{3} \; \log_2 \left ( \; \frac{ 2+3\alpha}{3} \right )
\;+\; 
\frac{1-3\alpha}{3} \log_2 \left ( \frac{ 1-3\alpha}{3} \right ) 
\;=\; 
\mathcal{C}_1^{-} (\alpha) 
$$

\vspace{0.2in}

\section{Non-Unital Channels}

Non-unital channels are generically more difficult to analyze due to
the fact that one or more of the $\{\,t_k\,\}$ can be non-zero.
This allows the average density 
matrix $\rho \,=\, \sum_i \, p_i \, \rho_i$ for
an optimal signal ensemble $\{\,p_i\,,\,\rho_i\,\}$
to move away from the Bloch sphere 
origin $\rho \,=\, \frac{1}{2} \, \mathcal{I} \;\equiv\;
\vec{\mathcal{V}} \;=\; \bmatrix{ 0 \cr 0 \cr 0 }$.
However, there still remains the symmetry present in the qubit form of the
relative entropy formula, namely that 
$\mathcal{D}(\,\rho\,\| \, \phi\,) \,=\,
\mathcal{D}(\,\vec{\mathcal{W}} \,\| \, \vec{\mathcal{V}} \,) \,=\,
f(\,r\,,\,q\,,\,\theta\,)$, 
where $r \;=\; \| \, \vec{\mathcal{W}} \, \|$, 
$q \;=\; \| \, \vec{\mathcal{V}} \, \|$, and
$\theta$ is the angle between 
$\vec{\mathcal{W}}$ and $\vec{\mathcal{V}}$. The fact that the qubit 
relative entropy depends only on $r$, $q$, and $\theta$ yields a
symmetry which can be 
used to advantage in analyzing non-unital channels, as our last example
will demonstrate.

\subsection{The Amplitude Damping Channel}

The amplitude damping channel is a non-unital channel with
$\{\, t_x\,=\, t_y\,=\, 0\, \}$ and
\linebreak
$\{\, t_z\,=\, 1 \, - \, \xi \, \}$.
The $\lambda_k$ are
$\{\lambda_x\,=\, \lambda_y\,=\, \sqrt{ \xi} \,\}$, and
$\{\lambda_z\,=\, \, \xi \,\}$, 
where $\xi$ is the channel parameter, $\xi \; \in \; [\,0\,,\,1\,]$. 
The amplitude damping channel is 
discussed in more detail
in Appendix D. 
The determination of the major axis/surface reduces to an analysis 
in either the X-Y or X-Z Bloch sphere {\em plane} because of 
symmetries of the channel ellipsoid {\em and} the relative entropy
formula for qubit density matrices. Since the relative entropy formula
depends only on the $r$, $q$ and $\theta$ quantities which were defined above,
by examining contour {\em curves} of relative entropy in
the X-Z plane, we can create a {\em surface} of constant relative entropy 
in the three dimensional X-Y-Z Bloch sphere space by the solid of revolution
technique. That is, we shall revolve our X-Z contour curves about the
axis of symmetry, here the Z-axis. Now the channel ellipsoid in this
case is also rotationally symmetric about the Z-axis, because 
$t_x \,=\, t_y \,=\,0$ and  
$\lambda_x \,=\, \lambda_y$.  Thus optimum signal {\em points} 
(points on the
channel ellipsoid surface which have maximal relative entropy distance
from the average signal density matrix), in the X-Z plane, will become
{\em circles} of optimal signals in the full three dimensional 
Bloch sphere picture
after the revolution about the Z - axis is completed. 
Therefore, due to the simultaneous rotational symmetry 
about the Bloch sphere Z axis of the relative
entropy formula (for qubits) and the channel ellipsoid, a full three
dimensional analysis of the amplitude damping channel reduces to a
much easier, yet equivalent, two dimensional analysis in the Bloch X-Z plane.

To illustrate these ideas, we take a specific instance of the
amplitude damping channel with $\xi \,=\, 0.36$. Then
$\{\, t_x\,=\, t_y\,=\, 0\, \}$ and
$\{\, t_z\,=\, 0.64\, \}$.
The $\lambda_k$ are
$\{\lambda_x\,=\, \lambda_y\,=\, 0.6 \,\}$, and
$\{\lambda_z\,=\, \, 0.36 \,\}$. In this case
$\mathcal{C}_1 \,=\, 0.3600$ is achieved with two equiprobable
signalling states. The optimum average density matrix has
Bloch vector $\vec{\mathcal{V}} \,=\,  \bmatrix{ 0 \cr 0.7126}$, and
is shown with an asterisk in the plots below.

In the first plot, we show the X(horizontal)-Z(vertical) Bloch sphere
plane. The outer 
bold dotted ring is the pure state boundary, with
Bloch vector magnitude equal to one. The inner dashed circle is the channel
ellipsoid. The middle solid contour is the curve of constant relative entropy,
equal to 0.3600, and
centered at $\vec{\mathcal{V}}$. This relative entropy contour in the X-Z
plane contacts the channel ellipsoid at two {\em symmetrical} points,
indicated in the plot as {\textbf O}. Note
that these two contact points, and the location of  $\vec{\mathcal{V}}$,
all lie on a perfectly horizontal line. The fact that the
line is horizontal is due to the fact that the two optimum signalling
states in the X-Z plane are symmetric about the Z axis. The point
$\vec{\mathcal{V}}$ is simply the two optimal output signal points
average. The corresponding optimal input signals are shown as  
{\textbf O}'s on the outer bold dotted pure state boundary 
semicircular curve.

\begin{center}
\includegraphics*[angle=-90,scale=0.6]
{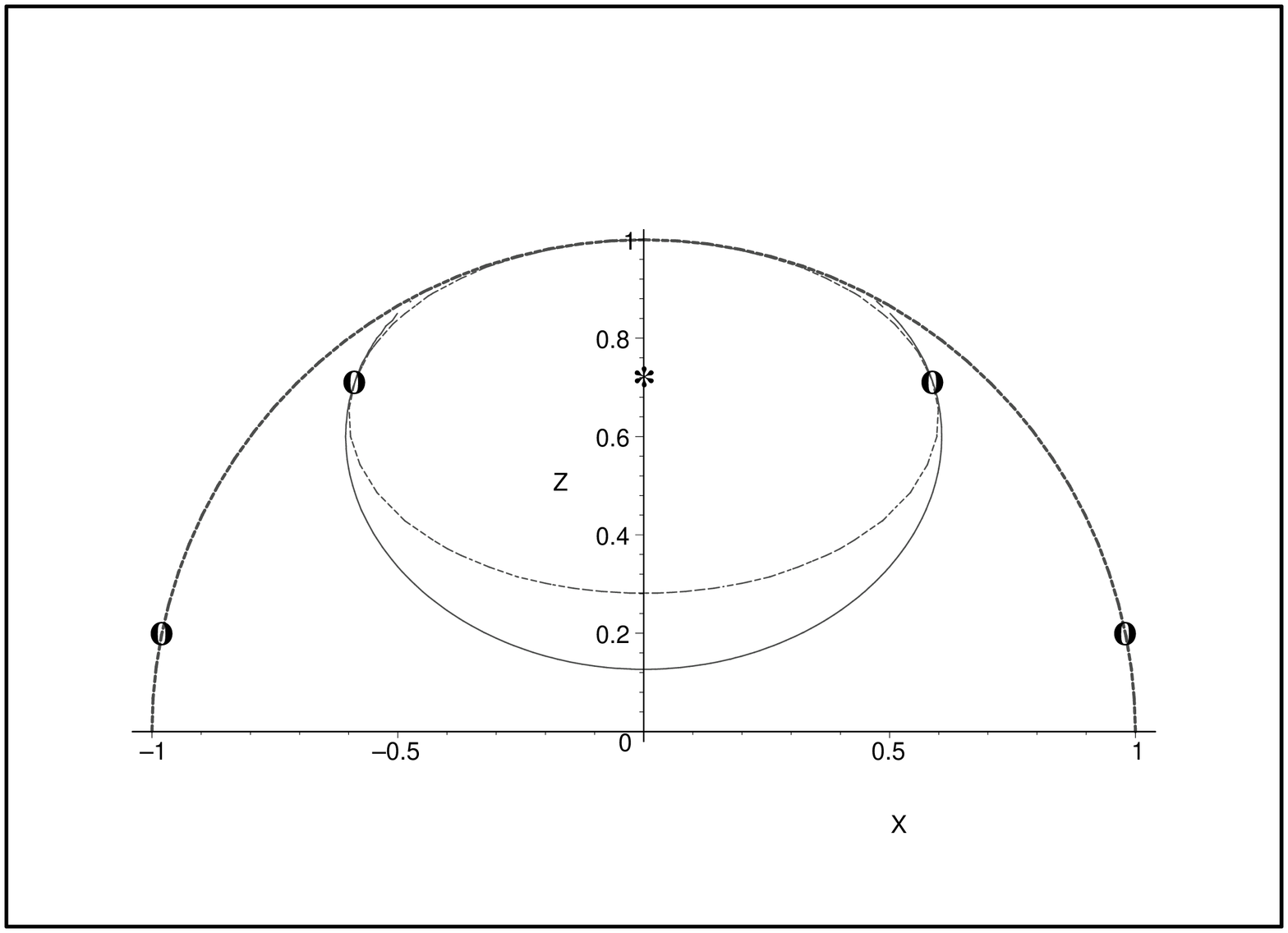}
\end{center}

\begin{center}
Figure 23: 
The intersection in the Bloch sphere X-Z plane of
the amplitude damping channel ellipsoid (the inner dashed curve)
and the optimum relative
entropy contour (the solid curve). The two
optimum input signal states (on the outer bold dashed
Bloch sphere boundary curve) and the two optimum output signal
states (on the channel ellipsoid  {\em and} the optimum relative entropy 
contour curve) are shown as {\textbf O}. 
\end{center}

Note that the optimum input signalling states are nonorthogonal. 
Furthermore, this analysis tells us that 
$C_2 \;=\; C_3 \;=\; C_4$. For the amplitude damping channel,
there is no advantage to using more than two signals 
in the optimum signalling ensemble. 

The following is a picture similar to those we have
done for the planar channels we 
examined. We plot the magnitude of the relative entropy as one
moves around the channel ellipsoid in the X-Z plane. The angle $\theta$ is
with respect to the Bloch sphere origin (ie : the X-Z plane origin).

\begin{center}
\includegraphics*[angle=-90,scale=0.6]
{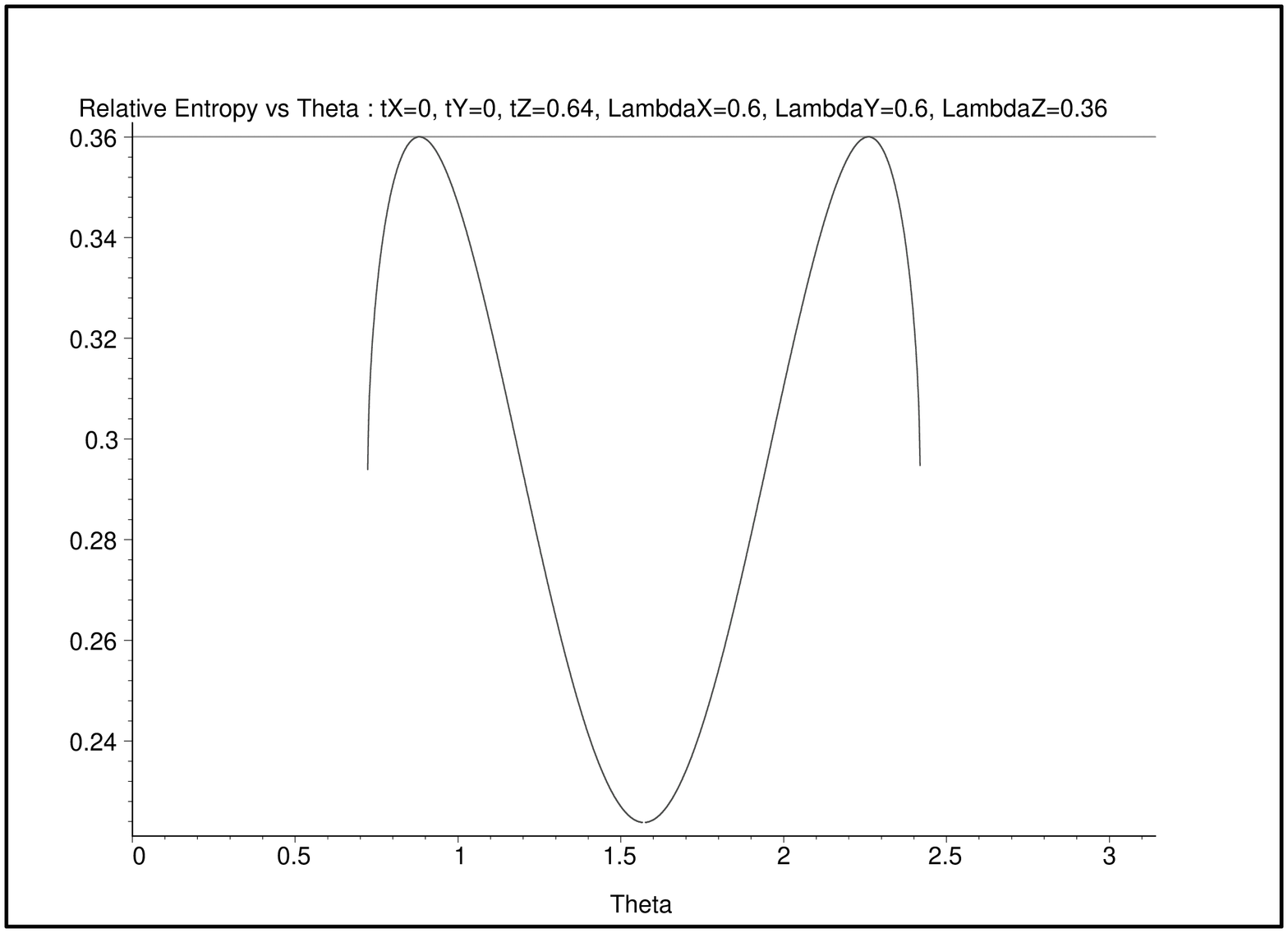}
\end{center}

\begin{center}
Figure 24: 
The change in $\mathcal{D}( \; \rho \; \| \; \phi \,\equiv \, $ 
{\textbf *} ) as we move $\rho$ around the channel ellipsoid.
The angle theta is with respect to the Bloch sphere origin. 
\end{center}

Thus, the rotational
symmetry about the Z-axis of the relative entropy formula, coupled with
the same Z - axis rotational symmetry of the amplitude damping channel 
ellipsoid, yields a complete 
understanding of the behavior of the amplitude damping channel
with just a simple two dimensional analysis.  

\section{Summary and Conclusions}

In this paper, we have derived a formula for the relative entropy 
of two single qubit density matrices. By combining our relative
entropy formula with the King-Ruskai et al. ellipsoid picture of qubit
channels, we can use the 
Schumacher-Westmoreland relative entropy approach to classical
HSW channel capacity to analyze unital and non-unital 
single qubit channels in detail. 

	The following observation also emerges from the examples and 
analyses above. In numerical simulations by this author and others,
it was noted that the a priori probabilities of the
optimum signalling states for non-unital qubit channels were in general,
approximately, but not exactly, equal. For example, consider the case of linear channels, 
where the optimum HSW channel capacity is achieved with two signalling states.
In our first linear channel example, one signalling
state had an a priori probability of 0.5156 and the other signalling
state had an a priori probability of 0.4844. Similarly, in our second
linear channel example, the respective a priori probabilities were 
0.5267 and 0.4733. These asymmetries in the a priori probabilities 
are due to the fact that $\mathcal{D}$ 
is not purely a radial function of distance from
$\vec{\mathcal{V}}_{optimum}$. The relative entropy contours shown in
Figures 2 through 11 are moderately, but not exactly, circular
about $\vec{\mathcal{V}}_{optimum}$. This slight radial asymmetry leads to
a priori signal probabilities that are approximately, but not exactly, equal. 
Thus, a graphical estimate of the a priori signal probabilities can be
made by observing the degree of asymmetry of the optimum relative entropy
contour about $\vec{\mathcal{V}}_{optimum}$. 

	In conclusion, the analysis above yields
a geometric picture which we hope will lead
to future insights into the transmission of classical information
over single qubit channels.

\vspace{0.2in}

\section{Acknowledgments}

The author would like to thank Patrick Hayden, Charlene Ahn, 
Sumit Daftuar and John Preskill 
for helpful comments on a draft of this paper. 
The author would also like to thank Beth Ruskai for many 
interesting conversations on the classical channel capacity of 
qubit channels. 

\vspace{0.2in}

\begin{appendix} 

\section{Appendix A - The Derivation Of The Bloch Sphere Relative
Entropy Formula }

The relative entropy of two density matrices $\varrho$ and $\psi$ is defined
to be

$$
\mathcal{D}  ( \; \varrho \, || \; \psi\;) \;=\; 
Tr [ \; \varrho \, ( \; \log_2(\varrho) \;-\; \log_2(\psi) \;) \; ]
$$

Our main interest is when both $\varrho$ and $\psi$ are qubit
density operators. In that case, $\varrho$ and $\psi$ can be written using 
the Bloch sphere representation.

$$
\varrho \; = \;  \frac{1}{2} \; \left ( \,  \mathcal{I} \; + \; \vec{\mathcal{W}} \bullet \vec{\sigma} \, \right ) \quad \quad \quad \quad \; \psi \; = \;  \frac{1}{2} \; \left ( \,  \mathcal{I} \; + \; \vec{\mathcal{V}} \bullet \vec{\sigma} \, \right)
$$

To simplify notation below, we define 

$$
r \;=\; \sqrt { \vec{\mathcal{W}} \bullet \vec{\mathcal{W}}} 
\quad \quad and \quad \quad q \;=\; \sqrt { \vec{\mathcal{V}} \bullet \vec{\mathcal{V}} }
$$

We shall also define $\cos(\theta)$ as :

$$
\cos(\theta) \;\;=\;\; 
\frac{\vec{\mathcal{W}}\bullet \vec{\mathcal{V}}}{ \; r \; q \; }
$$

where $r$ and $q$ are as above. 

The symbol $ \vec{\sigma}$ means the vector of 2 x 2 Pauli matrices 

$$  \vec{\sigma} \; = \; 
\bmatrix{ \sigma_x \cr \sigma_y \cr \sigma_z }   
\;\;\;\;\; where 
\;\;\;\;\; \sigma_x \;=\; \bmatrix{ 0  & 1 \cr 1  & 0 },
\;\;\;\;\; \sigma_y \;=\; \bmatrix{ 0  & -i \cr i & 0 },
\;\;\;\;\; \sigma_z \;=\; \bmatrix{ 1  & 0 \cr 0 & -1 }
$$

The Bloch vectors $\vec{\mathcal{W}}$ and $\vec{\mathcal{V}}$ are real,
three dimensional vectors
which have magnitude equal to one when representing a pure state density matrix,
and magnitude less than one for a mixed (non-pure) density matrix.

The density matrices for $\varrho$ and $\psi$ in terms of their Bloch
vectors are :

$$
\varrho \;=\;  \left[ \begin {array}{cc} \frac{1}{2}+\frac{1}{2}\,{\it w_3}&\frac{1}{2}\,{\it w_1}-\frac{1}{2}\,i{\it 
w_2}\\\noalign{\medskip}\frac{1}{2}\,{\it w_1}+\frac{1}{2}\,i{\it w_2}&\frac{1}{2}-\frac{1}{2}\,{\it w_3}
\end {array} \right] 
$$

$$
\psi \;=\;  \left[ \begin {array}{cc} \frac{1}{2}+\frac{1}{2}\,{\it v_3}&\frac{1}{2}\,{\it v_1}-\frac{1}{2}\,i{\it 
v_2}\\\noalign{\medskip}\frac{1}{2}\,{\it v_1}+\frac{1}{2}\,i{\it v_2}&\frac{1}{2}-\frac{1}{2}\,{\it v_3}
\end {array} \right] 
$$

We shall prove the following formula in two ways, an algebraic proof and 
a brute force proof. We conclude Appendix A with some alternate 
representations of
this formula.

$$
\mathcal{D}(\, \varrho \, \| \,  \psi \, ) \;=\; 
\mathcal{D}_1 \;-\; \mathcal{D}_2
\;=\; 
\frac{1}{2} \log_2 \left ( 1\;-\;r^2 \right ) \;+\; 
\frac{r}{2} \log_2 \left ( \frac{1\;+\;r}{1\;-\;r} \right )
\;-\; 
\frac{1}{2}
\;  \log_2 \left ( 1 \;-\; q^2\; \right ) 
\;-\; \frac{\; \vec{\mathcal{W}} \bullet \vec{\mathcal{V}} \; }
{\; 2 \; q\; } \; \log_2 \left ( \frac{  1 \;+\; q}{1 \;-\; q} \right)
$$

$$
\;=\; 
\frac{1}{2} \log_2 \left  ( 1\;-\;r^2 \right ) \;+\; 
\frac{r}{2} \log_2 \left ( \frac{1\;+\;r}{1\;-\;r} \right )
\;-\; 
\frac{1}{2}
\;  \log_2 \left ( 1 \;-\; q^2\; \right ) 
\;-\; \frac{r \; \cos ( \theta )  }
{\; 2 \; } \; \log_2 \left ( \frac{  1 \;+\; q}{1 \;-\; q} \right)
$$

where $\theta$ is the angle between $\vec{\mathcal{W}}$ 
and $\vec{\mathcal{V}}$.

\vspace{0.2in}

\subsection{Proof I : The Algebraic Proof }

$$
\mathcal{D}( \; \varrho \, || \; \psi\;) \;=\; Tr [ \; \varrho \, ( \; \log_2(\varrho) \;-\; 
\log_2(\psi) \;) \; ]
$$

Recall the following Taylor series, valid for $\| \, x\,\| \;\leq \, 1$.

$$
\ln( \;1\;+\; x\; ) \;=\; - \; \sum_{n=1}^{\infty} \; \frac{\;\left ( \; - x \;
\right ) ^n}{n} 
\;=\; x \;-\; \frac{x^2}{2} \;+\; \frac{x^3}{3} \;-\; \frac{x^4}{4} 
\;+\; \frac{x^5}{5} \;-\; \frac{x^6}{6} \;+\; \frac{x^7}{7} \;-\; \cdots
$$

$$
\ln( \;1\;-\; x\; ) \;=\; - \; \sum_{n=1}^{\infty} \; \frac{x^n}{n} 
\;=\; - \; x \;-\; \frac{x^2}{2} \;-\; \frac{x^3}{3} \;-\; \frac{x^4}{4} 
\;-\; \frac{x^5}{5} \;-\; \frac{x^6}{6} \;-\; \frac{x^7}{7} \;-\; \cdots
$$

Combining these two Taylor series yields another Taylor expansion
we shall be interested in :

$$
\frac{1}{2} \; \left \{ \; 
\ln( \;1\;+\; x\; ) \;-\;   \ln( \;1\;-\; x\; ) \; \right \}  \;=\; 
\frac{1}{2} \; \ln \left ( \;\frac{\; 1\;+\; x\;}{ \;1\;-\; x\; } \; \right )
$$

$$
\;=\; - \; \sum_{n=1}^{\infty} \; \frac{(-x)^n}{n} 
\quad-\quad \left ( \; -\; \sum_{n=1}^{\infty} \; \frac{x^n}{n} \; \right )
\;=\; x \;+\; \frac{x^3}{3} \;+\; \frac{x^5}{5} \;+\; \frac{x^7}{7} 
\;+\; \frac{x^9}{9} \;+\; \cdots
$$

A different combination of the first two Taylor series above yields 
yet another Taylor expansion we shall be interested in :

$$
\frac{1}{2} \; \left \{ \; 
\ln( \;1\;+\; x\; ) \;+\; \ln( \;1\;-\; x\; ) \; \right \}  \;=\; 
\frac{1}{2} \; \ln \left [ \; 1\;-\; x^2 \; \right ]
$$

$$
\;=\; - \; \sum_{n=1}^{\infty} \; \frac{(-x)^n}{n} 
\quad+\quad \left ( \; -\; \sum_{n=1}^{\infty} \; \frac{x^n}{n} \; \right ) \;
\;=\; - \; \frac{x^2}{2} \;-\; \frac{x^4}{4} \;-\; \frac{x^6}{6} 
\;-\; \frac{x^8}{8} \;-\; \cdots
$$

Consider $\log( \; \varrho \; )$ with the Bloch sphere representation for
$\varrho$.

$$
\varrho \; = \; \frac{1}{2} \; \left (
 \mathcal{I} \; + \; \vec{\mathcal{W}} \bullet \vec{\sigma} \right )
$$

We obtain, using the expansion given above for $\log ( \,1\,+\,x\,)$, 

$$
\log( \; \varrho \; ) \;=\; \log \left [  \; \frac{1}{2} \, \left ( \; \mathcal{I}
\;+\; \vec{\mathcal{W}} \bullet \vec{\sigma} \; \right  ) \; \right ]  
\;=\; \log \left [  \; \frac{1}{2} \; \right ]  \;+\; \log \left [ \; \mathcal{I}
\;+\; \vec{\mathcal{W}} \bullet \vec{\sigma}  \; \right ]
\;=\; \log \left [ \; \frac{1}{2} \; \right ]  \;-\; 
\sum_{n=1}^{\infty} \; \frac{
\;  \left ( \; - \; \vec{\mathcal{W}} \bullet \vec{\sigma}  \; \right )^n \;}
{n} 
$$

Recall that
$\left ( \; \vec{\mathcal{W}} \bullet \vec{\sigma}  \; \right )^2 \; =\;  r^2$,
where $r \;=\; \sqrt { \vec{\mathcal{W}} \bullet \vec{\mathcal{W}}}$.
Thus we have for even $n$,  
$\left ( \; \vec{\mathcal{W}} \bullet \vec{\sigma}  \; \right )^n 
\; =\;  r^{n}$, while for odd n we have  
$\quad \left ( \; \vec{\mathcal{W}} \bullet \vec{\sigma}  \; \right )^n
\; =\;  r^{n\,-\,1} \quad \vec{\mathcal{W}} \bullet \vec{\sigma}$.
The expression for $\log ( \;\varrho \; )$ then becomes 

$$
\log ( \;\varrho \; )  \;=\; 
\log \left [ \; \frac{1}{2} \; \right ]  \quad - \quad 
\sum_{n=1}^{\infty} \; \frac{
\;  \left ( \; - \; \vec{\mathcal{W}} \bullet \vec{\sigma}  \; \right )^n \;}{n} 
$$

$$
\;=\;
\log \left [ \; \frac{1}{2} \; \right ]  \quad + \quad 
\; \vec{\mathcal{W}} \bullet \vec{\sigma}
\;  - \; \frac{r^2}{2} \;+\; 
\; \frac{r^2}{3} \; \vec{\mathcal{W}} \bullet \vec{\sigma}  \; 
\;  - \; \frac{r^4}{4} \;+\; 
\; \frac{r^4}{5} \; \vec{\mathcal{W}} \bullet \vec{\sigma}  \; 
\;  - \; \frac{r^6}{6} \;+\; 
\; \frac{r^6}{7} \; \vec{\mathcal{W}} \bullet \vec{\sigma}  \; - \; \cdots  
$$

$$
\;=\;
\log \left [ \; \frac{1}{2} \; \right ]  \quad + \quad 
\; \frac{ \; \vec{\mathcal{W}} \bullet \vec{\sigma} \; }{r} \; \left ( \; 
\;  r
\;  + \; \frac{r^3}{3} 
\;  + \; \frac{r^5}{5} 
\;  + \; \frac{r^7}{7} \;+\;  \cdots \right ) 
\;+\; 
\left ( 
\;  - \; \frac{r^2}{2} \; 
\;  - \; \frac{r^4}{4} \; 
\;  - \; \frac{r^6}{6} \; 
\;  - \; \frac{r^8}{8} \; 
\;  - \; \frac{r^{10}}{10} \;-\; \cdots  \; \right ) 
$$

$$
\;=\; 
\log \left [ \; \frac{1}{2} \; \right ]  \quad + \quad 
\frac{ \; \vec{\mathcal{W}} \bullet \vec{\sigma} \;}{\; 2 \; r\; } \; 
\log \left [  \; \frac{ \;1\;+\; r\;}{ \;1\;-\; r\;} \; \right ] \; 
\;  + \; \frac{1}{2} \;  \log \left [  \;1\;-\; r^2\; \right ] \; 
$$

To evaluate $Tr \left [ \; \varrho \; \log ( \;\varrho \; )\; \right ]$ we again 
use the Bloch sphere representation for $\varrho$.

$$
\varrho \; = \; \frac{1}{2} \; \left ( \; \mathcal{I} \; + \; \vec{\mathcal{W}} \bullet \vec{\sigma} \; \right )
$$

We write 

$$
Tr \left [ \; \varrho \; \log ( \;\varrho \; ) \; \right ] \;=\; 
\frac{1}{2} \; Tr \left [ \;  \mathcal{I} \;\bullet \; \log  ( \;\varrho \; ) \;
\right ] 
\; + \; \frac{1}{2} \; 
Tr \left [ \; \left ( \; \vec{\mathcal{W}} \bullet \vec{\sigma} \; \right ) \; \log ( \;\varrho \; )\; \right ] 
$$

Using our results above, 

$$
\frac{1}{2} \; 
Tr \left [ \;  \mathcal{I} \;\bullet \; \log  ( \;\varrho \; ) \; \right ] 
\; =\; 
\log \left [  \;\frac{1}{2} \; \right ]  \;+\; 
\frac{1}{2} \; \log \left [  \;1\;-\; r^2\; \right ] 
$$

since 
$Tr[\;\mathcal{I}\;] \;=\; 2$ and 
$Tr[\;\sigma_x\;] \;=\; Tr[\;\sigma_y\;] \;=\; Tr[\;\sigma_z\;] \;=\; 0$. 

Similarly, 

$$
Tr \left [ \; \left ( \; \vec{\mathcal{W}} \bullet \vec{\sigma} \; \right ) \; \log ( \;\varrho \; )\; \right ] 
\; =\; 
\frac{\;  \left ( \; \vec{\mathcal{W}} \bullet \vec{\sigma} \; \right ) ^2 \; }{r}
\; \log \left [  \; \frac{ \;1\;+\; r\;}{ \;1\;-\; r\;} \; \right ] \; 
\; =\; r\; \log \left [  \; \frac{ \;1\;+\; r\;}{ \;1\;-\; r\;} \; \right ] \; 
$$

where we again used the fact 
$Tr[\;\mathcal{I}\;] \;=\; 2$ and 
$Tr[\;\sigma_x\;] \;=\; Tr[\;\sigma_y\;] \;=\; Tr[\;\sigma_z\;] \;=\; 0$. 

Putting all the pieces together yields :

$$
Tr \left [ \; \varrho \; \log ( \;\varrho \; ) \; \right ] \;=\; 
\frac{1}{2} \; Tr \left [ \;  \mathcal{I} \;\bullet \; \log  ( \;\varrho \; ) \;
\right ] 
\; + \; \frac{1}{2} \; 
Tr \left [ \; \left ( \; \vec{\mathcal{W}} \bullet \vec{\sigma} \; \right ) \; \log ( \;\varrho \; )\; \right ] 
$$

$$
\; =\; 
\log \left [ \,\frac{1}{2} \, \right ] \;+\; 
\frac{1}{2} \; \log \left [  \;1\;-\; r^2\; \right ] 
\; +\; \frac{r}{2} \; \log \left [  \; \frac{ \;1\;+\; r\;}{ \;1\;-\; r\;} \; \right ] \; 
$$

To evaluate $ Tr[ \; \varrho \; \log ( \;\psi \; ) \; ]$, we follow a similar
path and use the Bloch sphere representation for $\psi$ of 

$$
\psi \; = \; \frac{1}{2} \; \left ( \mathcal{I} \; + \; \vec{\mathcal{V}} 
\bullet \vec{\sigma} \right )
$$

The expression for $\log ( \;\psi\; )$ then becomes 

$$
\log ( \;\psi\; )  
\;=\; 
\log \left [  \;\frac{1}{2} \; \right ]  \;+\; 
\frac{ \; \vec{\mathcal{V}} \bullet \vec{\sigma} \;}{\; 2 \; q\; } \; 
\log \left [  \; \frac{ \;1\;+\; q\;}{ \;1\;-\; q\;} \; \right ] \; 
\;  + \; \frac{1}{2} \;  \log \left [  \;1\;-\; q^2\; \right ] \; 
$$

Using our results above, 

$$
\frac{1}{2} \;
Tr \left [ \;  \mathcal{I} \;\bullet \; \log  ( \;\psi \; ) \; \right ] 
\; =\; 
\log \left [  \;\frac{1}{2} \; \right ]  \;+\; 
\frac{1}{2} \; \log \left [  \;1\;-\; q^2\; \right ] 
\;=\; 
- \, \log [\,2\, ] \;+\; \frac{1}{2} \; \log \left [  \;1\;-\; q^2\; \right ] 
$$

$$
Tr \left [ \; \left ( \; \vec{\mathcal{W}} \bullet \vec{\sigma} \; \right ) \; 
\log ( \;\psi \; )\; \right ] 
\; =\; 
\frac{\;  \vec{\mathcal{W}} \bullet \vec{\mathcal{V}} \; }{q}
\; \log \left [  \; \frac{ \;1\;+\; q\;}{ \;1\;-\; q\;} \; \right ] \; 
\; =\; r \; \cos ( \; \theta \; ) \; \log \left [  \; \frac{ \;1\;+\; q\;}{ \;1\;-\; q\;} \; \right ] \; 
$$

where we again used the fact 
$Tr[\;\mathcal{I}\;] \;=\; 2$ and 
$Tr[\;\sigma_x\;] \;=\; Tr[\;\sigma_y\;] \;=\; Tr[\;\sigma_z\;] \;=\; 0$. 
We also used the fact that

$$
\left ( \; \vec{\mathcal{V}} \bullet \vec{\sigma} \; \right ) \; 
\left ( \; \vec{\mathcal{W}} \bullet \vec{\sigma} \; \right ) \; 
= \; 
\left ( \; \vec{\mathcal{V}} \bullet \vec{\mathcal{W}} \; \right ) \; \mathcal{I}
\;+\; 
\left ( \; \vec{\mathcal{V}} \times \vec{\mathcal{W}} \; \right ) \; 
\bullet \vec{\sigma} 
$$

and therefore 

$$
Tr \left [ \; \left ( \; \vec{\mathcal{V}} \bullet \vec{\sigma} \; \right ) \; 
\left ( \; \vec{\mathcal{W}} \bullet \vec{\sigma} \; \right ) \; \right ]
\; = \; 
Tr \left [ \; \left (  \; 
\vec{\mathcal{V}} \bullet \vec{\mathcal{W}} \; \right ) \; \mathcal{I} \; \right ]
\;+\; 
Tr \left [ \; \left ( \; \vec{\mathcal{V}} \times \vec{\mathcal{W}} \;
\right ) \; \bullet \vec{\sigma} \; \right ] 
$$

$$
\; = \; 
\left (  \; \vec{\mathcal{V}} \bullet \vec{\mathcal{W}} \; \right ) \;
Tr \left [ \; \mathcal{I} \; \right ]
\;+\; 
\left ( \; \vec{\mathcal{V}} \times \vec{\mathcal{W}} \;
\right ) \; \bullet \; Tr \left [ \; \vec{\sigma} \; \right ] 
\; = \; 
2 \; \vec{\mathcal{V}} \bullet \vec{\mathcal{W}} 
$$

Assembling the pieces :

$$
Tr[ \; \varrho \; \log ( \;\psi \; ) \; ] \;=\; 
\frac{1}{2} \; Tr[\;  \mathcal{I} \;\bullet \; \log  ( \;\psi \; ) \;]\;
+ \; \frac{1}{2} \; 
Tr \left [ \; \left ( \; \vec{\mathcal{W}} \bullet \vec{\sigma} \; \right ) \; \log ( \;\psi \; )\; \right ] 
$$

$$
\; =\; 
\log \left [ \; \frac{1}{2} \; \right ]  \quad + \quad 
\frac{1}{2} \; \log \left [  \;1\;-\; q^2\; \right ] 
\; +\; \frac{r}{2} \; \cos ( \; \theta \; ) \; \log \left [  \; \frac{ \;1\;+\; q\;}{ \;1\;-\; q\;} \; \right ] \; 
$$

Using these pieces, we obtain our final formula :
 
$$
\mathcal{D}( \; \varrho \, || \; \psi\;) \;=\; Tr [ \; \varrho \, ( \; \log_2(\varrho) \;-\; \log_2(\psi) \;
) \; ]
$$

$$
\; =\; 
\log_2 \left [ \; \frac{1}{2} \; \right ]  \quad + \quad 
\frac{1}{2} \; \log_2 \left [  \;1\;-\; r^2\; \right ] 
\; +\; \frac{r}{2} \; \log_2 \left [  \; \frac{ \;1\;+\; r\;}{ \;1\;-\; r\;} \; \right ] \;
$$

$$
- \; \log_2 \left [ \; \frac{1}{2} \; \right ] 
\; -\; \frac{1}{2} \; \log_2 \left [  \;1\;-\; q^2\; \right ] 
\; -\; \frac{r}{2} \; \cos ( \; \theta \; ) \; \log_2 \left [  \; \frac{ \;1\;+\; q\;}{ \;1\;-\; q\;} \; \right ] \; 
$$

$$
\; =\; 
\frac{1}{2} \; \log_2 \left [  \;1\;-\; r^2\; \right ] 
\; +\; \frac{r}{2} \; \log_2 \left [  \; \frac{ \;1\;+\; r\;}{ \;1\;-\; r\;} \; \right ]
\; -\; \frac{1}{2} \; \log_2 \left [  \;1\;-\; q^2\; \right ] 
\; -\; \frac{r}{2} \; \cos ( \; \theta \; ) \; \log_2 \left [  \; \frac{ \;1\;+\; q\;}{ \;1\;-\; q\;} \; \right ] \; 
$$

which is our desired formula. 

$\bigtriangleup$ - {\em End of Proof I}. 

\vspace{0.2in}

\subsection{Proof II : The Brute Force Proof }

The density matrices for $\varrho$ and $\psi$ in terms of their Bloch
vectors are :

$$
\varrho \;=\;  \left[ \begin {array}{cc} \frac{1}{2}+\frac{1}{2}\,{\it w_3}&\frac{1}{2}\,{\it w_1}-\frac{1}{2}\,i{\it 
w_2}\\\noalign{\medskip}\frac{1}{2}\,{\it w_1}+\frac{1}{2}\,i{\it w_2}&\frac{1}{2}-\frac{1}{2}\,{\it w_3}
\end {array} \right] 
$$

$$
\psi \;=\;  \left[ \begin {array}{cc} \frac{1}{2}+\frac{1}{2}\,{\it v_3}&\frac{1}{2}\,{\it v_1}-\frac{1}{2}\,i{\it 
v_2}\\\noalign{\medskip}\frac{1}{2}\,{\it v_1}+\frac{1}{2}\,i{\it v_2}&\frac{1}{2}-\frac{1}{2}\,{\it v_3}
\end {array} \right] 
$$

The eigenvalues of these two  density matrices are :

$$
\lambda_{\varrho}^{(1)} \;=\; \frac{1}{2}\; +\; \frac{1}{2}\,\sqrt {{{\it w_2}}^{2}+{{\it w_3}}^{2}+{{\it w_1}}^{2}} \;=\; \frac{1 \;+\; r}{2}
$$

$$
\lambda_{\varrho}^{(2)} \;=\; \frac{1}{2}\;- \;\frac{1}{2}\,\sqrt {{{\it w_2}}^{2}+{{\it w_3}}^{2}+{{\it w_1}}^{2}} \;=\; \frac{1 \;-\; r}{2}
$$

$$
\lambda_{\psi}^{(1)} \;=\; \frac{1}{2}\; +\; \frac{1}{2}\,\sqrt {{{\it v_2}}^{2}+{{\it v_3}}^{2}+{{\it v_1}}^{2}} \;=\; \frac{1 \;+\; q}{2}
$$

$$
\lambda_{\psi}^{(2)} \;=\; \frac{1}{2}\;- \;\frac{1}{2}\,\sqrt {{{\it v_2}}^{2}+{{\it v_3}}^{2}+{{\it v_1}}^{2}} \;=\; \frac{1 \;-\; q}{2}
$$

\vspace{0.2in}

We shall also be interested in the two eigenvectors of $\psi$. These are :
$$
\ket{e_1} \;=\; N_1 \;  
\bmatrix{1 \cr -{\frac {-2\, \left( \frac{1}{2}+\frac{1}{2}\,\sqrt {{{\it w_2}}^{2}+{{\it w_3}}^{2}+
{{\it w_1}}^{2}} \right) {\it w_1}-2\,i \left( \frac{1}{2}+\frac{1}{2}\,\sqrt {{{\it w_2}
}^{2}+{{\it w_3}}^{2}+{{\it w_1}}^{2}} \right) {\it w_2}+{\it w_1}+i{\it 
w_2}+{\it w_3}\,{\it w_1}+i{\it w_3}\,{\it w_2}}{{{\it w_1}}^{2}+{{\it w_2}}^
{2}}}}
$$

where $N_1$ is the normalization constant given below.

$$
N_1 \;=\; \sqrt{ 2\,{\frac {{{\it w_1}}^{2}+\sqrt {{{\it w_2}}^{2}+{{\it w_3}}^{2}+{{\it
w_1}}^{2}}{
\it w_3}+{{\it w_2}}^{2}+{{\it w_3}}^{2}}{{{\it w_1}}^{2}+{{\it w_2}}^{2}}}}
$$

Similarly,

$$
\ket{e_2} \;=\; N_2 \; 
\bmatrix{ {\frac {2\, \left( \frac{1}{2}-\frac{1}{2}\,\sqrt {{{\it w_2}}^{2}+{{\it w_3}}^{2}+{{\it
w_1}}^{2}}
 \right) {\it w_1}-2\,i \left( \frac{1}{2}-\frac{1}{2}\,\sqrt {{{\it w_2}}^{2}+{{\it
w_3}}^{2}+{{\it 
w_1}}^{2}} \right) {\it w_2}-{\it w_1}+i{\it w_2}+{\it w_3}\,{\it w_1}-i{\it
w_3}\,{\it 
w_2}}{{{\it w_1}}^{2}+{{\it w_2}}^{2}}} \cr 1}
$$

$$
N_2 \; =\; 
\sqrt{2\,{\frac {{{\it w_1}}^{2}+\sqrt {{{\it w_2}}^{2}+{{\it w_3}}^{2}+{{\it
w_1}}^{2}}{
\it w_3}+{{\it w_2}}^{2}+{{\it w_3}}^{2}}{{{\it w_1}}^{2}+{{\it w_2}}^{2}}}}
$$

We wish to derive a formula for $\mathcal{D}(\varrho \, \| \,  \psi)$ in terms of
the Bloch sphere vectors $\vec{\mathcal{W}}$ and 
$\vec{\mathcal{V}}$. We do this by 
breaking  $\mathcal{D}(\varrho \, \| \, \psi)$
up into two terms, $\mathcal{D}_1$ and 
$\mathcal{D}_2$.

$$
\mathcal{D}(\varrho \, \| \,  \psi) \;=\; \mathcal{D}_1 \;-\; \mathcal{D}_2
$$

We expand $\mathcal{D}_1$ using our knowledge of the eigenvalues of $\varrho$.

$$
\mathcal{D}_1\;=\; Tr [ \;\varrho \; \log_2(\varrho) \; ]
$$

$$
 \;=\; 
\lambda_{\varrho}^{(1)} \; \log_2( \lambda_{\varrho}^{(1)} ) \; + \; 
\lambda_{\varrho}^{(2)} \; \log_2( \lambda_{\varrho}^{(2)} ) 
$$

$$
\;=\; 
\left ( \frac{ \;1\;+\;r\;}{\;2\;} \right ) 
\log_2 \left ( \frac{ \;1\;+\;r\;}{\;2\;} \right ) \;\; + \;\; 
\left ( \frac{ \;1\;-\;r\;}{\;2\;} \right ) 
\log_2 \left ( \frac{ \;1\;-\;r\;}{\;2\;} \right ) 
$$

$$
\;=\; -\;\; 1 \;+\; 
\left ( \frac{ \;1\;+\;r\;}{\;2\;} \right ) 
\log_2 \left ( 1\;+\;r \right ) \;\; + \;\; 
\left ( \frac{ \;1\;-\;r\;}{\;2\;} \right ) 
\log_2 \left ( 1\;-\;r \right ) 
$$

One notes that $\mathcal{D}_1\;=\; -\mathcal{S}(\varrho)$, where 
$\mathcal{S}(\varrho)$ is the von
Neumann entropy of the density matrix $\varrho$.
The second term, $\mathcal{D}_2$, is 
$\mathcal{D}_2\;=\; Tr [ \;\varrho \; \log_2(\psi) \; ]$.
We evaluate $\mathcal{D}_2$ in the basis which diagonalizes $\psi$.

$$
\mathcal{D}_2\;=\; Tr [ \;\varrho \; \log_2(\varrho) \; ] \;=\; 
\log_2( \lambda_{\varrho}^{(1)} ) \; 
Tr \left [ \; \varrho \; \ket{e_1}\bra{e_1} \; \right ]\; + \; 
\log_2( \lambda_{\varrho}^{(2)} ) \; 
Tr \left [ \; \varrho \; \ket{e_2}\bra{e_2} \; \right ]
$$

We use the Bloch sphere representation for $\varrho$ in the expression
for $\mathcal{D}_2$.

$$
\psi \; = \;  \frac{1}{2} \; 
( \; \mathcal{I} \; + \; \vec{\mathcal{V}} \bullet \vec{\sigma} \; )
$$

$$
\mathcal{D}_2\;=\; Tr [ \;\varrho \; \log_2(\psi) \; ] \;=\; 
$$

$$
\frac{1}{2} \; \log_2( \lambda_{\psi}^{(1)} ) \left [ 
\; Tr[ \; \ket{e_1}\bra{e_1} \; ]\; + \; 
\; \sum_{i} \; w_i \; Tr[ \; \sigma_i \; \ket{e_1}\bra{e_1}  \; ] \right ] 
\;+\; 
$$

$$
\frac{1}{2} \; \log_2( \lambda_{\psi}^{(2)} ) \left [ 
\; Tr[ \; \ket{e_2}\bra{e_2} \; ]\; + \; 
\; \sum_{i} \; w_i \; Tr[ \; \sigma_i \; \ket{e_2}\bra{e_2} \; ] \right ]
$$

First note that
$Tr[ \; \ket{e_1}\bra{e_1} \; ]\; = \; Tr[ \; \ket{e_2}\bra{e_2} \; ]\; = \;1$
since the $\ket{e_{j}}$ are projection operators. 

Next define 

$$
\alpha_{i}^{(j)} \;=\; Tr[ \sigma_i \; \ket{e_j}\bra{e_j} ] 
\;=\; \bra{e_j} \, \sigma_i \; \ket{e_j}
$$

Evaluating these six ( i = 1,2,3 and j = 1,2 ) constants yields :

$$
\alpha_1^{(1)} \;=\; 
{\frac {{\it v_1}\, \left( \sqrt {{{\it v_2}}^{2}+{{\it v_3}}^{2}+{{\it
v_1}}^{2}
}+{\it v_3} \right) }{{{\it v_1}}^{2}+\sqrt {{{\it v_2}}^{2}+{{\it
v_3}}^{2}+{{\it v_1}
}^{2}}{\it v_3}+{{\it v_2}}^{2}+{{\it v_3}}^{2}}}
\;=\; 
\frac{ v_1 \; ( \; q \;+\; v_3\;)}{q^2 \;+\;
q\,v_3} \;=\; \frac{v_1}{q}
$$

$$
\alpha_2^{(1)} \;=\; 
{\frac {{\it v_2}\, \left( \sqrt {{{\it v_2}}^{2}+{{\it v_3}}^{2}+{{\it
v_1}}^{2}
}+{\it v_3} \right) }{{{\it v_1}}^{2}+\sqrt {{{\it v_2}}^{2}+{{\it
v_3}}^{2}+{{\it v_1}
}^{2}}{\it v_3}+{{\it v_2}}^{2}+{{\it v_3}}^{2}}}
\;=\; 
\frac{ v_2 \; ( \; q \;+\; v_3\;)}{q^2 \;+\;
q\,v_3} \;=\; \frac{v_2}{q}
$$

$$
\alpha_3^{(1)} \;=\; 
{\frac { \left( \sqrt {{{\it v_2}}^{2}+{{\it v_3}}^{2}+{{\it
v_1}}^{2}}+{\it v_3}
 \right) {\it v_3}}{{{\it v_1}}^{2}+\sqrt {{{\it v_2}}^{2}+{{\it v_3}}^{2}+{{\it
v_1}}^
{2}}{\it v_3}+{{\it v_2}}^{2}+{{\it v_3}}^{2}}}
\;=\; 
\frac{ v_3 \; ( \; q \;+\; v_3\;)}{q^2 \;+\;
q\,v_3} \;=\; \frac{v_3}{q}
$$

$$
\alpha_1^{(2)} \;=\; 
{\frac {{\it v_1}\, \left( \sqrt {{{\it v_2}}^{2}+{{\it v_3}}^{2}+{{\it
v_1}}^{2}
}-{\it v_3} \right) }{-{{\it v_1}}^{2}+\sqrt {{{\it v_2}}^{2}+{{\it
v_3}}^{2}+{{\it v_1
}}^{2}}{\it v_3}-{{\it v_2}}^{2}-{{\it v_3}}^{2}}}
\;=\; 
- \frac{ v_1 \; ( \; q \;-\; v_3\;)}{q^2 \;-\;
q\,v_3} \;=\; -\frac{v_1}{q}
$$

$$
\alpha_2^{(2)} \;=\; 
{\frac {{\it v_2}\, \left( \sqrt {{{\it v_2}}^{2}+{{\it v_3}}^{2}+{{\it
v_1}}^{2}
}-{\it v_3} \right) }{-{{\it v_1}}^{2}+\sqrt {{{\it v_2}}^{2}+{{\it
v_3}}^{2}+{{\it v_1
}}^{2}}{\it v_3}-{{\it v_2}}^{2}-{{\it v_3}}^{2}}}
\;=\; 
- \frac{ v_1 \; ( \; q \;-\; v_3\;)}{q^2 \;-\;
q\,v_3} \;=\; - \frac{v_2}{q}
$$

$$
\alpha_3^{(2)} \;=\; 
{\frac { \left( \sqrt {{{\it v_2}}^{2}+{{\it v_3}}^{2}+{{\it
v_1}}^{2}}-{\it v_3}
 \right) {\it v_3}}{-{{\it v_1}}^{2}+\sqrt {{{\it v_2}}^{2}+{{\it
v_3}}^{2}+{{\it v_1}}
^{2}}{\it v_3}-{{\it v_2}}^{2}-{{\it v_3}}^{2}}}
\;=\; 
- \frac{ v_3 \; ( \; q \;-\; v_3\;)}{q^2 \;-\;
q\,v_3} \;=\; - \frac{v_3}{q}
$$

Putting it all together yields : 

$$
\mathcal{D}_2\;=\; Tr [ \;\varrho \; \log_2(\psi) \; ]
$$

$$
 \;=\; \frac{1}{2} \; \log_2 \left ( \lambda_{\psi}^{(1)} \right ) \left [ 
\; 1\; + \; 
\; \sum_{i} \; w_i \, \alpha_{i}^{(1)}  \; \right ] 
\;+\; 
\frac{1}{2} \; \log_2 \left ( \lambda_{\psi}^{(2)} \right ) \left [ 
\; 1\; + \; 
\; \sum_{i} \; w_i \, \alpha_{i}^{(2)}  \; \right ] 
$$

$$
 \;=\; \frac{1}{2} \; \left [ \; 1\; + \; 
\; \sum_{i} \; w_i \, \frac{v_i}{q}  \; \right ] 
\; \log_2 \left ( \lambda_{\psi}^{(1)} \right ) 
\;+\; 
\frac{1}{2} \; \left [ \; 1\; + \; 
\; \sum_{i} \; w_i \, \frac{-\,v_i}{q}  \; \right ] 
\; \log_2 \left ( \lambda_{\psi}^{(2)} \right ) 
$$

$$
\;=\; \frac{1}{2}
\; \left [ 
\; 1\; + \; \; \frac{ \vec{\mathcal{W}} \bullet \vec{\mathcal{V}}}{q} \; \right ] 
\; \log_2 \left ( \lambda_{\psi}^{(1)} \right ) 
\;+\; 
\frac{1}{2}
\; \left [ 
\; 1\; - \; 
\; \frac{ \vec{\mathcal{W}} \bullet \vec{\mathcal{V}}}{q} \; \right ] 
\; \log_2 \left ( \lambda_{\psi}^{(2)} \right ) 
$$

Plugging in for the eigenvalues $\lambda_{\psi}^{(1)}$ and 
$\lambda_{\psi}^{(2)}$ which we found above yields :

$$
\mathcal{D}_2\;=\; Tr [ \;\varrho \; \log_2(\psi) \; ]
$$

$$
\;=\; \frac{1}{2}
\; \left [ 
\; 1\; + \; \; \frac{ \vec{\mathcal{W}} \bullet \vec{\mathcal{V}}}{q} \; \right ] 
\; \log_2 \left ( \frac{ 1 \;+\; q}{2} \right ) 
\;+\; 
\frac{1}{2}
\; \left [ 
\; 1\; - \; 
\; \frac{ \vec{\mathcal{W}} \bullet \vec{\mathcal{V}}}{q} \; \right ] 
\; \log_2 \left ( \frac{ 1 \;-\; q}{2} \right ) 
$$

$$
\;=\; \frac{1}{2}
\;  \log_2 ( 1 \;-\; q^2\; ) \;-\; 1
\;+\; \frac{\; \vec{\mathcal{W}} \bullet \vec{\mathcal{V}} \; }
{\; 2 \; q\; } \; \log_2 \left ( \frac{ 1 \;+\; q }{1 \;-\; q} \right)
$$

Putting all the pieces together to 
obtain $\mathcal{D}( \, \varrho \,\| \,  \psi \, )$, we find 

$$
\mathcal{D}( \, \varrho \,  \| \,  \psi \, ) \;=\; 
\mathcal{D}_1 \;-\; \mathcal{D}_2
\;=\; 
\frac{1}{2} \log_2 ( 1\;-\;r^2 ) \;+\; 
\frac{r}{2} \log_2 \left ( \frac{1\;+\;r}{1\;-\;r} \right )
\;-\; 
\frac{1}{2}
\;  \log_2 ( 1 \;-\; q^2\; ) 
\;-\; \frac{\; \vec{\mathcal{W}} \bullet \vec{\mathcal{V}} \; }
{\; 2 \; q\; } \; \log_2 \left ( \frac{  1 \;+\; q}{1 \;-\; q} \right)
$$

$$
\;=\; 
\frac{1}{2} \, \log_2 ( 1\;-\;r^2 ) \;+\; 
\frac{r}{2} \, \log_2 \left ( \frac{1\;+\;r}{1\;-\;r} \right )
\;-\; 
\frac{1}{2}
\;  \log_2 ( 1 \;-\; q^2\; ) 
\;-\; \frac{r \; \cos ( \theta )  }
{\; 2 \; } \; \log_2 \left ( \frac{  1 \;+\; q}{1 \;-\; q} \right ) .
$$

where $\theta$ is the angle between $\vec{\mathcal{W}}$ 
and $\vec{\mathcal{V}}$.

$\bigtriangleup$ - {\em End of Proof II}. 

Ordinarily, $\mathcal{D}( \rho \| \phi ) \;\neq \;\mathcal{D}( \phi \| \rho )$. 
However, when $r$ = $q$, we can see from the above formula that 
$\mathcal{D}( \rho \| \phi ) \;=\;\mathcal{D}( \phi \| \rho )$.  

A few special cases of 
$\mathcal{D}( \rho \| \phi )$ are worth examining.
Consider the case when $\phi \;=\; \frac{1}{2} \; \mathcal{I}$.
In this case, $q$ = 0, and 
$$
\mathcal{D}( \; \rho \; || \;  \phi \;  ) \;=\; 
\frac{1}{2} \; \log_2 \left (  1 - r^2 \right ) \;+ \;  
\frac{r}{2} \; \log_2 \left (  \frac{1 + r}{1 - r } \right )  
$$

$$
\;=\; \frac{ 1 \;+\; r}{2} \; \log_2 \left ( \frac{ 1 \;+\; r}{2} \; \right  )
\;+\; \frac{ 1 \;+\; r}{2} \; 
\;+\; \frac{ 1 \;-\; r}{2} \; \log_2 \left ( \frac{ 1 \;-\; r}{2} \; \right  )
\;+\; \frac{ 1 \;-\; r}{2} \; 
\;=\; 1 \;-\; \mathcal{S}(\rho)
$$

Thus, 
$\mathcal{D}(\; \rho \; || \;  \frac{1}{2} \; \mathcal{I} \; ) \;=\; 1 \;-\; \mathcal{S}(\rho)$,
where $\mathcal{S}(\rho)$ is the von Neumann entropy of $\rho$, the first density matrix
in the relative entropy function. 

\section{Appendix B - The Derivation Of The Linear Channel
Transcendental Equation}

In this appendix, we derive the transcendental equation for determining the
optimum position of the average density matrix for a linear channel.
The picture of the quantities we shall define shortly is below.

\begin{center}

\setlength{\unitlength}{0.00033300in}%
\begingroup\makeatletter\ifx\SetFigFont\undefined
\def\x#1#2#3#4#5#6#7\relax{\def\x{#1#2#3#4#5#6}}%
\expandafter\x\fmtname xxxxxx\relax \def\y{splain}%
\ifx\x\y   
\gdef\SetFigFont#1#2#3{%
  \ifnum #1<17\tiny\else \ifnum #1<20\small\else
  \ifnum #1<24\normalsize\else \ifnum #1<29\large\else
  \ifnum #1<34\Large\else \ifnum #1<41\LARGE\else
     \huge\fi\fi\fi\fi\fi\fi
  \csname #3\endcsname}%
\else
\gdef\SetFigFont#1#2#3{\begingroup
  \count@#1\relax \ifnum 25<\count@\count@25\fi
  \def\x{\endgroup\@setsize\SetFigFont{#2pt}}%
  \expandafter\x
    \csname \romannumeral\the\count@ pt\expandafter\endcsname
    \csname @\romannumeral\the\count@ pt\endcsname
  \csname #3\endcsname}%
\fi
\fi\endgroup
\begin{picture}(9000,6738)(1201,-7015)
\thicklines
\put(1801,-3961){\line( 2, 1){7780}}
\put(1801,-3961){\line( 5,-2){7758.621}}
\put(1801,-3961){\line( 6, 1){7783.784}}
\put(9571,-101){\line( 0,-1){7000}}
\put(5401,-2761){\makebox(0,0)[lb]{\smash{\SetFigFont{12}{14.0}{rm}$\theta_{+}$}}}
\put(1201,-4561){\makebox(0,0)[lb]{\smash{\SetFigFont{12}{14.0}{rm}Bloch}}}
\put(1201,-4956){\makebox(0,0)[lb]{\smash{\SetFigFont{12}{14.0}{rm}Sphere}}}
\put(1201,-5351){\makebox(0,0)[lb]{\smash{\SetFigFont{12}{14.0}{rm}Origin}}}
\put(10201,-2761){\makebox(0,0)[lb]{\smash{\SetFigFont{12}{14.0}{rm}q}}}
\put(10201,-361){\makebox(0,0)[lb]{\smash{\SetFigFont{12}{14.0}{rm}$r_{+}$}}}
\put(10201,-6961){\makebox(0,0)[lb]{\smash{\SetFigFont{12}{14.0}{rm}$r_{-}$}}}
\put(5401,-4561){\makebox(0,0)[lb]{\smash{\SetFigFont{12}{14.0}{rm}$\theta_{-}$}}}
\end{picture}

\end{center}

\begin{center}
Figure 25: 
Definition of the Bloch vectors 
$\vec{r}_{+}$, 
$\vec{q}$, and  
$\vec{r}_{-}$ used in the derivation below. 
\end{center}

We assume that in general all the $\{\,t_k\,\neq\,0\}$. We also assume the
linear channel is oriented in the z direction, so that
$\lambda_x \,=\, \lambda_y \,=\, 0$, but 
$\lambda_z \,\neq\, 0$. We define

$$
A \;=\; t_x^2 \,+\, t_y^2 \,+\, ( \, t_z \,+\, \lambda_z \,)^2 \;=\; r_{+}^2 
$$

$$
B \;=\; t_x^2 \,+\, t_y^2 \,+\, ( \, t_z \,+\, \beta \, \lambda_z \,)^2
\;=\; q( \, \beta \, )^2 
$$

$$
C \;=\; t_x^2 \,+\, t_y^2 \,+\, ( \, t_z \,-\, \lambda_z \,)^2 \;=\; r_{-}^2 
$$

The three quantities above refer respectively 
to the distance from the Bloch
sphere origin to $r_{+}$, the optimum point {\em q} we seek, and 
$\,r_{-}$. 
We define the three Bloch vectors 
$\vec{r}_{+}$,  $\vec{q}$ and $\vec{r}_{-}$ in Figure 25 above, and
refer to their respective magnitudes as 
$r_{+}$,  $q$, and $r_{-}$.  
Here $\beta \, \in [-1,1]$, so that $q$ can range along the entire line
segment between $r_{+}$ and $r_{-}$. 

As discussed in the Linear Channels section of this paper, 
the condition on $q$ is that 
$\mathcal{D}(\, r_{+} \, \| \, q \, ) \;= \; \mathcal{D}(\, r_{-} \, \| \, q \, )$.  

Now recall that 

$$
\mathcal{D}( \, r\, \|\,q\,)\;=\; 
\;=\; 
\frac{1}{2} \, \log_2 ( 1\;-\;r^2 ) \;+\; 
\frac{r}{2} \, \log_2 \left ( \frac{1\;+\;r}{1\;-\;r} \right )
\;-\; 
\frac{1}{2}
\;  \log_2 ( 1 \;-\; q^2\; ) 
\;-\; \frac{r \; \cos ( \theta )  }
{\; 2 \; } \; \log_2 \left ( \frac{  1 \;+\; q}{1 \;-\; q} \right)
$$

where $\theta$ is the angle between $r$ and $q$. To determine $\theta$, we use
the law of cosines. If $\theta $ is the angle between sides a and b of 
a triangle with sides a, b and c, then we have :

$$
\cos(\theta) \;=\; \frac{  a^2 \;+\; b^2 \; -\; c^2}{2 \, a \, b }
$$

Our condition 
$\mathcal{D}(\, r_{+} \, \| \, q \, ) \;= \; \mathcal{D}(\, r_{-} \, \| \, q \, )$ becomes :

$$
\frac{1}{2} \, \log ( 1\;-\;r_{+}^2 ) \;+\; 
\frac{r}{2} \, \log \left ( \frac{1\;+\;r_{+}}{1\;-\;r_{+}} \right )
\;-\; \frac{r_{+} \; \cos ( \theta_{+} )  }
{\; 2 \; } \; \, \log \left ( \frac{  1 \;+\; q}{1 \;-\; q} \right)
$$
$$
\;=\; 
\frac{1}{2} \, \log ( 1\;-\;r_{-}^2 ) \;+\; 
\frac{r}{2} \, \log \left ( \frac{1\;+\;r_{-}}{1\;-\;r_{-}} \right )
\;-\; \frac{r_{-} \; \cos ( \theta_{-} )  }
{\; 2 \; } \; \log \left ( \frac{  1 \;+\; q}{1 \;-\; q} \right)
$$

where we canceled the term which was identically a function of $q$ from both
sides, and converted all logs from base 2 to natural logs by multiplying
both sides by $\log(2)$. 

Determining $\theta_{+}$ and $\theta_{-}$, we find :

$$
\cos(\theta_{+} ) \;=\; \frac{  r_{+}^2 \;+\; q^2 \; -\; ( \, ( \, 1 \, -\,
\beta \, ) \, \lambda_z \, ) ^2}{2 \, q\,  r_{+} \,  }
$$

$$
\cos(\theta_{-} ) \;=\; \frac{  r_{-}^2 \;+\; q^2 \; -\; ( \, ( \, 1 \, +\,
\beta \, ) \, \lambda_z \, ) ^2}{2 \, q \, r_{-} \,  }
$$

Next, recall the identity 

$$
\tanh^{(-1)} [ \; x \; ] \;=\;  
\frac{1}{2} \, \log \left ( \frac{1\;+\;x}{1\;-\;x} \right )
$$

Using this identity for arctanh, our relative entropy equality relation
between the two endpoints of the linear channel becomes :

$$
\frac{1}{2} \log ( 1 \,-\, A ) \, +\, 
\sqrt{A} \, \tanh^{(-1)} \left ( \, \sqrt{A} \, \right ) 
\,-\, \frac { \sqrt{A} \; ( \,  A \,  + \,  B \,  - \, 
(( 1 \, - \, \beta)\, \lambda_z)^2\, )}{2 \, \sqrt{A \, B } } \; 
\tanh^{(-1)}\left ( \sqrt{B} \right ) 
$$

$$
\;=\; 
\frac{1}{2} \, \log ( 1 \,-\, C ) \, +\, 
\sqrt{C} \, \tanh^{(-1)} \left ( \, \sqrt{C} \, \right ) 
\,-\, \frac { \sqrt{C} \, ( \,  C  \,+  \, B  \, -  \, ((
1\, + \, \beta)\, \lambda_z)^2\, ) }{2\sqrt{B \, C } } \;
\tanh^{(-1)}\left ( \sqrt{B} \right ) 
$$

We can cancel several terms to obtain 

$$
\frac{1}{2} \, \log ( 1 \,-\, A ) \, +\, 
\sqrt{A} \, \tanh^{(-1)} \left ( \, \sqrt{A} \, \right ) 
\,-\, \frac { ( \,  A \,  + \,  
\, 2 \,  \beta\, \lambda_z^2\, )}{2 \, \sqrt {B} } \; 
\tanh^{(-1)}\left ( \sqrt{B} \right ) 
$$

$$
\;=\; 
\frac{1}{2} \, \log ( 1 \,-\, C ) \, +\, 
\sqrt{C} \, \tanh^{(-1)} \left ( \, \sqrt{C} \, \right ) 
\,-\, \frac { ( \,  C  \, -  \, 2 \,  \beta\, \lambda_z^2\, ) }{2 \, \sqrt{B} } \;
\tanh^{(-1)}\left ( \sqrt{B} \right ) 
$$

which in turn becomes :

$$
\frac{1}{2} \, \log ( 1 \,-\, A ) \, +\, 
\sqrt{A} \, \tanh^{(-1)} \left ( \, \sqrt{A} \, \right ) 
\,-\, \frac { ( \,  A \,  - \, C \, + \,  
\, 4 \,  \beta\, \lambda_z^2\, )}{2 \, \sqrt {B} } \; 
\tanh^{(-1)}\left ( \sqrt{B} \right ) 
$$

$$
\;=\; 
\frac{1}{2} \, \log ( 1 \,-\, C ) \, +\, 
\sqrt{C} \, \tanh^{(-1)} \left ( \, \sqrt{C} \, \right ) 
$$

Using our definitions above for A and C, we find that
A - C $=\; 4 \, \lambda_z \, t_z $. Substituting this into 
the relation immediately above yields :

$$
\frac{1}{2} \, \log ( 1 \,-\, A ) \, +\, 
\sqrt{A} \, \tanh^{(-1)} \left ( \, \sqrt{A} \, \right ) 
\,-\, \frac { ( \,   4 \, \lambda_z \, t_z \, + \,  
\, 4 \,  \beta\, \lambda_z^2\, )}{2 \, \sqrt {B} } \; 
\tanh^{(-1)}\left ( \sqrt{B} \right ) 
$$

$$
\;=\; 
\frac{1}{2} \log ( 1 \,-\, C ) \, +\, 
\sqrt{C} \, \tanh^{(-1)} \left ( \, \sqrt{C} \, \right ) 
$$

which we adjust to our final answer :

$$
\frac {   4 \, \lambda_z \,( \,  t_z \, + \,  
\, \beta\, \lambda_z\, )}{\sqrt {B} } \; 
\tanh^{(-1)}\left ( \sqrt{B} \right ) 
$$

$$
=\; 
\log ( 1 \,-\, A ) 
\;-\; 
\log ( 1 \,-\, C ) 
\; +\;
2 \, \sqrt{A} \, \tanh^{(-1)} \left ( \, \sqrt{A} \, \right ) 
\; -\; 
2 \, \sqrt{C} \, \tanh^{(-1)} \left ( \, \sqrt{C} \, \right ) 
$$

Note that B is a function of $\beta$, so the entire functionality of
$\beta$ lies to the left of the equality sign in the expression above.
All terms on the right hand side are functions of the $\{t_k\}$
and $\{\lambda_z\}$, so the right hand side is a constant while we vary
$\beta$. 
Since all the functions of $\beta$ on the left hand side are smooth
functions, the search for the optimum $\beta\, \equiv\, q$,
although transcendental, is well behaved and fairly easy.

\vspace{0.2in}

\section{Appendix C - Donald's Equality}

We prove Donald's Equality below\cite{Donald}. Let $\rho_i$ be a set
of density matrices with a priori probabilities $\alpha_i$, so that 
$\alpha_i \;\geq \;0 \; \forall \; i$ and $\sum_i \; \alpha_i \;=\; 1$.
Let $\phi$ be any density matrix, and define
$\sigma\;=\; \sum_i\; \alpha_i \; \rho_i$. Then :

$$
\sum_i \; \alpha_i \; D(\, \rho_i \, \| \, \phi \, ) \;=\; 
D(\,\sigma\,\|\,\phi\,) \;+\; \sum_i \; \alpha_i \; D(\, \rho_i \, \| \, \sigma\,)
$$

{\em Proof :}

$$
\sum_i \; \alpha_i \; D(\, \rho_i \, \| \, \phi \, ) \;=\; 
\sum_i \; \alpha_i \; \left \{  \; Tr [ \, \rho_i \, \log( \, \rho_i \,)\, ] \;-\; 
Tr [ \, \rho_i \, \log( \, \phi \,) \, ] \; \right \} 
$$

$$
\;=\; \sum_i \; \alpha_i \; \left \{  \; Tr [ \, \rho_i \, \log( \, \rho_i \,)\, ]
\; \right \}  \;-\; Tr [ \, \sigma \, \log( \, \phi \,) \, ] 
$$

$$
\; =\; \left \{ \; Tr [ \, \sigma \, \log(  \, \sigma \,) \, ] \;-\; 
\; Tr [ \, \sigma \, \log(  \, \sigma \, ) \; \right \} \;
-\; Tr [ \, \sigma \, \log( \, \phi \,) \, ] \; +\; 
\sum_i \; \alpha_i \; Tr [ \, \rho_i \, \log( \, \rho_i \,)\, ] \; 
$$

$$
=\; 
D(\, \sigma \, \|  \, \phi \,) \; 
\;-\; Tr [ \, \sigma \, \log(  \, \sigma \, ) \,  ] \;
\;+\; 
\sum_i \; \alpha_i \; Tr [ \, \rho_i \, \log( \, \rho_i \,)\, ] \; 
$$

$$
\;=\; 
D(\, \sigma \, \| \, \phi \,) \; 
\;+\; 
\sum_i \; \alpha_i \; \left \{ \;
Tr [ \, \rho_i \, \log( \, \rho_i \,)\, ] \; 
\;-\; Tr [ \, \rho_i \, \log( \, \sigma \, ) \,  ] \; \right \} 
$$

$$
\; =\; 
D(\, \sigma \, \| \, \phi \,) \; 
\;+\; 
\sum_i \; \alpha_i \; D(  \, \rho_i \, \| \, \sigma \, ) \;  
$$

$\bigtriangleup$ - {\em End of Proof}. 

\section{Appendix D - Quantum Channel Descriptions}

The Kraus quantum channel representation is given by the set of
Kraus matrices $\mathcal{A} \;=\; \{\;A_i\;\}$
which represent the channel dynamics via the relation :

$$
\mathcal{E}(\rho) \;=\; \sum_i \; A_i \; \rho \; A_{i}^{\dagger}
$$ 

The normalization requirement for the Kraus matrices is :

$$\sum_i \; A_{i}^{\dagger} \; \rho \; A_i \;=\; I $$ 

A channel is unital if it maps the identity to the identity.
This requirement becomes, upon setting $\rho$ = I  :

$$\sum_i \; A_i \; \rho \; A_{i}^{\dagger} \;=\;
\sum_i \; A_i \; A_{i}^{\dagger} \;=\; I $$ 

Each set of Kraus operators, $\{\;\mathcal{A}_i\;\}$ can mapped 
to a set of King-Ruskai-Szarek-Werner ellipsoid channel 
parameters $\{\;t_k\,,\,\lambda_k\;\}$, where $k=1,2,3$.

\vspace{0.2in}

\newpage

{\bf The Two Pauli Channel Kraus Representation}

$$ A_1 \;=\; \bmatrix{ \sqrt{x}  \quad 0 \cr 0  \quad \sqrt{x} } \qquad 
A_2 \;=\; \sqrt { \; \frac{1\,-\,x}{2} \; }  \; \sigma_x \;=\; 
\bmatrix{ \; 0  \qquad  \qquad \sqrt{ \frac{1 \, - \, x}{2}} \cr
\sqrt{ \frac{1 \, - \, x}{2}} \qquad \qquad 0 \;  } $$

$$ A_3 \;=\; - \;i\; \sqrt { \; \frac{1\,-\,x}{2} \; }  \quad \sigma_y \;=\; 
\bmatrix{ 0  \qquad  \qquad -\; \sqrt{ \frac{1 \, - \, x}{2}} \cr
\sqrt{ \frac{1 \, - \, x}{2}} \qquad \qquad 0 } $$ 

In words, the channel leaves the qubit transiting the channel 
alone with probability $x$, and does a $ \sigma_x$ 
on the qubit with probability $\frac{1 \, - \, x}{2}$ or 
does a $ \sigma_y$ 
on the qubit with probability $\frac{1 \, - \, x}{2}$.  
The Two Pauli channel is a unital channel.  
The corresponding King-Ruskai-Szarek-Werner ellipsoid channel 
parameters are 
$t_x \;=\; t_y \;=\; t_z \;=\;0$,  and 
$\lambda_x \;=\; \lambda_y \;=\; x$, while
$\lambda_z \;=\; 2\,x\,-\,1$. \cite{Ruskai99a}
Here $x \; \in \; [\,0\,,\,1\,]$.  

\vspace{0.2in} 

{\bf The Depolarization Channel Kraus Representation} 

$$ A_1 \;=\; \bmatrix{ \sqrt{x}  \quad 0 \cr 0  \quad \sqrt{x} } \qquad 
A_2 \;=\; \sqrt { \; \frac{1\,-\,x}{3} \; }  \quad \sigma_x \;=\; 
\bmatrix{ 0  \qquad  \qquad \sqrt{ \frac{1 \, - \, x}{3}} \cr
\sqrt{ \frac{1 \, - \, x}{3}} \qquad \qquad 0 } $$

$$ A_3 \;=\; - \;i\; \sqrt { \; \frac{1\,-\,x}{3} \; }  \quad \sigma_y \;=\; 
\bmatrix{ 0  \qquad  \qquad -\; \sqrt{ \frac{1 \, - \, x}{3}} \cr
\sqrt{ \frac{1 \, - \, x}{3}} \qquad \qquad 0 } $$

$$ A_4 \;=\; \sqrt { \; \frac{1\,-\,x}{3} \; }  \quad \sigma_z \;=\; 
\bmatrix{ \sqrt{ \frac{1 \, - \, x}{3}} \qquad \qquad 0 \cr
 0 \qquad \qquad - \; \sqrt{\frac{1 \, - \, x}{3}} } $$

In words, the channel leaves the qubit transiting the channel 
alone with probability $x$, and does a $ \sigma_x$ 
on the qubit with probability $\frac{1 \, - \, x}{3}$ or 
does a $ \sigma_y$ 
on the qubit with probability $\frac{1 \, - \, x}{3}$.
or does a $ \sigma_z$ 
on the qubit with probability $\frac{1 \, - \, x}{3}$.  
The Depolarization channel is a unital channel.  
The corresponding King-Ruskai-Szarek-Werner ellipsoid
channel parameters are 
$t_x \;=\; t_y \;=\; t_z \;=\; 0$,  and 
$\lambda_x \;=\; \lambda_y \;=\; \lambda_z \;=\; \frac{4\,x \;-\;1}{3}$. 
\cite{Ruskai99a}
Again $x \; \in \; [\,0\,,\,1\,]$.  

\vspace{0.2in} 

{\bf The Amplitude Damping Channel Kraus Representation} 

$$ A_1 \;=\; \bmatrix{ \sqrt{x}  \quad 0 \cr 0  \quad 1 } \qquad 
A_2 \;=\; 
\bmatrix{ 0  \qquad  \qquad 0 \cr
\sqrt{ 1 \, - \, x} \qquad \qquad 0 } $$

In this scenario, the channel leaves untouched a spin down qubit.
For a spin up qubit, with probability $x$ it leaves the qubit alone, while
with probability 1 - $x$ the channels flips the spin from up to down.
Thus, when $x$ = 0, every qubit emerging from the channel is in the spin
down state. The Amplitude Damping channel is $not$ a unital channel.  
The corresponding King-Ruskai-Szarek-Werner ellipsoid channel
parameters are $t_x \;=\; 0$, 
$t_y \;=\; 0$, 
$t_z \;=\;1\,-\, x$, 
$\lambda_x \;=\; \sqrt{x}$,
$\lambda_y \;=\; \sqrt{x}$, and 
$\lambda_z \;=\; x$. \cite{Ruskai99a}
Again $x \; \in \; [\,0\,,\,1\,]$.  

\vspace{0.3in} 

\section{Appendix E - Numerical Analysis Of Optimal Signal Ensembles Using MAPLE and MATLAB}

The iterative, relative entropy based 
algorithm outlined above was implemented
in MAPLE, and provided the plots and numbers cited in this paper. 
In addition, numerical answers were verified using a brute force
algorithm based on MATLAB's Optimization Toolbox. The MATLAB optimization
criterion was the channel output Holevo $\chi$ quantity. Input qubit
ensembles of two, three and four states were used. After channel
evolution, the output ensemble Holevo $\chi$ was calculated. With this
function specified as to be maximized, the MATLAB Toolbox varied the
parameters for the ensemble qubit input pure states and the states
corresponding a priori probabilities. Pure state qubits
were represented as :

$$
| \, \psi \, \rangle \;=\; \bmatrix{ \; \alpha \;  \cr
\; \sqrt{1 \,-\,\alpha^2} \; e^{i\, \theta} \; }
$$

thereby requiring two parameters, $\{ \, \alpha \, , \, \theta \, \}$,
for each input qubit state.
Thus a two state input qubit ensemble required an optimization over a space
of dimension five, when the a priori probabilities are included. Three and
four state ensembles required optimization over spaces of dimension
eight and eleven respectively.  

\end{appendix}

\vspace{0.2in}


\end{document}